\documentclass[10pt]{iopart}
\usepackage{epsf,epsfig}

\def\fp {fundamental polyhedron}
\def\uc {universal covering}
\def\flz {Fang \cite{Fan90}}
\def\sos {Sokoloff \& Shvartsman \cite{Sok74}}

\def\ks {Kantowski--Sachs}

\newcommand{\eqn}[1]{eq.(\ref{#1})}
\newcommand{\secn}[1]{\S \ref{#1}}

\def\etc {{\it etc.}}
\def\apriori {{\it a priori~}}
\def\hmpc {h^{-1}~\mbox{Mpc}}
\def\etal {  et al.~}

\def\deg {^\circ}

\def\sol {_{\odot}}
\def\bb {big--bang}
\def\frl{Friedmann--Lema\^\i tre}
\def\hz {Harrison--Zeldovich}
\def\rw{Robertson--Walker}

\def\sw{Sachs--Wolfe}

\def\cobe{{\it COBE}}
\def\rms {{\it rms}}
\def\dtt {\frac{\Delta T}{T}}

\def\apj#1#2{{Ap. J.}, {#1}, #2}
\def\apjs#1#2{{Ap. J. S}, {#1}, #2}
\def\apjl#1#2{{Ap. J. Letters}, {#1}, #2}
\def\aa#1#2{{A\&A}, {#1}, #2}
\def\grg#1#2{{Gen. Rel. Grav.}, {#1}, #2}
\def\jetpl#1#2{{JETP~letters}, {#1}, #2}
\def\mn#1#2{{MNRAS}, {#1}, #2}

\def\prl#1#2{{Phys. Rev. Lett.}, {#1}, #2}
\def\physrevl#1#2{{Phys. Rev. Lett.}, {#1}, #2}
\def\physrevd#1#2{{Phys. Rev. D}, {#1}, #2}

\def\nat#1#2{{Nature}, {#1}, #2}

\def\cqg#1#2{{Class. Quantum Grav.}, {#1}, #2}

\def\th {$^{th}$}
\def\eg {{e.g.~}}
\def\ie {{i.e.}}
\def\los {{ line of sight}}
\def\lss {last scattering surface}
\def\gr {general relativity}
\def\rec {recombination}
\def\mw {Milky Way}
\def\cmb {Cosmic Microwave Background}

\def\xx {{\bf x}}
\def\yy {{\bf y}}

\def\ee {{\bf e}}

\def\kk {{\bf k}}

\def\xx {{\bf x}}

\def\bbbr{{\rm I\!R}} 

\def\bbbh{{\rm I\!H}}

\def\bbbp{{\rm I\!P}}
\def\tvi(#1,#2){\vrule height #1pt depth #2pt width 0pt}

\def\bbbc{{\mathchoice {\setbox0=\hbox{$\displaystyle\rm C$}\hbox{\hbox
to0pt{\kern0.4\wd0\vrule height0.9\ht0\hss}\box0}}
{\setbox0=\hbox{$\textstyle\rm C$}\hbox{\hbox
to0pt{\kern0.4\wd0\vrule height0.9\ht0\hss}\box0}}
{\setbox0=\hbox{$\scriptstyle\rm C$}\hbox{\hbox
to0pt{\kern0.4\wd0\vrule height0.9\ht0\hss}\box0}}
{\setbox0=\hbox{$\scriptscriptstyle\rm C$}\hbox{\hbox
to0pt{\kern0.4\wd0\vrule height0.9\ht0\hss}\box0}}}}
\def\bbbq{{\mathchoice {\setbox0=\hbox{$\displaystyle\rm
Q$}\hbox{\raise
0.15\ht0\hbox to0pt{\kern0.4\wd0\vrule height0.8\ht0\hss}\box0}}
{\setbox0=\hbox{$\textstyle\rm Q$}\hbox{\raise
0.15\ht0\hbox to0pt{\kern0.4\wd0\vrule height0.8\ht0\hss}\box0}}
{\setbox0=\hbox{$\scriptstyle\rm Q$}\hbox{\raise
0.15\ht0\hbox to0pt{\kern0.4\wd0\vrule height0.7\ht0\hss}\box0}}
{\setbox0=\hbox{$\scriptscriptstyle\rm Q$}\hbox{\raise
0.15\ht0\hbox to0pt{\kern0.4\wd0\vrule height0.7\ht0\hss}\box0}}}}
\def\bbbt{{\mathchoice {\setbox0=\hbox{$\displaystyle\rm
T$}\hbox{\hbox to0pt{\kern0.3\wd0\vrule height0.9\ht0\hss}\box0}}
{\setbox0=\hbox{$\textstyle\rm T$}\hbox{\hbox
to0pt{\kern0.3\wd0\vrule height0.9\ht0\hss}\box0}}
{\setbox0=\hbox{$\scriptstyle\rm T$}\hbox{\hbox
to0pt{\kern0.3\wd0\vrule height0.9\ht0\hss}\box0}}
{\setbox0=\hbox{$\scriptscriptstyle\rm T$}\hbox{\hbox
to0pt{\kern0.3\wd0\vrule height0.9\ht0\hss}\box0}}}}
\def\bbbs{{\mathchoice
{\setbox0=\hbox{$\displaystyle     \rm S$}\hbox{\raise0.5\ht0\hbox
to0pt{\kern0.35\wd0\vrule height0.45\ht0\hss}\hbox
to0pt{\kern0.55\wd0\vrule height0.5\ht0\hss}\box0}}
{\setbox0=\hbox{$\textstyle        \rm S$}\hbox{\raise0.5\ht0\hbox
to0pt{\kern0.35\wd0\vrule height0.45\ht0\hss}\hbox
to0pt{\kern0.55\wd0\vrule height0.5\ht0\hss}\box0}}
{\setbox0=\hbox{$\scriptstyle      \rm S$}\hbox{\raise0.5\ht0\hbox
to0pt{\kern0.35\wd0\vrule height0.45\ht0\hss}\raise0.05\ht0\hbox
to0pt{\kern0.5\wd0\vrule height0.45\ht0\hss}\box0}}
{\setbox0=\hbox{$\scriptscriptstyle\rm S$}\hbox{\raise0.5\ht0\hbox
to0pt{\kern0.4\wd0\vrule height0.45\ht0\hss}\raise0.05\ht0\hbox
to0pt{\kern0.55\wd0\vrule height0.45\ht0\hss}\box0}}}}
\def\bbbz{{\mathchoice {\hbox{$\sf\textstyle Z\kern-0.4em Z$}}
{\hbox{$\sf\textstyle Z\kern-0.4em Z$}}
{\hbox{$\sf\scriptstyle Z\kern-0.3em Z$}}
{\hbox{$\sf\scriptscriptstyle Z\kern-0.2em Z$}}}}

\begin{document} 

\title {{\bf COSMIC TOPOLOGY}}

\author{{\bf Marc Lachi\`eze-Rey}\\
CNRS UPR-182 \\
CEA, DSM/DAPNIA/ Service d'Astrophysique \\
CE Saclay, F-91191 Gif--sur--Yvette CEDEX, France \\
and \\
{\bf Jean-Pierre Luminet } \\
CNRS UPR-176 \\
D\'epartement d'Astrophysique Relativiste et de  Cosmologie \\
Observatoire de Paris--Meudon, F-92195 Meudon Cedex, France}

\baselineskip = .85 truecm

\begin{abstract} 
General relativity does not allow to specify the topology of space, leaving the
possibility  that space is multi-- rather than simply--connected. We
review the main mathematical properties of multi--connected spaces, and
the different tools to classify them and to analyse their properties. 
Following  the mathematical classification, we  describe the different
possible muticonnected spaces which may be used to construct universe
models. We briefly discuss some implications of multi--connectedness  for
quantum cosmology, and its consequences concerning quantum field theory
in the early universe.   We  consider in details  the properties of the
cosmological models  where space is multi--connected, with emphasis
towards     observable effects. We then review the analyses of
observational results obtained in this context, to search for a
possible signature of multi--connectedness, or to constrain the models.
They may concern the distribution of images of cosmic objects like
galaxies, clusters, quasars,..., or more global effects, mainly those
concerning the \cmb, and the present limits resulting from them.  
 
\end{abstract} 

\newpage
\tableofcontents
\newpage

\begin{verse}
{\sl For in and out above, about, below\\
It is nothing but a Magic Shadow-Show\\
Play'd in a Box whose candle is the Sun\\
Round which we Phantom Figures come and go.}\\
Omar Khayyam, $XII^{th} century$
\end{verse}

\section {Introduction}

\indent
Topology plays to differential geometry
a role somewhat like quantum theory to classical physics \cite{Ati88}.
Both lead from continuous to the discrete, and at their levels relationships are
more global and less local. 

Topology can be applied in particular to cosmology. The
purpose of relativistic cosmology is to deduce from the Einstein's field
equations some physically plausible models of the universe as a whole.
However, such a program cannot be completed within the framework of
general relativity only : Einstein's equations being partial differential
equations,  they describe only {\sl local} properties of  spacetime. The
latter are entirely contained in the metric tensor $g_{ij}~(i,j = 0,1,2,3)$,
or  equivalently in the   infinitesimal  distance element $ds$~such that
$ds^2 = g_{ij}dx^idx^j$.  But Einstein's equations  do not fix the {\sl
global} structure -- namely the topology -- of the spacetime : to a  given
local metric element correspond several -- generally an infinite number --
of topologically distinct universe models.\footnotemark[1] \footnotetext[1]{~The
expression ``cosmic topology" is occasionally used by some authors, e.g.
\cite{Rho94}, to discuss the large scale distribution of matter in the
universe. Here we place at the more fundamental level of spacetime
global structure.}

As soon as 1917, after Einstein found \cite{Ein17} the first cosmological
solution of general relativity -- namely a static model with
three--dimensional spheres $\bbbs^3$ as spatial sections~-- de Sitter
\cite{deS17a} had already noticed that the solution could also fit with a
variant form of spherical space : the  three-dimensional projective (or
elliptical) space $\bbbp^3$, constructed from the 3--sphere $\bbbs^3$ by
identifying diametrically opposite points. 
The projective space has the same metric than the spherical space, but a
different topology, with half the volume.

The discovery of non static cosmological solutions of general
relativity \cite{Fri22,Lem31} enriched considerably the
field of modelisation. According to the well known
picture, the spatially homogeneous, isotropic 
Friedmann-Lema{\^\i}tre universe models (hereafter
denoted FL) admit spatial  sections of the spherical, 
Euclidean or hyperbolic type according when the (constant spatial)
curvature is positive, zero or  negative. Although it was
soon recognized by Friedmann \cite{Fri24}, Lema{\^\i}tre 
\cite{Lem58} and a few others \cite{Hec62} that the FL metrics with zero
or negative curvature admitted spatially closed topologies, the idea of
multi-connectedness has not attracted much support. Pioneering work in
cosmic topology by Ellis \cite{Ell71}, Sokoloff and Schvartsman
\cite{Sok74}, Zeldovich \cite{Zel77}, Fang and Sato \cite{Fan85},
Fagundes \cite{Fag85} and some others have remained widely ignored,
and in almost all cosmological studies and classical textbooks, e.g. 
\cite{Wei72}, it is implicitly assumed that the topology of space is
simply--connected, namely that of the finite  hypersphere $\bbbs^3$, of the
infinite Euclidean space $\bbbr^3$ or of the infinite hyperbolic space
$\bbbh^3$, without even mentionning the multi--connected alternatives. 
 This arbitrary simplification is	at the origin of a common belief of
modern cosmology according to  which, in order to know if space is
finite or infinite, it would be sufficient to determine  the sign of its
spatial curvature, or equivalently to compare  its energy density to the
critical ``closure" value\footnotemark[1] \footnotetext[1]{~The denominations ``closed" and
``open" commonly used for the spherical and the Euclidean/hyperbolic FL
universes contribute to the confusion : they apply correctly to {\sl time},
not to space.}.
  Present astronomical data indicate that the
energy density parameter  in the observable Universe is less
than the critical value, but this {\sl does not} exclude closed
space in FL solutions,  with or without a cosmological constant.

 Now one can ask why the universe should not have the simplest
topology. Some authors use the philosophical  ``principle of
economy'' to exclude complicated topologies, but this  principle is so
vague that it can also be  invoked to promote the contrary, for instance
the topology which gives the smallest volume \cite{Hay90}~! Indeed, 
quantum cosmology provides some new insights on this question. For
instance, the  spontaneous birth of the universe from quantum vacuum
requires the universe to have compact spacelike hypersurfaces (see e.g. 
\cite{Zel84}), and the probability is bigger for spaces of smaller volume. 
Since the observations suggest that the universe  is locally Euclidean or
hyperbolic, then its spatial topology must be non trivial.  More generally,
the closure of space is considered as a necessary condition in quantum
theories of gravity \cite{Haw84}. 

This review will be mainly pedagogical. Since many cosmologists 
 are unaware of how topology and cosmology can fit together and
provide new highlights in universe models, we aim to present here the
"state-of-the-art" of cosmic topology in a non--technical way. The review is
organized in the following manner.

 In section \ref{stor}, we examine whether there are any
physical arguments suggesting that realistic universe models
must be time-oriented and/or space-oriented. 

The section \ref{topo} is devoted to the mathematical aspects of the
topological classification of manifolds. Some elementary techniques
are supplied to the reader and are applied in section \ref{class2} to the
classification of Riemannian surfaces.

Section \ref{homog} is devoted to 3-dimensional homogeneous
manifolds, and sections \ref{flatm} - \ref{positivem} - \ref{negativem}
describe the topological classification of spaces of constant curvature
--those directly involved in realistic universe models. 

In sections \ref{Cosmic models} - \ref{sec10} we discuss the 
properties of multi-connected cosmological models, both at a quantum and
at a classical level. The last two sections are devoted to the possible
observational effects of multi-connectedness in the distribution of
discrete sources (section \ref{sec11}) and in the distribution of
continuous fields and backgrounds, in particular the Cosmic Microwave
Background (section \ref{cmb}).

\section {Spacetime orientability}\label{stor}

The solutions of the equations of general relativity are 
spacetimes $({\cal M}_4,{\bf  g})$, namely 4-dimensional  manifolds 
endowed with a Lorentzian metric\,\footnotemark[1] \footnotetext[1]{~That is, a
pseudo-riemannian metric with signature $(-+++)$} $g_{ab}$. 
This condition
is not very  restrictive, due to the following theorems (see, e.g., \cite{Ger79})~:

\noindent - any {\sl non-compact} 4-manifold admits a Lorentzian metric

\noindent - a {\sl compact} 4-manifold admits a Lorentzian metric if and
only if its Euler-Poincar\'e characteristic\,\footnotemark[1] \footnotetext[1]{~The Euler-Poincar\'e
number for ${\cal M}_4$ is $\chi({\cal M}_4) = \sum_{n=0}^{4} (-1)^n
B_n$, where $B_n \ge 0$ is the $n^{th}$ Betti number of ${\cal M}_4$
\cite{Ste51}} is zero.

The range of possible topological structures compatible with a given
metric solution of Einstein's equations thus remains huge, but it is clear
that most of the Lorentzian 4-manifolds have no
physical  relevance : the building of realistic universe models sets
additional restrictions. To begin, spacetime manifolds $\cal M$$_4$  with a
boundary, or manifolds which are non--connected are
likely to be eliminated. We shall assume also the manifolds to be
inextendible, to ensure that all non-singular points of spacetime are
included. Next come into play the conditions of time and space orientability,
that we discuss in the following. More technical definitions are available
elsewhere, for instance \cite{Haw73}.

\subsection {Time orientability}

In the Minkowski spacetime of special relativity,   particles follow 
worldlines from the past to the future. At any event one can define a 
class of future--oriented vectors and a class of past--oriented
vectors. This property of {\sl local} time orientability still holds
in the  curved  spacetimes of general relativity, because special
relativity remains locally  valid. 
However, in order to define a {\sl global} time orientation, that is, valid 
throughout the entire spacetime, the choices of local time
orientations  must be consistent. Namely they must vary continuously
along the trajectories and, for closed trajectories, the final
orientation must remain the same as the initial  one.

Fortunately, to ensure the required consistency it is sufficient to test it
only along certain classes of closed curves. Let us consider an arbitrary point
$p \in {\cal M}_4$ with a closed curve $\gamma$ passing through $p$.
Let fix an initial time orientation at $p$ and carry it continuously along
$\gamma$. If, when returned at $p$, the orientation has not changed, the
curve $\gamma$ is said to be time--preserving. By definition, a
spacetime $({\cal M}_4,\bf g)$ is time-orientable if and only if
every closed curve is time-preserving. 

\subsection {Causality} 

The notion of causality, which intuitively requires that the cause
precedes the effect, is a rule imposed by logics and common sense,
not by the theory of  relativity.  Causality is implicit in special relativity,
 because the travel into the past is strictly
equivalent to a motion along spacelike curves, which is forbidden
for real particles. On the contrary, in general relativity, certain subtle
distortions imprinted on curved spacetime by particular gravitational fields --
for instance the one generated by a rotating black hole or a wormhole -- could
in principle authorize the exploration of the past while remaining inside the
future light cones (\cite{Mor88,Got91}, and for a semi-popular
account \cite{Lum92}). 

\begin{figure}[tb]
  \begin{center}
    \leavevmode
    \includegraphics{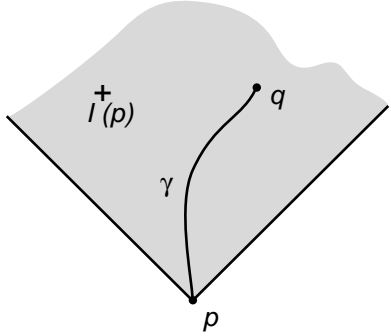}
\caption{\label{Fig1}{\it Chronological future in spacetime}}
  \end{center}
\end{figure}

However the common experience \footnotemark[1] \footnotetext[1]{At a classical level. When
quantum physics is involved, see e.g. \cite{Deu91}} shows that, locally,
different observers perceive a same preferred time--direction. In order to
construct step-by-step a global time orientation, the physicist first defines a
chronology. Given two events $p$ and $q$ in ${\cal M}_4$, $p$ is said to
precede $q$ $( p \prec q)$ if there exists a continuous, timelike, future-oriented
curve $\gamma$ from $p$ to $q$. The chronological future ${\cal I}^+(p)$
(resp. past ${\cal I}^-(p)$) of $p \in {\cal M}_4$ is the set of points $\{q\in
{\cal M}_4, p \prec q \}$ (resp. $\{q\in {\cal M}_4, q \prec p \}$). For instance,
in Minkowski spacetime, ${\cal I}^+(p)$ is merely the interior of the future
light-cone at $p$ (figure \ref{Fig1}). 

However the chronological past and future sets may be quite
pathological. This is for instance the case with the portion of
Minkowski spacetime obtained by the temporal identification 
$(-1,x_1,x_2,x_3) \equiv (+1,x_1,x_2,x_3)$, where every event both belongs
to its own future and past sets. More generally, if  $p \in {\cal I}^+(p)$ for
some $p \in  {\cal M}_4$, the spacetime manifold contains closed
timelike curves. This is the case for {\sl any compact}
spacetime,  and also for some non-compact spacetimes such as 
the G\"odel and the Taub-NUT solutions \cite{Haw73}. 
From the point of view of physics, all these manifolds are causally
misbehaved and are generally ruled out as realistic universe models,
although not as solutions of general relativity.  

In fact, the absence of anomalies in causality is expressed fairly well by
the condition of {\sl stable causality}~:  a spacetime is stably
 causal if it admits a cosmic time function, that is a continuous real
function $T : {\cal M}_4 \mapsto \bbbr$, whose gradient $\nabla T$
is everywhere timelike~:  $ {\bf g}(\nabla T(p),\nabla T(p)) < 0,~ 
\forall p \in {\cal M}$$_4$.

 The usual spacetimes (Minkowski,
Schwarzschild, Friedmann) are stably causal. Stable
causality implies global time orientability, because the time function
$T$ must necessarily increase along future-oriented, null or
timelike curves, and prohibits the changes of orientation along
closed curves. It also allows to ``slice'' the spacetime into 
hypersurfaces of constant time function, and thus to split 
the spacetime metric into

\begin{equation}\label{3+1}
g_{ab} = - n_a n_b + h_{ab},
\end{equation}

where $n_a$  is the future directed normal to the hypersurface of constant
time.

 \subsection {Global hyperbolicity}

The structure of physical laws generally requires that the evolution
of a  system can be determined from the knowledge of its state at a
given  time. This is the case in classical
mechanics, where the  trajectory of a point mass is entirely specified
by its initial position and velocity, or in quantum
mechanics, where the  Schr\"odinger equation calculates the future
states knowing the present wave  function.

General relativity theory also possess  such a property. It is 
convenient to introduce the notion of domain of dependence. Given
an  initial spatial hypersurface $\Sigma$, its future domain of dependence
${\cal D}^+(\Sigma)$
 (resp. past domain of dependence ${\cal D}^-(\Sigma)$) is the set of
points $p$ such that any timelike curve reaching $p$ (resp. starting
from $p$) intersects $\Sigma$.  The union  ${\cal D}(\Sigma) = {\cal
D}^+(\Sigma) \cup {\cal D}^-(\Sigma)$ is thus the region of spacetime which
is entirely  determined by the ``information'' on $\Sigma$. The problem of
initial data in  general relativity \cite{Cho62,Fis79} is  reduced to the 
question of knowing the nature of the data on $\Sigma$ that specify the physics
in  ${\cal D}(\Sigma)$. These required initial data  are determined by fixing
the induced spatial metric $h_{ab}$ on $\Sigma$  and its normal derivative
$K_{ab} = \frac{1}{2} L_n h_{ab}$, called the extrinsic curvature
of $\Sigma$ in ${\cal M}$$_4$.

An hypersurface $\Sigma$ whose domain of dependence
$D(\Sigma)$ is the whole manifold ${\cal M}$$_4$   is
called a Cauchy surface. For instance, the hyperplane  $\{ t = 0\}$ in
Minkowski spacetime is a Cauchy surface.  A  spacetime which admits a
Cauchy surface is said to be {\sl globally hyperbolic}. 
A globally hyperbolic spacetime is necessarily stably causal and
time--oriented (i.e., has a global time function which increases on any
timelike or null curve). It is diffeomorphic to $\bbbr \times {\cal
M}_3$, where ${\cal M}_3$ is a 3-dimensional  riemannian manifold (with
positive definite metric).

The condition of global hyperbolicity sets
severe constraints on spacetime,  but it is difficult to justify it on
physical grounds, except if we believe in strong determinism, \ie,
the wish that the entire spacetime can be calculated from the
information on a single hypersurface. However we shall assume it thereafter.

\subsection {Space orientability and CPT invariance}

\begin{figure}[tb]
  \begin{center}
    \leavevmode
    \includegraphics{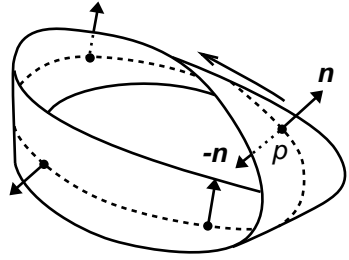}
\caption{\label{Mobius}{\it The non-oriented M\"obius band}}
  \end{center}
\end{figure}

Assuming global hyperbolicity, the
search for the topology of the real spacetime reduces to the exploration  of 
the possible topologies of the spatial hypersurfaces $\cal M$$_3$ of constant
time function \cite{Sok71}. May we impose additional restriction on the
topology of $\cal M$$_3$ by assuming space orientability ?
The latter can be defined in
a variety of ways. It has its origine in the simple observation of surfaces :
two--sided surfaces are called orientable because we can use their
two--sidedness to define an orientation or a direction in $\bbbr^3$. This is
not possible with one--sided surfaces. The simplest example, the
notorious M\"obius strip, is obtained by taking a rectangle and joining two ends
having first twisted one of the ends. If one takes a normal
${\bf n}$ to the surface at a point $p$ and moves it continuously around
the surface until it returns to $p$, it will then point in the
direction ${\bf - n}$  (fig. \ref{Mobius}). This is a sign of
non--orientability. The following definition arises : any two-dimensional
manifold lying in $\bbbr^3$ is orientable if and only if it is two-sided. 

This can be generalized to higher dimensions. At  any point of the spacetime,
one can define two classes of spacelike triads~:  the left--handed class
and the right--handed class. The spacetime is space--orientable  if
every closed curve preserves the spatial parity.

If  time orientability can be justified on physical grounds such as  the
existence of an ``arrow of time'', the requirement of spatial
orientability is less  stringent. In particle physics, the CPT theorem
\cite{Str64} states that any relativistic quantum field theory must be
invariant under the combination of charge conjugation C, space
inversion P and time reversal T. The CPT invariance is also
satisfied in some versions of quantum cosmology, for instance the
no--boundary proposal \cite{Har83}, although some authors
\cite{Pen79} have questioned whether a full quantum gravity theory
might not violate it. 
Until three decades ago it was commonly 
believed that the laws of physics were separately invariant under C, P
and T transformations. Then it was experimentally
discovered \cite{Wu57} that the weak interaction  violated the
parity  symmetry P (or, equivalently, the CT product).  Next,  CP  appeared to
be violated in the decay of the $K_{0}$ meson \cite{Chr64}, and other CP
violations are now currently researched \cite{Win93}.  
 This series of results thus suggested that the laws of
physics were not even invariant under time reversal T.  

CT non invariance allows one to distinguish two possible orientations of
${\cal M}_3$ \cite{Zel67,Zel73}, whereas CP non invariance allows one to
distinguish two possible orientations of timelike vectors given on  ${\cal
M}_3$ \cite{Sok71}.
This line of arguments leads to the strong conclusion that our spacetime
must be  total--orientable. Since we have assumed, from  global
hyperbolicity, that spacetime is already time--orientable, we
conclude that the physical space must be orientable. Also, the splitting ${\cal
M}_3 \times \bbbr$ with ${\cal M}_3$ spacelike and orientable ensures well
defined spinor fields, required by elementary particle theories to describe the
variety of species of particles in the Universe \cite{Pen84}.  We shall
thus adopt this simplification in the remaining of the article.

\newpage
\section {Basic topology for Riemannian manifolds}\label{topo}

\subsection {What is topology ?} 

In simple words,  topology is the mathematical framework within
which to study continuity : the topological properties are  those which
remain insensitive to continuous transformations. Thus, size and
distance are in some sense ignored in topology :  stretching, squeezing
or ``kneading'' a manifold change the metric but not the topology; 
cutting, tearing or making holes and handles change the latter. As a
consequence, a topologist does not distinguish a triangle, a square and a
circle; or a soccer ball and a rugby ball; even worse, a coffee cup
and a curtain ring are the same topological entity. However, he is able
to recognize the difference between a bowl and a beer mug~: due
to its handle, the mug  cannot be continuously deformed into the bowl or
into the 2--sphere $\bbbs ^2$.

\begin{figure}
\centerline{\epsfig{file = 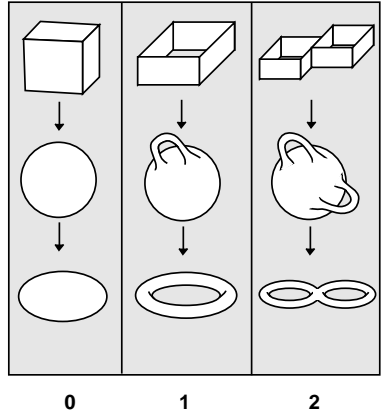,width=5cm}}
\caption{\label{F3}{\it Classes of homeomorphic surfaces. The digits
 below the columns denote the number of holes, a topological invariant.}}
\end{figure}

Continuous transformations are mathematically depicted
 by {\sl homeomorphisms}.  If we
consider two manifolds
 $\cal M$$_1$ and $\cal M$$_2$, a homeomorphism
is a continuous map $\Phi : \cal M$$_1  \mapsto \cal M$$_2$ which
has an inverse also continuous. Homeomorphisms
allow one to divide the set of all possible manifolds into topologically
equivalent classes : two manifolds $\cal M$$_1$ and $\cal M$$_2$ belong to
the same topological class if they are homeomorphic (figure \ref{F3}).  

The topologist's work is to fully characterize 
all equivalence classes defined from homeomorphisms and to place the
manifolds  in their appropriate classes. However this task is still  unachieved,
excepted in some restricted cases such as two--dimensional closed
surfaces (section 4), three--dimensional flat (section 6) and spherical
(section 7) spaces.

It is often possible to visualize two--dimensional manifolds by
representing them as embedded in three--dimensional Euclidean
space (such a  mapping does not necessarily
exist however, see below). Three--dimensional manifolds require the
introduction of more abstract representations, like for instance the {\sl
fundamental domain}. For the sake of clarity let us illustrate
our purpose by some elementary examples \cite{Sei30,Hil52}. 

\subsection {Stories of tori}

\subsubsection{The two--dimensional simple torus}\label{2torus}

\begin{figure}
\centerline{\epsfig{file = 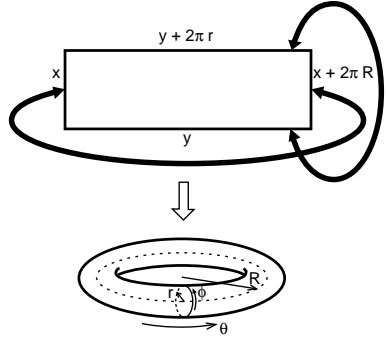,width=5cm}}
\caption{\label{tor}{\it Construction of the flat torus}}
\end{figure}

\begin{figure}
\centerline{\epsfig{file = 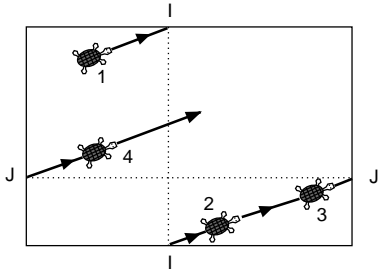,width=5cm}}
\caption{\label{turtle}{\it The turtle's walk on a flat torus}} 
\end{figure}

It has been shown since the nineteenth century that the different topological 
surfaces can be represented by polygons
whose edges are suitably identified by pairs. Identifying one pair of opposite
edges of a square gives a portion of a cylinder; then, stretching the portion of
cylinder and gluing together the two circular ends generates a simple torus, a
{\sl closed} surface (figure~\ref{tor}). The torus is thus topologically equivalent to a rectangle
with opposite edges identified. The rectangle is
called a {\sl fundamental domain} of the torus.  From a topological
point of view (namely without reference to  size),
the fundamental domain can be chosen in different ways (a
rectangle, a square, \ldots). If a turtle
moves on a fundamental domain of the torus,  as soon as it crosses
the upper edge of the domain at a given point, it reappears on  the
lower opposite edge at a so-called ``equivalent point'' (figure
\ref{turtle}). Many computer games where
the screen plays the role of a fundamental domain are indeed played onto the
surface of a torus. 

This illustrates  the difference between the metric and
the topology. The torus $\bbbs ^1 \times \bbbs^1$, obtained by
identification of the opposite edges of a square, is geometrically different
from an usual torus $T_1$ (the surface of a ring for instance), which is a 
subset of $\bbbr^3$. The latter is not flat and has varying curvature, whereas 
$\bbbs ^1 \times \bbbs^1$  is flat everywhere and cannot be properly
visualized because it cannot be immersed in  $\bbbr ^3$.  It is only
topologically speaking that these two tori are the same because there is an
homeomorphism between them.

As food for thought we provide a more precise, although
elementary, statement of this. The usual torus $T_1$ can be endowed with a
natural riemannian metric $g_{ij}$ by taking the Euclidean metric in $\bbbr
^3$ and imposing the restriction that the points in $\bbbr ^3$ lie on the torus. In
polar coordinates we obtain for instance 
\begin{equation}
ds^2 = (R+rcos\phi)^2 (d\theta ^2 + r^2 d\phi ^2) 
\end{equation}
 (for the torus obtained by rotation of a circle of radius r along a
circle of radius R). With respect to the metric $g_{ij}$ we have a
curved torus, with a Gauss curvature
 \begin{equation}
K = \frac{\cos \phi}{r (R+ r \cos \phi)}. 
\end{equation}
 On the other
hand, the same manifold can also be given a different metric by
defining a new distance between two points  $t = (x,y)$ and $t' = (x',y')$ 
 as  $d(t,t') = \bigl[(x-x')^2 +(y-y')^2\bigr]^{1/2}$.
 The metric becomes 
 \begin{equation}
ds'^2 = ad\theta ^2 + 2b d\theta d\phi +c d\phi ^2~~~(a, b, c~ constants).
\end{equation}
With respect to this metric, the torus is flat.
But it cannot lie in  $\bbbr ^3$, because any
two-dimensional compact connected surface in $\bbbr ^3$
must have at least one point of non zero curvature.

\subsubsection {The two--dimensional g--torus} 

\begin{figure}[tb]
  \begin{center}
    \leavevmode
    \includegraphics{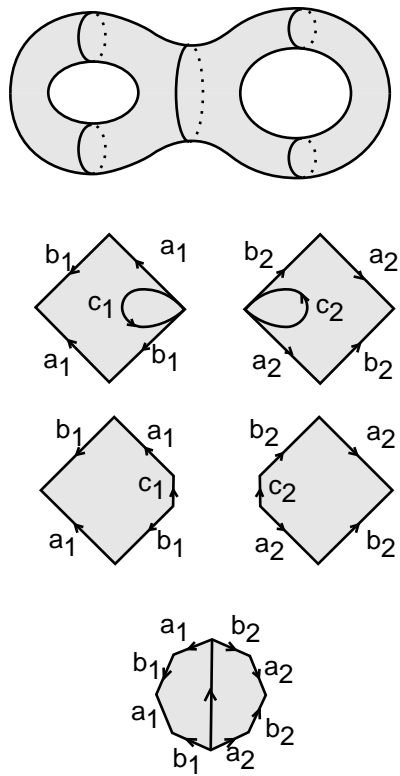}
\caption{\label{gtorus}{\it The two-torus as the connected sum of two
simple tori.}}
  \end{center}
\end{figure}

The gluing method described above becomes extremely fruitful
when the surfaces are more complicated. A two--dimensional g--torus
$T_g$ is a torus with $g$ holes. The term ``pretzel" is sometimes used in the
English litterature, but the French ``fougasse" (a delicious kind of bread
from Provence) is still more picturesque.  $T_g$ can be constructed as
the {\sl connected sum}\footnotemark[1] \footnotetext[1]{~More
generally, a connected sum of two n--dimensional manifolds $\cal
M$$_1$ and $\cal M$$_2$ is formed by cutting out a n--ball from each
manifold and identifying the resulting boundaries to get $\cal M$$ =
\cal M$$_1 \# \cal M$$_2$.} of $g$ simple tori (figure
\ref{gtorus}). The g--torus is therefore topologically equivalent to a
connected sum of $g$ squares whose opposite edges have been
identified. This sum is itself topologically equivalent to a 4g--gone
where all the vertices are identical with each other and the sides are
suitably identified by pairs .  

It would be tempting to visualize the g--torus by gluing together
equivalent  edges, like for the simple torus. But such an operation is not 
straightforward when $g \ge 2$. All the vertices
of the polygon correspond to the same point of the surface. Since
the polygon has at least 8 edges, it is necessary to make the
internal angles thinner in order to fit them suitably around a single
vertex. This can only be achieved if the polygon is represented in the
{\sl hyperbolic plane} $\bbbh^2$ instead of the Euclidean plane $\bbbr^2$ :
this increases the area and decreases the angles. The more
angles to adjust, the thinner they have to be and the greater the surface. The
g-torus $(g \geq 2)$ is therefore a {\sl compact surface of negative
curvature}.

More generally -- as we shall detail in section 4 --, the sphere
$\bbbs^2~ (g = 0)$ and the g--torus $T_g ~(g \ge 1)$ are the only
possible  compact oriented (two--sided) surfaces. $g$ is called the
{\sl genus} of the surface. The non--oriented surfaces are similarly
defined by their genus. The major triumph of topology in the nineteenth
century was the complete classification of all compact surfaces in terms
of two and only two items of data : the number of holes $g$ and the
orientability / non--orientability property.

It may be useful here to recall the link between the genus and
the  Poincar\'e-Euler characteristic (whose general definition was given in 
footnote 4). Any compact surface can be triangulated by a polyhedron.  If V is
the number of vertices, S the number of edges and F the number of faces, then
the Poincar\'e-Euler characteristic reduces to $\chi = V - S + F$. It is a
topological invariant, related to the genus by $g = 1 - \chi/2$.

\subsubsection {The three--dimensional torus}

When one deals with more than two dimensions, the gluing method
remains the simplest way to visualize spaces. By analogy with the
two-dimensional case, the three-dimensional simple torus $\bbbs ^1 \times
\bbbs^1 \times \bbbs^1$ (also referred to as the {\sl hypertorus}) is
obtained by identifying the opposite faces of a parallelepiped such that $x =
x+L_1,~ y = y+L_2, ~z = z+L_3$. The resulting volume is finite, equal to $L_1
\times L_2 \times L_3$. Let us imagine a light source at our position,
immersed in such a structure. The light emitted backwards crosses the face
of the parallelepiped behind us and reappears on the opposite face in front  of
us; therefore, looking forward we can see our back (as in the spherical
Einstein's universe model). Similarly,  we
see in our right our left profile, or upwards the  bottom of
our feet. In fact, as light propagates in all  directions, we would observe an
infinity of ghost images of any object  viewed under all angles. The
resulting visual effect would be comparable (although not identical) to
what  could be seen from inside a parallelepiped whose internal sides
are covered with mirrors. Thus one would have the visual
impression  of infinite space, although the real space is closed.  The beautiful
popular article by Thurston and Weeks  \cite{Thu84} provides a striking
illustration of such a space.

More generally, any three dimensional compact manifold
can be represented as a polyhedron -- what we define later
more precisely as the fundamental polyhedron (hereafter $FP$) -- whose
faces  are suitably identifyied by pairs. But, as soon as the number of faces of a
$FP$ exceeds 6, the compact manifold resulting from
identifications  cannot be developed into the Euclidean space $\bbbr^3$ : the
$FP$ must be built  in hyperbolic space $\bbbh^3$ in order to adjust all the
angles at vertices.

\subsection {Metric, Curvature and Homogeneity}

\subsubsection {Metric tensor}

In a n-dimensional manifold $\cal{M}$, points
are represented in a general coordinate system $x^i  ~(i=1,2,..., n)$. A
coordinate line passing through 
 a given point $P$ is obtained by varying a coordinate $x^k$
while keeping the  other ones constant. The set $\{ {\bf e}_k \}$ of vectors
tangent to the $n$  coordinate lines at $P$ constitute a
 basis called the natural frame at $P$. A point $P'$ infinitesimally
close to $P$ is separated by a distance  $ds = \vert P'- P \vert$ 
such that $ds^2 = g_{ij} ~dx^i ~dx^j$. The  $g_{ij}$, which
depend on
 coordinates  $x^k$, are the components of the metric tensor, which
is symmetric $(g_{ij} =  g_{ji})$. 

In the natural frame
$\{ {\bf e}_k \} $ at $P$, the infinitesimal displacement vector is ${\bf 
P'}- {\bf P}  = {\bf e}_k ~dx^k$ with ${\bf e_i.e_j} = g_{ij}$. The
natural frame $\{ {\bf e'}_k \}$ at $P'$ can be deduced from the natural
frame  $\{ {\bf e}_k \}$ at $P$  by ${\bf e'}_i  = {\bf e}_i +
\Gamma^j_{ik} dx^k e_j$.  The coefficients $\Gamma^j_{ik}$ (called 
Christoffel symbols) are functions of the partial  derivatives of the
metric tensor components, given  by 
\begin{equation}
\Gamma^i_{jk} = {1\over 2} g^{il} \Bigl( {\partial
g_{lj}\over \partial x^k} + {\partial g_{lk}\over \partial x^j} - {\partial
g_{jk}\over \partial x^l}\Bigr) 
\end{equation}
where $g^{ik}g_{kj} \equiv \delta ^i_j.$

\subsubsection {Curvature}

In any metric space, one can define the quantities 
\begin{equation}
R^l_{ijk} = {\partial \Gamma^l_{ik} \over\partial x^j} - 
{\partial \Gamma^l_{ij}\over\partial x^k }+ \Gamma^m_{ik}
\Gamma^l_{mj} -  \Gamma^m_{ij} \Gamma^l_{mk}
\end{equation}
which constitute the components of the Riemann curvature tensor.
The  latter contains all the information on the local geometry of the
space at any given point.  In Euclidean space, 	all the $R^l_{ijk}$
vanish identically at every point, which means that the construction of the
natural frame in  $P'$ does not depend  on the path from $P$ to $P'$. 

Describing the curvature involves ``contractions" of the Riemann tensor : 
the Ricci tensor and the scalar curvature are respectively given by :  
\begin{equation}
	R_{ij} = R^k_{ikj}~, ~~~ÊR = g^{ij}R_{ij}.
\end{equation}
The components of the curvature tensor are not all independent.  The number
of independent components of $R^l_{ijk}$ is ${1\over 12} n^2(n^2 - 1)$,
where $n \ge 2$ is the dimension of the manifold. 

Thus for surfaces there is only one independent component, say $R_{1212}$.
The Ricci tensor and the scalar curvature are respectively
 \begin{equation}
R_{ij} = g_{ij} {R_{1212}\over det(g_{ij})}~, ~~~ÊR = 2
{R_{1212}\over det(g_{ij})}
\end{equation}
R is just  (to a $-~\frac{1}{2}$ historical factor) the usual Gaussian
curvature of the surface.

In three dimensions there are six independent components. However they do
not describe the curvature in an invariant manner, that is independent of the
chosen coordinate system. The invariant characterization must be
formulated in terms of 3 scalars constructed from  $R^l_{ijk}$ and $ g_{ij}$.
At any point P of the
space one can define the Ricci principal directions or {\sl sectional
curvatures}, given by the roots $K_p$ of the characteristic equation
$det(R_{ij} - \lambda g_{ij}) = 0$, namely~:
\begin{equation}\label{Kp}
K_p \equiv \Bigl(~R, ~~R_{ij}R^{ij},~~det(R_{ij})/ det(g_{ij}) \Bigr)
\end{equation}

In any dimension, a space where the relation 
\begin{equation}
R_{ij} = \lambda~ g_{ij},  ~~~\lambda = const
\end{equation}
 holds everywhere is said  to have a {\sl constant curvature}. In
dimension 3, the sectional curvatures (\ref{Kp}) are then all equal :
they depend only on the point, not on the directions.

\subsubsection {Homogeneous spaces}\label{homogen}

We have seen as an introductory example that the
two--dimensional torus $\bbbs^1 \times \bbbs^1$ has the  topology 
of a square with opposite edges fitted together. It is thus a locally Euclidean
space with constant zero curvature. Generally speaking, spaces with constant
curvature (zero, positive or negative) have ``nice" metrics in the sense
that an observer will see the same picture wherever he stands and in
whichever direction it looks. It was shown in the
last century that any connected closed surface is homeomorphic to a
Riemannian surface of constant curvature (ref. \cite{Gug63}, chap.11). This
major result implies that there are only {\sl three} types of
two--dimensional geometries, corresponding to the possible signs of
their curvature~: locally spherical, Euclidean, or hyperbolic. 

The situation is more complicated in 3 dimensions. Obviously there are still
the three constant curvature geometries $\bbbs^3$, $\bbbr^3$ and
$\bbbh^3$. But  the three--dimensional cylinder $\bbbs^2 \times
\bbbr$ is not homeomorphic to $\bbbs^3$ or $\bbbr^3$. It can be endowed
with a natural metric which is the product of the standard metrics
of $\bbbs^2$ and $\bbbr$, but this metric is {\sl anisotropic}~: for
an observer at a given point of $\bbbs^2 \times \bbbr$, the manifold 
appears different in different directions; however the metric is still
homogeneous in the sense that the manifold will look the same at different
points. 
This simple example clearly shows that the three-dimensional spaces of
constant curvature are just a very special case of more general  {\sl
homogeneous} spaces. As we shall see in more details in
section 5,   there are  {\sl eight} types of homogeneous
three--dimensional ``geometries", five of them not admitting a metric of
constant curvature \cite{Thu79,Thu82}.

Let us give mathematical substance to these notions. Quite generally, to any 
manifold $({\cal M},{\bf g})$ is associated a group G of {\sl
isometries},  i.e., transformations which leave the meric invariant. The
manifold  ${\cal M}$ is said {\sl homogeneous} if  G is non
trivial\,\footnotemark[1] \footnotetext[1]{~As we shall see below, the concept of homogeneity used
in relativistic cosmology requires $dim(G)~\ge~3$}. 

The group G is said to act transitively on $\cal M$  if, for any points $\bf x$
and $\bf y$ in $\cal M$ there is an isometry $g \in G$ such that $g({\bf x}) =
{\bf y}$. The set $H$ of all points $\bf y$ in $\cal M$ such that $g({\bf
x}) = {\bf y}$ for some $g \in G$ is called the
orbit of $\bf x$.  The subgroup of isometries  which leave a point
$\bf x$ fixed (for instance a rotation
in Euclidean space) is the {\sl isotropy group} $I$  at $\bf x$.

 We have (theorem)~:
\begin{equation}
dim(G) = dim(H) + dim(I).
\end{equation}
If $dim(H) = dim(G)$,  G is called {\sl simply transitive} on $H$ (the
transformation $g$ such that $g({\bf x}) = {\bf y}$ is unique for any
$\bf x$ in $\cal M$).  

If $dim(H) < dim(G)$, G is called {\sl multiply
transitive}.

For a n--dimensional manifold, the dimension of its full isometry
group G must be $dim(G) \leÊ\frac{n(n+1)}{2}$ \cite{Eis26}. Thus, for
surfaces, $dim(G) \le 3$, and for three--dimensional Riemannian
spaces, $dim(G) \le 6$. When the dimension of the isometry group is
maximum, the space is called maximally symmetric \cite{Wei72}. The
following theorem holds~: a n--dimensional manifold is maximally
symmetric iff it has constant curvature. 
 
In general relativity, manifolds are
spacetimes  ${\cal M}_4$, so that their full isometry
group $G$ has necessarily $dim(G) \le 10$.

\begin{itemize}

\item A spacetime with $dim(G) = 10$ (that is, with constant
spacetime curvature) is not physically realistic  (if the curvature is zero it is the
Minkowski spacetime).

\item For $6 < dim(G) \le 10$, G is necessarily transitive on ${\cal M}_4$.
Such groups have been classified by Petrov \cite{Pet69}, but due to their
high dimension they do not provide a realistic basis for cosmological
models.

\item For $dim(G) \le 6$, the group may act transitively on ${\cal M}$
or else act on lower dimensional submanifolds. 
\begin{itemize}
\item If $G$ is simply transitive on all of
${\cal M}_4$, then $dim(G) = 4$ and the manifold is called {\sl
homogeneous
 in space and time}. The Einstein static universe and the de Sitter
universe (with positive curvature),  the anti--de Sitter universe (with
negative 
 curvature) are such cases \cite{Haw73}. But such universe
models, in which the spatial metric remains the same in time, do not
admit expansion and contradict the cosmological observations.
\item If  G admits a subgroup acting transitively on spacelike hypersurfaces
(and not on the spacetime itself), the spacetime is said {\it spatially
homogeneous}. There are still three subcases: 
 \begin{itemize}
\item $dim(G) = 6$, decomposed into  a $G_3$ simply transitive on spacelike
hypersurfaces  and a $G_3$ isotropy group. We have the {\sl spatially
homogeneous and isotropic} spacetimes, admitting spacelike
hypersurfaces of {\sl constant curvature} (the celebrated
Friedmann--Lema{\^\i}tre universe models).
Other homogeneous spacetimes are anisotropic.
 \item $dim(G) = 4$ and G is multiply--transitive on 3--dimensional
subspaces. Corresponding spacetimes have been considered by Kantowski
and Sachs \cite{Kan66}. For more details, see \cite{MaC79} and
  \secn{Bianchi} below.
 \item  $dim(G) = 3$ and G is simply transitive on 3-dimensional
subspaces. The corresponding groups have been classified by Bianchi
\cite{Bia97}. See also \cite{Rya75,MaC79} and   \secn{Bianchi} below.
 \end{itemize}
\end{itemize}
\end{itemize}

\subsection{Basic tools for the topological classification of  spaces} 

\subsubsection {Connectedness, homotopy and fundamental group}

 The mathematical  notions involved in  the study and the
classification of topological  structures are those of
multi--connectedness, homotopy, fundamental group,
universal covering, holonomy, and fundamental
polyhedron. All these concepts have very  formal and abstract
definitions that can be found in classical textbooks in topology
(for instance, \cite{Mas87,Nas83} and, in the particular context of
Lorentzian manifolds used in relativity, \cite{Pen72,Ger79}).  In
this primer we just provide  pictorial definitions -- with no lack of
rigour, we hope -- illustrated mostly in the cases of locally
Euclidean surfaces.

The strategy for characterizing spaces is to
produce invariants which capture the key features of the topology and
uniquely specify each equivalence class. The topological invariants can take
many forms. They can be just numbers, such as the dimension of the
manifold, the degree of connectedness or the
Poincar\'e -- Euler characteristic. They can also be whole
mathematical structures, such as the homotopy groups.  

Let us introduce first the concept of {\sl homotopy}. A loop at ${\bf x} \in
\cal{M}$  is any path which starts at $\bf x$ and ends at $\bf x$. Two loops
 $\gamma$ and $\gamma ' $ are homotopic if $\gamma$ can be continuously
deformed into  $\gamma '$ .
The manifold $\cal{M}$ is {\sl simply--connected} if, for any $\bf x$,
two any  loops through $\bf x$ are homotopic -- or, equivalently, if every loop
is  homotopic to a point.  If not, the manifold is said to be {\sl
multi--connected}. 
Obviously, the Euclidean spaces $\bbbr ^1$, $\bbbr ^2$,\ldots, $\bbbr ^n$,
and the  spheres $\bbbs ^2,~ \bbbs ^3,\ldots, \bbbs ^n$ are
simply--connected, whereas the circle $\bbbs ^1$, the  cylinder $\bbbs
^1 \times$$ \bbbr $ or the torus $\bbbs ^1 \times \bbbs ^1$ are
multi--connected. 

\begin{figure}
\centerline{\epsfig{file = 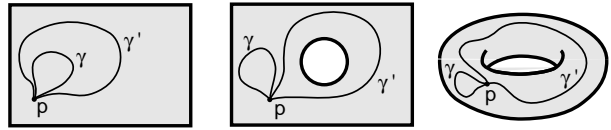,width=8cm}}
\caption{\label{homot2}{\it Classes of homotopy. 
\underline{ Left : the Euclidean plane}. Any loop can be shrunk
to a point. The fundamental group reduces to
identity. 
\underline{Center : a plane with a hole}.  $ \gamma$ is
not homotopic to $ \gamma'$ . Every homotopy class $h_n$ is
associated to an integer n~: $ \gamma \in h_n$ if it winds n times
round the hole in clockwise direction $(n  > 0)$, $n < 0$ in the
anticlockwise direction, $n = 0$ if it does not wind.  Thus the
fundamental group is the infinite cyclic group of integers
$\bbbz$. 
\underline{Right : a torus $\bbbs^1 \times \bbbs^1$}.
Loops can wind m times around the central hole and n times
around the body of the torus. Thus the fundamental group consists
of pairs $(m,n)$ of integers with addition $(m,n) + (p,q) = (m+p,
n+q)$. In other words it is isomorphic to $\bbbz \oplus \bbbz$.}}
\end{figure}

The study of homotopic loops in a manifold $\cal M$ is a way of
detecting holes or handles. Moreover the equivalence classes of
homotopic loops can be endowed with a group structure, essentially
because loops can be added by joining them end to end. For
instance in the  Euclidean plane, joining a loop winding $m$ times
around a hole
 to another loop winding $n$ times gives a
loop winding $m+n$ times.  The group of loops is called the first
homotopy group at $\bf x$ or, in the terminology originally
introduced by Poincar\'e \cite{Poi53}, the {\sl fundamental group}
$\pi _1($$\cal M$,${\bf x})$. If $\cal M$ is (arcwise) connected,
then for any $\bf x$ and $\bf x'$ in $\cal M$, $\pi _1({\cal M},{\bf
x})$ and $\pi _1({\cal M},{\bf x'})$ are
isomorphic\,\footnotemark[1] \footnotetext[1]{~Two groups are isomorphic if they have the
same structure, namely, if their elements can be put into one-to-one
correspondance which is preserved under their respective
combination laws. In fact, two isomorphic groups are {\sl the same}
(abstract) groups.}; the fundamental group is thus independent on
the base point : it is a topological invariant of the manifold. 
Figure \ref{homot2} depicts some elementary examples. 

For surfaces, it was shown in the last century that multi--connectedness
means that the fundamental group is non trivial : loosely
speaking, there is at least one loop that cannot be shrunk to a point.
But in higher dimensions the problem is more complex because
loops, being only one--dimensional structures, are not sufficient to
capture all the topological features of the manifolds. The purpose of
algebraic topology, extensively developed during the twentieth
century, is to generalise the concept of homotopic loops and to
define higher homotopy groups. However the fundamental group
(the first homotopy group) remains essential. In 1904, Poincar\'e
\cite{Poi53} had conjectured that any connected closed
n--dimensional manifold having a trivial fundamental group must be
topologically equivalent to the sphere $\bbbs^n$. The conjecture
was proved by steps during the last 80 years; curiously enough the
most difficult case was for n = 3  \cite{Reg86}.

\subsubsection {Universal Covering Space}

\begin{figure}[tb]
  \begin{center}
    \leavevmode
    \includegraphics{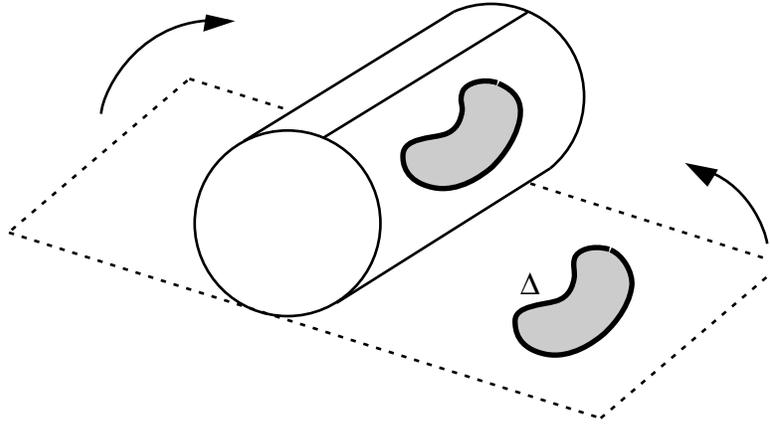}
\caption{\label{dev}{\it  Development of a simply-connected
domain of the cylinder.}}
  \end{center}
\end{figure}

The cylinder $\bbbs ^1 \times
\bbbr$, embedded in $\bbbr ^3$, is a locally Euclidean space whose metric
can be written   $ds^2 = R^2 d\theta^2 + dz^2$. Its geodesics are helices. 
Any domain $\cal{D}$ bounded by a  closed curve that does not intersect all
the generatrices of the cylinder  is simply-connected. If we unroll
once the cylinder on the Euclidean plane $\bbbr ^2$,
 the domain $\cal{D}$ leaves an imprint domain $\Delta$ called its
{\sl development} (figure \ref{dev}).
There is a one-to-one correspondence between the points of $\cal{D}$
and those of $\Delta$, and all  the distances remain unchanged.
Inside $\cal{D}$, all the properties of Euclidean  geometry are valid :
the sum of the angles of a triangle is 180 degrees;  one and only one
geodesic joins two any distinct points; and so on \ldots

\begin{figure}[tb]
  \begin{center}
    \leavevmode
    \includegraphics{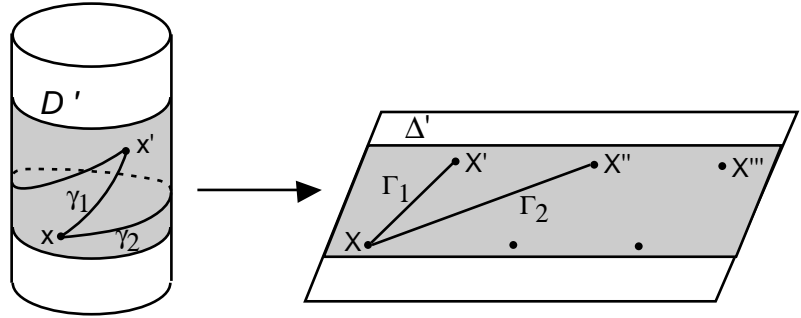}
\caption{\label{dev2}{\it Development of a multi-connected
domain of the cylinder.}}
  \end{center}
\end{figure}

Now consider the domain $\cal{D'}$ bounded by two circular
sections of the  cylinder  (figure~\ref{dev2}). $\cal{D'}$ is obviously
multi--connected because between two arbitrary  points $P$ and
$P'$ can now pass an infinite number of geodesics, which are helices of
different pitch. Furthermore, the development $\Delta '$ of
 $\cal{D'}$ in the plane $\bbbr ^2$ is no more a one--to--one
correspondance. If we unroll
 the cylinder on $\bbbr ^2$,  every point of $\cal{D'}$ generates an
infinite number of imprinted points in $\Delta '$.  Therefore, although
the metric properties of Euclidean space remain valid in $\cal{D'}$ 
(such as the value of the sum of the angles of a triangle), the
topological properties (such as the unicity of geodesics) do not.

The development can be extended step by step. A
point $\bf x$ and  a path $\gamma$ from $\bf x$ to $\bf x'$ on the cylinder
can be developed into the point $\bf X$ and the path $\Gamma$   from
$\bf X$ to $\bf X'$ in $\bbbr^2$. $\bf X'$ and $\Gamma$ are unique if
$\bf x'$ and $\gamma$ lie in a simply-connected domain $\cal{D}$ of the
cylinder. In the other case, if $\cal{D}$ is  multi-connected, there are several
paths $\gamma_1, \gamma_2, \ldots$ from $\bf x$ to  $\bf x'$ such that
their developments $\Gamma_1, \Gamma_2$ \ldots generate the distinct
points ${\bf X'}, {\bf X''}$, \ldots in $\bbbr ^2$. The Euclidean
plane appears as the {\sl Universal Covering Space} of the cylinder.

\begin{figure}[tb]
  \begin{center}
    \leavevmode
    \includegraphics{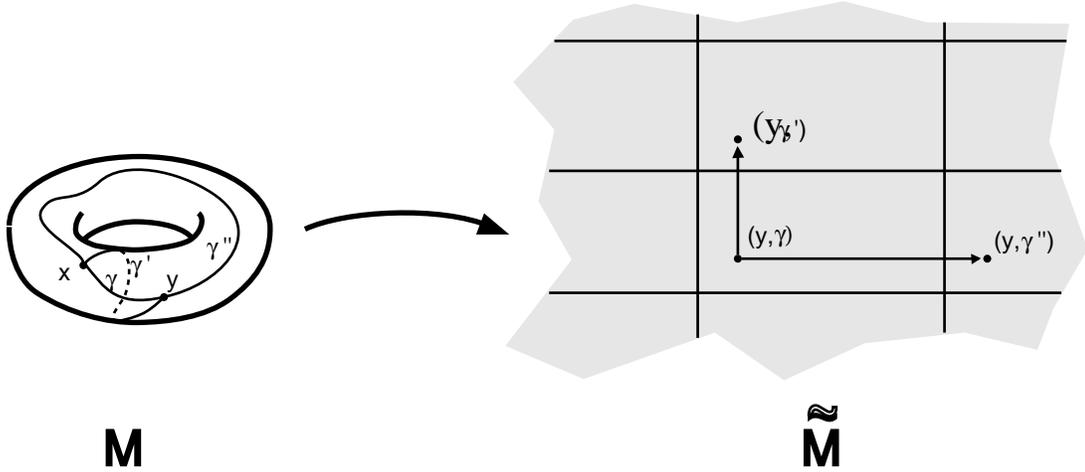}
\caption{\label{devtor}{\it Universal covering space of the
torus.}}
  \end{center}
\end{figure}

Such a procedure can be generalized to any manifold.
Start with a manifold $\cal M$ with metric $\bf g$. Choose a base
point {\bf x} in $\cal M$ and consider the differents paths from {\bf
x} to an other point {\bf y}. Each path belongs to a homotopy class
$\gamma$ of loops at {\bf x}. We construct the universal covering
space as the new manifold  ($\widetilde{ \cal M}$,$\tilde{\bf g}$) such
that each point $\tilde{\bf y}$ of $\widetilde{ \cal M}$ is obtained as a
pair (${\bf y},\gamma$), {\bf y} varying over the whole of $\cal M$
while {\bf x} remains fixed . The metric $\tilde{\bf g}$ is obtained by
defining the interval from $\tilde{\bf x} = ({\bf x},\gamma)$ to a
nearby point $\tilde{\bf x'}= ({\bf x'},\gamma)$ in $\widetilde{ \cal
M}$ to be equal to the interval from $\bf x$ to $\bf x'$ in $\cal M$.
By construction, ($\widetilde{ \cal
M}$,$\tilde{\bf g}$) is locally indistinguishable from (${ \cal
M},{\bf g}$). But its global -- namely topological -- properties
can be quite different. It is clear that, 
 when $\cal M$ is simply--connected, it is  identical to its universal
covering space $\widetilde{ \cal M}$. When $\cal M$ is
multi--connected, each point of $\cal M$ generates an infinite number of
points in $\widetilde{ \cal M}$. The universal covering space can thus
be thought of as an ``unwrapping" of the original manifold  (see figure
\ref{devtor}).

\subsubsection{Holonomy group}

Consider a point ${\bf x}$ and a loop $\gamma$ at {\bf x} in $\cal M$.
If $\gamma$ lies entirely  in a simply-connected domain of
$\cal M$, (${\bf x},\gamma$) generates a single point $\tilde{\bf x}$ 
in $\widetilde{ \cal M}$. 
Otherwise, it generates additional points $\tilde{\bf x'},\tilde{\bf x''}
$, \ldots which are said to be {\sl homologous} to $\tilde{\bf x}$.
The displacements $\tilde{\bf x} \mapsto \tilde{\bf x'}$, $\tilde{\bf
x} \mapsto \tilde{\bf x''}$, \ldots  are isometries and form the  so-called
{\sl holonomy group} $\Gamma$ in $\widetilde{ \cal M}$. This group
is discontinuous, i.e., there is a non zero  shortest distance between any
two homologous points, and the generators of the group (except the
identity) have no fixed point. This last property is very restrictive (it
excludes for instance the  rotations) and allows to classify all the
possible groups of holonomy.

Equipped with such properties, the holonomy group is said to
act freely and discontinuously on $\widetilde{ \cal M}$. The holonomy group is isomorphic
to the fundamental group $\pi_1(\widetilde{\cal M})$ (see for
instance \cite{Boo75}).

\subsubsection {Fundamental polyhedron}

The geometrical properties of a manifold $\cal M$  within a
simply--connected domain are the same as those of its development in
the universal covering $\widetilde{\cal M}$.  It may be asked what is
the largest simply--connected domain containing a given point
$x$ of $\cal M$, namely the set $ \bigl\{ y \in {\cal M},
d(y,x) \le d(y,\gamma (x)),  \forall \gamma \in \Gamma 
\bigr\}$.  Its  development in $\widetilde{ \cal M}$ is called the
{\sl fundamental polyhedron} ($FP$). 
 
The $FP$ is always convex and has a finite number of faces (due
to the fact that the holonomy group  is discrete). These faces are
homologous by pairs : to every face $\cal F$ corresponds one
 and only one face $\cal {F'}$, such that, for any point ${\bf X}
\in ~{\cal F}$ there exists a point ${\bf X'} \in ~{\cal F'}$, which
are two developments of the same point $\bf x$ in $\cal M$. The
displacements carrying  $\cal F$ to $\cal {F'}$ are the generators
of the holonomy group $\Gamma$.

 Note that in dimension 2, the $FP$ is a surface whose boundary is
constituted by lines, thus a polygon. In dimension 3, it is
a volume bounded by  faces, thus a polyhedron.

 The configuration formed by the fundamental polyhedron $\cal P$ and its
images $\gamma $$\cal P$ ($\gamma \in \Gamma$) is called a
{\sl tesselation} of $\widetilde{ \cal
M}$, each image $\gamma $$\cal P$ being a cell of the tesselation. 

\begin{figure}[tb]
  \begin{center}
    \leavevmode
    \includegraphics{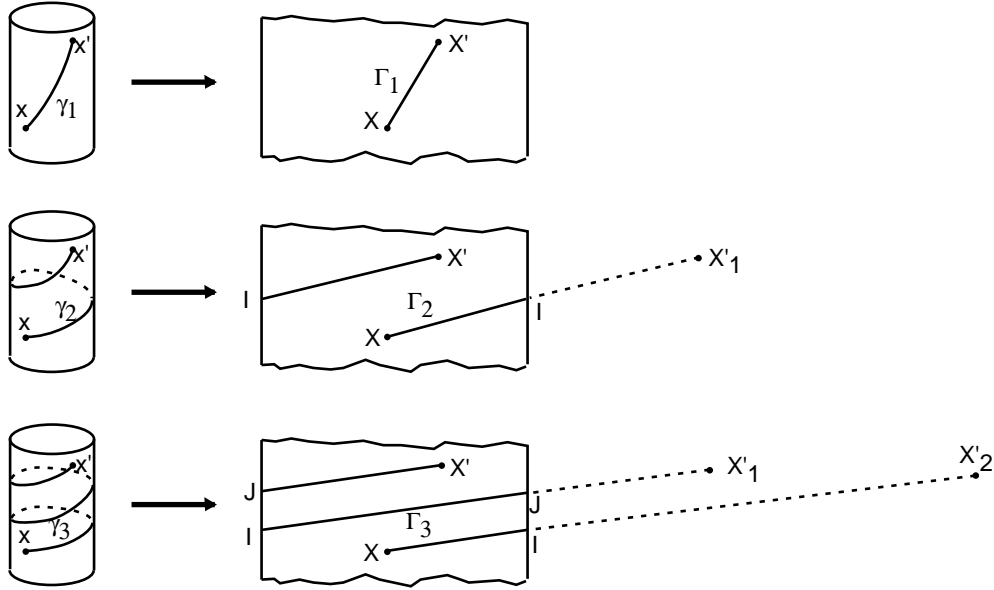}
\caption{\label{geodcyl}{\it Development of geodesics of the
cylinder.}}
  \end{center}
\end{figure}

The $FP$ presents two major interests:
\begin{itemize}
\item The fundamental group of a given topological
manifold $\cal M$ is isomorphic to the fundamental group of the
$FP$. Since routine methods are available to determine
the holonomy group of as a polyhedron, the
problem is considerably simplified. 
\item The $FP$ allows one to  represent any curve in
$\cal M$, since any portion of a curve lying  outside the $FP$ can
be carried inside it by appropriate holonomies
(figure \ref{geodcyl}). 
\end{itemize}

As a general conclusion of this section, the method for classifying the
topologies of a given manifold $\cal M$  is :
\begin{itemize}
\item to determine its universal covering space $\widetilde{ \cal
M}$ 
\item  to find the fundamental polyhedron $FP$
\item  to calculate the holonomy group acting on the $FP$.
\end{itemize}

 In sections 4 to 8 this is done for the two-- and three--dimensional
homogeneous manifolds.

\section {Classification of Riemannian surfaces}\label{class2}

In addition to pedagogical and illustrative interest, the
classification of two--dimensional Riemannian surfaces plays an
important role in physics for understanding (2+1)--dimensional
gravity, a toy model to gain insight into the real world of
(3+1)--dimensional  quantum gravity
\cite{Sta63,Got84,Gid84,Des84,Wit88,Fuj93}. Also, from a mathematical
point of view, three--dimensional spaces can be constructed from surfaces. 

\begin{figure}
\centerline{\epsfig{file=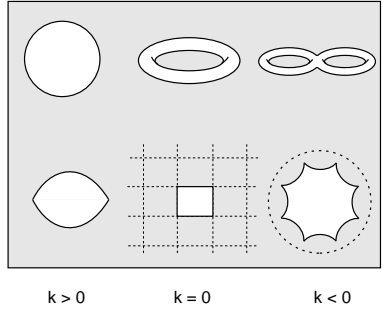, width=5cm}}
\caption{\label{UCsurf}{\it The three kinds of geometries for
Riemannian surfaces, with their universal covering space.}} \end{figure}

 As we have seen in section 3.3.3, any Riemannian surface is
homeomorphic to a surface admitting a metric with constant curvature
k. Thus any Riemannian surface can be expressed as the quotient
  $\cal{M}$ $= \widetilde{\cal M}/\Gamma$,
 where the universal covering space $\widetilde{\cal M}$ is
either (figure \ref{UCsurf})~:
 \begin{itemize}
 \item the Euclidean plane $\bbbr ^2$ if $k = 0$
 \item the sphere $\bbbs ^2$ if $k > 0$
\item  the hyperbolic plane $\bbbh ^2$ if $k < 0$.
\end{itemize}
and $\Gamma$ is a discrete group of isometries without fixed point of 
$\widetilde{\cal M}$ (figure \ref{UCsurf}).

To characterise the quotient spaces we adopt the following
abbreviations~:

 C = closed, O = open, SC = simply--connected, MC =
multi--connected, OR = orientable, NOR = non-orientable.

\subsection {Locally Euclidean surfaces}

The UC space is the Euclidean plane $\bbbr^2$ with standard metric
$d\sigma^2 = dx^2 + dy^2$ or, in polar coordinates, $d\sigma^2 = dr^2 +
r^2d\phi^2$.
The full isometry group of $\bbbr^2$ (the Galilean
group) is composed of translations, rotations, reflections and glide
reflections (a glide reflection is a translation composed with a reflection in a
line parallel to the translation; more pictorially, the correspondance
between two successive footprints on a straight snowy path is a glide
reflection) . 

\begin{figure}[tb]
  \begin{center}
    \leavevmode
    \includegraphics{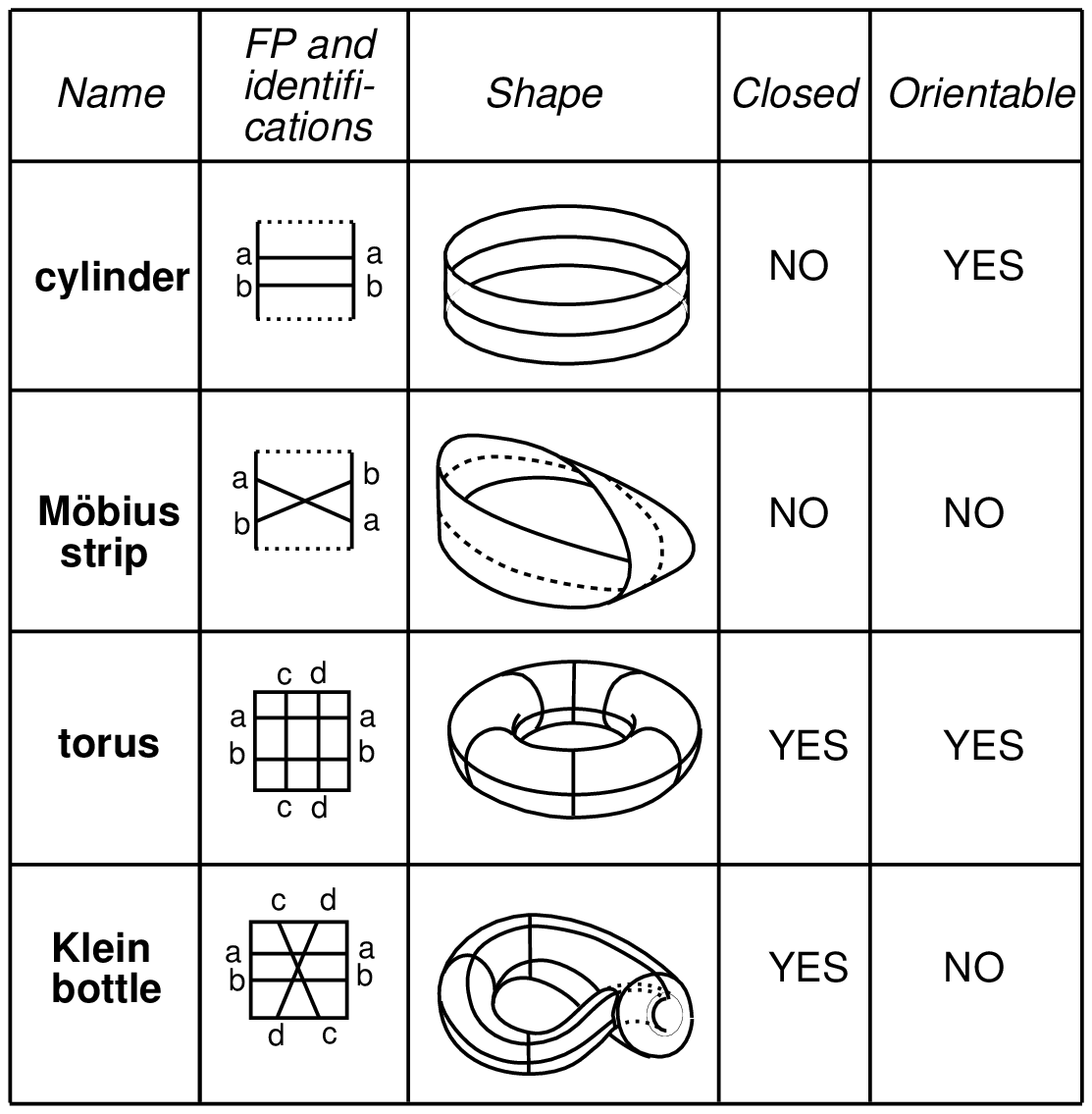}
\caption{\label{surf}{\it The four types of multi-connected Euclidean surfaces.}}
  \end{center}
\end{figure}

The subgroups of discrete isometries {\sl without
fixed point} contain only translations and glide reflections. This allows
one to classify locally Euclidean surfaces into only 5 types  : the
simply--connected Euclidean plane itself $\bbbr$,  the multi--connected
cylinder $\bbbr \times \bbbs^1$, the M\"obius band,  the torus $\bbbs^1
\times \bbbs^1$ and the Klein bottle. Their characteristics are summarized
in figure \ref{surf}. It is well known that the M\"obius band and the Klein
bottle are not  orientable. The torus and the Klein bottle are closed spaces.
We point out that the projective plane, obtained by identifying the
opposite faces of a square under the action of two independent
translations, has a  strictly positive curvature and is therefore locally
spherical.

\subsection {Locally spherical surfaces}

The sphere $\bbbs^2$ admits a homogeneous metric induced from its
embedding in $\bbbr^3$, namely the surface of equation $x^2 + y^2 +
z^2 = 1$. Introducing coordinates $(\theta,\phi)$ by 
$$x~=~sin\theta~cos\phi, ~y~=~sin\theta~sin\phi, ~z~=~cos\theta,$$
the induced metric on $\bbbs^2$ becomes
 \begin{equation}
d\sigma^2 = d\theta^2 + sin^2\theta d\phi^2~~~~~~ ~(0 \le \phi \le 2\pi , ~ 0 \le
\theta \le \pi).
\end {equation}

The full isometry group of $\bbbs^2$ is the group of $3 \times 3$
orthogonal matrices $O(3)$ (with determinant $\pm 1$). But there is
only one non--trivial discrete subgroup without fixed point, namely the
group $\bbbz_2$, of order\footnotemark[1] \footnotetext[1]{The order of a group is the number
of elements in the group} two. It is generated by the antipodal map of
$\bbbs^2$ which identifies diametrically opposite points on the
surface of the sphere.

 As a result there are only two spherical surface forms\footnotemark[1] \footnotetext[1]{This result has been generalized to any constant positive
curvature manifold of {\it even} dimension \cite{Wol84}.}~:
 \begin{itemize}
 \item the sphere $\bbbs^2$ iself : C, SC, OR
 \item the projective plane $\bbbp^2 \equiv \bbbs^2/\bbbz_2$ (also called
the elliptic plane) : C, MC, NOR.
\end{itemize}

Whereas the surface of the unit sphere is $4\pi$, the surface of the unit
projective plane is only $2\pi$, and its diameter, i.e., the distance
between the most widely separated points, only $\pi/2$. 

\subsection {Locally hyperbolic surfaces}

\subsubsection{The geometry of $\bbbh^2$} \label{gh2}

The hyperbolic plane
$\bbbh^2$, historically known as the Lobachevski space, is
difficult to visualize because it cannot be isometrically imbedded in
$\bbbr^3$. Nevertheless it can be thought of as a surface with a saddle point
at every point. 

Consider  the surface of equation  
$- z^2 + x^2 + y^2  = 1$
in the pseudo-Euclidean three-dimensional space with metric 
$d\sigma^2 = - dz^2 +  dx^2 + dy^2$ . 

\begin{figure}
\centerline{\epsfig{file=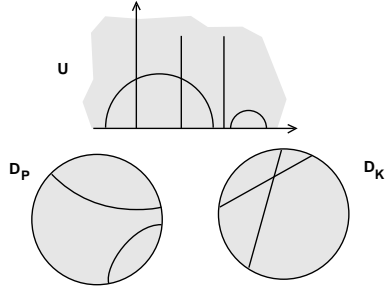, width=5cm}}
\caption{\label{PKH2}{\it Three representations of $\bbbh^2$ : the
upper-half space $U$, the Poincar\'e model $D_P$  and the Klein
model $D_K$. Representative geodesics are shown. }} \end{figure}

If we introduce coordinates $(\chi, \phi)$ by
$$z = cosh \chi, ~~x = sinh \chi~cos \phi, ~~y = sinh \chi~sin \phi, 
~~~0 \le \chi  < \infty, ~Ê0 \le \phi \le 2\pi,$$ 
the induced metric on $\bbbh^2$ is written as 
\begin{equation}\label{h2m}
d\sigma ^2 = d\chi ^2 + sinh^2 \chi ~d\phi ^2 .
\end{equation}
Other  representations of $\bbbh^2$ are well-known (figure
\ref{PKH2}) : 
\begin{itemize}
\item The upper--half plane $U \equiv \{(x,y) \in \bbbr^2, y
>0\}$ equipped with the metric 
\begin{equation}\label{h2U} 
 d\sigma^2 = (dx^2 + dy^2)/y^2,~~y > 0.
 \end{equation}  

The hyperbolic geodesics correspond to the Euclidean semi-circles,
which orthogonally intersect the boundary $\partial U$. The metric
(\ref{h2U}) is conformally flat, so that the angles between the hyperbolic
lines coincide with the Euclidean ones.

\item The Poincar\'e model represents $\bbbh^2$ as the open unit disc
$D_P \equiv \{(x',y') \in \bbbr^2,~ x'^2 + y'^2 < 1\}$. It is obtained from
the upper--half space by a coordinate transformation $U \mapsto D_P$
which maps $z = x+iy$ into $z^{\prime} = x^{\prime}+iy^{\prime}$ such
that $z^{\prime} = -i~\frac{z-i}{z+i}$. The metric  becomes 
\begin{equation}\label{zmetric} 
d \sigma ^2
=\frac{4~dz^{\prime}~d\bar{z}^{\prime}}{(1-z^{\prime}\bar{z}^{\prime})^2}.
\end{equation}
 or, introducing  polar
coordinates $(r,\phi)$ such that $z^{\prime} = r~e^{i\phi}$ :
 \begin{equation}
d\sigma^2 = \frac {4(dr^2 + r^2 d\phi ^2)}{(1 - r^2)^2}
\end{equation} 

The model is also conformally flat, and the hyperbolic geodesics are
mapped onto arcs of circle which meet the frontier of $D_P$  at right
angles. 

\item  The Klein representation also represents $\bbbh^2$ in an unit
open disc $D_K \equiv \{(x^{\prime\prime},y^{\prime\prime}) \in \bbbr^2,
x^2{}^{\prime\prime} + y^2{}^{\prime\prime} < 1\}$ with a mapping
$D_P \mapsto D_K$ such that $z^{\prime\prime} \equiv
x^{\prime\prime}+iy^{\prime\prime} = \frac{2
z^{\prime}}{1+z^{\prime}\bar{z}^{\prime}}$. 

 The hyperbolic geodesics are
mapped onto Euclidean lines, but this model is not conformally flat.
 \end{itemize} 

\subsubsection{The holonomies of $\bbbh^2$}\label{holoh2}
 The full
isometry group of $\bbbh^2$ is $PSL(2,\bbbr) \equiv
SL(2,\bbbr)/\bbbz_2$, where $SL(2,\bbbr)$ is the group of real $2 \times 2$
matrices with unit determinant.  In metric (\ref{zmetric}), any 
isometry of $\bbbh^2$ can be expressed by a M\"obius transformation 
$$z \mapsto \frac{az + b}{cz + d}$$
where $z$ is complex, $a,~b~,c~,d$ real, $ad-bc > 0$.   

Discrete subgroups without fixed point $\Gamma$ are described for instance
in \cite{Mag74}.
The topological classification of locally hyperbolic surfaces follows.
It  is complete only for the {\sl compact} $\bbbh^2/ \Gamma$, which
fall into one of the following categories : 

\begin{itemize} 

\item g--torus $T_g$, $g \ge 2$ (connected sum of g simple toruses) : C,
MC, OR

\item connected sum of n projective planes : C, MC, NOR

\item connected sum of a compact orientable surface ($\bbbs^2$ or $T_g$) and
of a projective plane or  a Klein bottle : C, MC, NOR 
\end{itemize}

All of these surfaces have a finite area bounded below by $2\pi$, and a
diameter greater than $ch^{-1}(4) \approx 2.06$ \cite{Ber80}.

In addition, there are an infinite number of {\sl non--compact} locally
hyperbolic surfaces, but their full classification is not achieved. Anyway,
it is clear that ``almost all" Riemannian surfaces are hyperbolic, since :

  - Any open surface other than the Euclidean plane, the cylinder and the
M\"obius band is homeomorphic to a locally hyperbolic surface, for example an
hyperbolic plane with or without handles.

- Any closed surface which is not the sphere, the projective plane, the torus
or the Klein bottle is homeomorphic to a locally hyperbolic surface. 

\subsubsection{Examples} 
\label{flatsexamples}
\begin{figure}[tb]
  \begin{center}
    \leavevmode
    \includegraphics{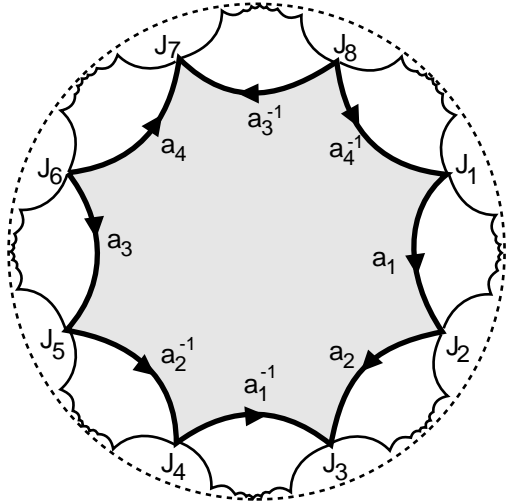}
\caption{\label{esch}{\it Tesselation of $\bbbh^2$ by octogons in the Poincar\'e
representation. The vertices have the coordinates $z(J_k) =
0.841 ~exp~\frac{(3-2k)\pi i}{8}$.}}
  \end{center}
\end{figure}

 The  best known example of a compact hyperbolic
surface is the 2-torus $T_2$ ( see section 3.2.2).  In  this case, the
FP is a regular octogon with pairs of sides identified.  In the
Poincar\'e representation of $\bbbh^2$, the FP appears
curvilinear.  The pavement of the unit disk by homologous octogons
(which appear distorded in this representation)  corresponds to the
tesselation of $\bbbh^2$ by regular octogons (figure \ref{esch}).  
The famous Dutch artist M.--C. Escher \cite{Cox86} has designed
fascinating  drawings and prints using such tilings of the hyperbolic
plane.

More generally, any compact Riemannian surface $\Sigma_g$ with
genus $g \ge 2$ can be modelled in $\bbbh^2$. It is representable
by the interior of a regular polygon with 4g edges. The
length $R$ of an edge is determined by the curvature $k = -1/R^2$ of the
hyperbolic plane. The angles are $\pi /2 g$. The
fundamental group is generated by the $2g$ elements $a_1, a_2,\ldots,
a_{2g}$ such that  $a_1a_2a_1^{-1}a_2^{-1}\ldots
a_{2g-1}a_{2g}a_{2g-1}^{-1}a_{2g}^{-1} = Id$ (figure \ref{esch}). 
The Poincar\'e--Euler characteristic is $\chi (\Sigma_g) = 2(1-g)$. But
from Gauss--Bonnet theorem it is also given by $\chi (\Sigma_g) = {1
\over {2\pi}} \int \int k d\sigma$. One deduces that the area of the surface
is $4\pi (g-1) R^2$.  

\newpage
\section  {Three-dimensional homogeneous spaces}\label{homog}

We now consider a three-dimensional Riemannian
manifold $\cal M$ admitting at least a 3--dimensional discrete
isometry group $\Gamma$ simply transitive on $\cal M$ (cf. section 3.3.3).
Such a (locally) homogeneous manifold can be written as the quotient 
$\widetilde{\cal M} / \Gamma$, where  $\widetilde{ \cal M}$ is the
universal covering space of $\cal M$. Let  $G$ be the full group
of isometries of $\cal M$ (containing $\Gamma$ as a discrete subgroup).
In the terminology used in the theory of classification of compact
three--manifolds, $\cal M$ is said to admit a {\sl geometric structure}
modelled on $(\widetilde {\cal M}, G)$.

On one hand, Thurston \cite{Thu82} has classified the homogeneous
three--dimensional geometries into eight distinct types, generally
used by mathematicians.

On the other hand, the Bianchi types are defined  from the
classification of all simply-transitive 3--dimensional Lie
groups\footnotemark[1] \footnotetext[1]{ A Lie group is a differentiable manifold with a group
structure such that $(a,b) \mapsto ab^{-1}$ is differentiable.}.  Since the
 isometries of a Riemannian manifold form a Lie group, the Bianchi
classification is used  by workers in relativity and cosmology for the
description of spatially homogeneous spacetimes \cite{Rya75}.

We present below the two approaches and then discuss their mutual
relationships \cite{Osi73,Fag85,Fag92}.

\subsection {The Thurston's eight geometries}\label{Thur}

Any simply-connected
3--dimensional geometry which admits a compact quotient is equivalent
to one of the eight geometries ($\widetilde{\cal M}$$, G$), where
$\widetilde {\cal M}$ is 
$\bbbr^3$, $\bbbs^3$, $\bbbh^3$, $\bbbs^2 \times \bbbr$, $\bbbh^2 \times
\bbbr$, $\widetilde{SL_2\bbbr}$, $Nil$ or $Sol$, that we shortly
describe now (for full details we refer
the interested reader to \cite{Thu82,Sco83}). 

 \begin{itemize}
\item $\widetilde{\cal M}$$ = \bbbr^3$ : Euclidean geometry 
 
$G = ISO(3) \equiv R^3 \times SO(3)$, i.e.  the product of the group of
translations in $\bbbr^3$ by the group of $3 \times 3$ special orthogonal
matrices.

This is the geometry of constant zero curvature. More details are given
in section~6

\item $\widetilde{\cal M}$$ = \bbbs^3$ : spherical geometry 
 
$G = SO(4)$

This is the geometry of constant positive curvature. More details are
given in section 7.

\item $\widetilde{\cal M}$$ = \bbbh^3$ : hyperbolic geometry 

$G = PSL(2,\bbbc) \equiv SL(2,\bbbc)/\bbbz_2$
 
This is the geometry of constant negative curvature. More details are
given in section 8.

\item $\widetilde{\cal M}$$ = \bbbs^2 \times \bbbr$

$G = SO(3) \times \bbbr$, the products of the corresponding groups. 

The subgroups $\Gamma$ of G acting freely and
discontinuously are given by \cite{Sco83}. Only seven
3--manifolds without boundary  can be modelled on  $\bbbs^2 \times
\bbbr$. Three are non--compact (including $\bbbs^2 \times \bbbr$ itself),
four are compact, including the ``three--handle"\,\footnotemark[1] \footnotetext[1]{Also called
``closed wormhole"} $\bbbs^2 \times \bbbs^1$ \cite{Gow74} and the
connected sum of projective spaces $\bbbp^3 \oplus \bbbp^3$.

 Metric :  $d\sigma^2 = dr^2 + sin^2rd\phi^2 + dz^2$

\item $\widetilde{\cal M}$$ = \bbbh^2 \times \bbbr$

$G = PSL(2,\bbbr) \times \bbbr$,  the product of the corresponding
groups for $\bbbh^2$ and $\bbbr$. 

The $\widetilde{\cal M} / \Gamma$ include for instance the product of
any compact hyperbolic surface $T_g$ (the g--torus or g--handled sphere)
with $\bbbs^1$
 or $\bbbr$ (see the example below).

 Metric :  $d\sigma^2 = dr^2 + sinh^2rd\phi^2 + dz^2$

\item $\widetilde{\cal M}$$ =\widetilde{SL2\bbbr}$

$\widetilde{SL2\bbbr}$ is the universal covering of $SL(2,\bbbr)$, the
3-dimensional Lie group of all $2 \times 2$ real matrices with
determinant 1. It is more geometrically described by a fiber bundle
whose basis is  the hyperbolic plane. Its isometry group is thus the
product of translations by $PSL(2,\bbbr).$

 Metric :  $d\sigma^2 = dx^2 + cosh^2xdy^2 +( dz + sinhxdy)^2$

 \item $\widetilde{\cal M}$$ = Nil$

$Nil$ is the 3-dimensional Lie group composed of all $3 \times 3$ 
 Heisenberg matrices of the form 
$\left(
\begin{array}{ccc} 
1 & x & z\\
0 & 1 & y\\
0 & 0 & 1
\end{array}
\right)~ x, y, z \in \bbbr$.
It is more geometrically described by a bundle with the Euclidean
plane as base and lines as fibers, or by a bundle with a circle as base and
tori as fibers. See \cite{Sco83} for G, which is too complex to be
described here.

Metric :
$$d\sigma^2 = dx^2 + dy^2 + (dz-xdy)^2$$

 \item $\widetilde{\cal M}$$ = Sol$

$Sol$ is a Lie group which can be represented by $\bbbr^3$ with the
multiplication law $$(x,y,z)(x',y',z') = (x+e^{-z}x', y+e^zy', z+z').$$
It is more geometrically described by a bundle over one--dimensional
base and two--dimensional fibers.  See \cite{Sco83} for G.

Metric : $d\sigma^2 = e^{2z}dx^2 + e^{-2z}dy^2 + dz^2$
 
\end{itemize}
Thus, in addition to the three geometries of constant curvature
$\bbbr^3$, $\bbbs^3$ and $\bbbh^3$, there exist five additional
homogeneous geometries.  In the three first types, $dim(G) = 6$ and we
have the spaces of constant curvature. In the other types (except $Sol$),
$dim(G) = 4$ and the corresponding spaces are called locally rotational
symmetric.

\subsection {Bianchi types}\label{Bianchi}

 The
original work of Bianchi \cite{Bia97} was improved by theoretical
cosmologists \cite{Est68,Ell69}, because of some redundancy between
types. 
For a three-dimensional Lie group, let $\{\xi_i\}_{i = 1,2,3}$
be a basis of infinitesimal generators, called Killing vectors. The
commutation relations
$[\xi_i,\xi_j] = C^k_{ij}~ \xi_k$ define the 
structure constants of the Lie group $C^k_{ij}$, which fully
characterize its algebraic structure. The classification of
3--dimensional Lie groups involves the following decomposition of
$C^k_{ij}$  : 

\begin{equation} 
	C^k_{ij} = \varepsilon_{lij}N^{kl} +
\delta^k_j A_i - \delta^k_i A_j 
\end{equation}

where $\delta^k_{ij}$ is the Kronecker symbol and
$\varepsilon_{lij}$ the completely antisymmetric form with
$\varepsilon_{123}= 1$. The Jacobi identity yields $N^{jk}A_k = 0$. By a
change of basis, $N^{jk}$ can be reduced to the diagonal form
$(N_1,N_2,N_3)$ with each $N_i = \pm 1, 0$ and $A_i = (a,0,0)$.
It follows a natural division into two large classes :
\begin{itemize}
\item class A : $a = 0$
\item class B : $a \ne 0$
\end{itemize}

The table \ref{BB} shows the resulting Bianchi--Behr types. 

\begin{table}
\caption{\label{BB}
{\bf The Bianchi-Behr classification of groups}}
\begin{center}
\vspace{1.cm}
\begin{tabular}{|c|c|ccc|c|}
\hline
class  & type & & N &  & a    \\ \hline
      & I       &  0 &  0 &  0  & 0 \\
      & II      &  1 &  0 &  0  & 0 \\
  A   & VI$_0$  &  0 &  1 &  -1 & 0 \\
      & VII$_0$  &  0 &  1 &  1 & 0 \\
      & VIII  &  1 &  1 &  -1 & 0 \\
      & IX  & 1 & 1 &  1 & 0 \\
\hline
      & V  &  0 &  0 &  0 & 1 \\
      & IV  &  0 &  0 &  1 & 1 \\
  B   & VI$_a$, $a<0$  &  0 & 1 &  -1 & $\sqrt{-a}$ \\
      & (III=VI$_{-1}$)& & & & \\
      & VII$_a$, $a>0$  &  0 &  1 &  1 & $\sqrt{a}$\\
\hline
\end{tabular}
\end{center}
\end{table}

These groups may also be characterized by invariant bases of
1--forms $\{\omega^j\}$, in terms of which the ``standard" metric of a
given Bianchi type is written as:

\begin{equation}
 d\tilde{\sigma}^2 = (\omega^1)^2 + (\omega^2)^2 + (\omega^3)^2.
\end{equation}

Now the most general Riemannian space invariant under a Bianchi
group ( thus called a Bianchi space) has a metric 
 \begin{equation}\label{Biam}
 d\sigma^2 =  \gamma_{ab}~ \omega^a\omega^b, 
\end{equation}
where the symmetrical coefficients $\gamma_{ab}$ are constant.

Finally, any spacetime with metric 
 \begin{equation} \label{homo}
 ds^2 =  dt^2 - \gamma_{ab}(t) ~\omega^a \omega^b
\end{equation}
admits a Bianchi group acting
transitively on its spacelike sections. According to our
definitions, it is thus a spatially homogeneous universe model.  

The Bianchi type $I$ spaces have a group isomorphic to the
3--dimensional translation group of the Euclidean space. They include 
locally Euclidean spaces. The flat FL
(Einstein-de Sitter) spacetime model is invariant under a simply transitive
group of type $I$ (defining the spatial homogeneity) and also an isotropy
group of type $VII_0$. The very-well studied
Kasner spacetime \cite{Kas21,Rya75}  has only a simply--transitive
isometry group $G_3$ of type $I$. It is therefore homogeneous but
anisotropic. Its metric is written as ~:
 \begin{equation} 
ds^2 = dt^2 -
t^{2p_1}~dx^2 + t^{2p_2}~Êdy^2 + t^{2p_3}~dz^2,
\end{equation}
with the $p_i$ are constants satisfying 
$p_1 + p_2 + p_3 = (p_1)^2 + (p_2)^2 + (p_3)^2 = 1$.
Each hypersurface $\{ t = constant\}$ of the Kasner spacetime is a flat
three-dimensional space, but the spacetime expands anisotropically.

The Bianchi type $V$ contains locally hyperbolic spaces. The hyperbolic
($k = -1$) FL model is invariant under a simply transitive group of type
$V$  (spatial homogeneity) and also an isotropy group of type $VII_a$.

The Bianchi type $IX$ has a group isomorphic to the 3--dimensional
rotation group SO(3). It therefore contains locally spherical spaces. The
spherical ($k = +1$)  FL model is invariant under a simply transitive
group of type $IX$ (spatial homogeneity) and also an isotropy group of
type $IX$ also. The anisotropic ``mixmaster" universe 
\cite{Mis69b,Bel70} has only a simply-transitive isometry group $G_3$
of type $IX$. 

\subsection {Correspondance between Thurston's
geometries and BKS types}

One one hand, all the homogeneous 3--dimensional metrics are described
by the Bianchi metrics (\ref{Biam}) and the additional Kantowski--Sachs
metric
 \begin{equation}
d\sigma^2 = a^2dx^2 + b^2(dy^2 + sin^2ydz^2),
\end{equation}
for which the isometry group has dimension 4 and is
multiply transitive on 3--dimensional spaces (cf. \secn{homogen}). They
are called collectively BKS metrics. 

On the other hand, if a closed 3-space $\cal M$ (not necessarily
homogeneous) admits a given BKS metric, then it
possesses a geometric structure modelled on $(\widetilde {\cal M}, G)$,
where $\widetilde {\cal M}$ is the universal covering space and $G$ the
corresponding BKS group. 

This allows to establish a correspondance between  the Thurston's
geometries described in   \secn{Thur} and
the Bianchi-Kantowski-Sachs types, summarized in Table \ref{BKST}. 

\begin{table}
\caption{\label{BKST}{\bf Relation between Thurston's
geometries and BKS types}}
\vspace{1.cm}
\begin{center}
\begin{tabular}{|c|c|c|c|} 
\hline
Thurston's geometries& BKS types  & class & sectional curvature    \\
\hline 
\tvi(15,15)   $\bbbr^{3}$ &I, ~VII$_0$      & A& $0,0,0 $  \\  \hline 
\tvi(15,15)   $\bbbs^{3}$ &IX      & A&$ 1,1,1 $  \\ \hline 
\tvi(15,15)    $\bbbh^{3}$ &V       &B&$ -1,-1,-1  $ \\
\tvi(15,15)     & VII$_a$, $a>0$        & & $-a^2,-a^2,-a^2$   \\ \hline
\tvi(15,15)  $\bbbs^{2} \times \bbbr$ & K.S.       & & $1,0,0 $  \\ \hline
\tvi(15,15)  $\bbbh^{2}\times \bbbr$ &III=VI$_{-1}$    & B&$-1,0,0$ \\
\hline  
 \tvi(15,15)  $\widetilde{SL2\bbbr}$ &   VIII       &A& $
-{5\over 4}$,$-{1\over 4}$,$-{1\over 4}$ \\ \hline
 \tvi(15,15)  $Nil$ &  II    &A& $-{1\over 4},{1\over 4},{1\over 4}$  \\ \hline
\tvi(15,15)    $Sol$ &VI$_0$       &A& $1,-1,-1 $  \\    \hline
\end{tabular}
\end{center}
\end{table}

The following remarks must be done.
\begin{itemize} 

\item Within the Bianchi types ($I$, $V$, $VII_a$, $IX$) admitting a
constant curvature space as universal covering, spaces are generally
anisotropic.  More generally, within a given type, the change of topology
obtained by quotienting the universal covering space by $\Gamma$
 lowers the dimension of the full group of isometries, because
the isotropy group is broken\,\footnotemark[1] \footnotetext[1]{~For instance, the perpendiculars
to the boundaries of the FP define preferred directions}. A theorem
\cite{Kob63} states that the only three--dimensional Riemannian
spaces having the full six-dimensional group of isometries  are 
$\bbbr^3,~Ê \bbbs^3,~ \bbbp^3$ and $\bbbh^3$. Thus, whereas the
universal covering spaces and the projective space  are {\sl globally}
isotropic, the quotient spaces are only {\sl locally} isotropic
\cite{Ell71,Sok77}. 

\item The Bianchi types $IV$ and $VI_a~ (a \ne 0,1)$ are not in
correspondance with Thurston's geometries because
they do not admit closed spaces. This may be related to the fact that their
sectional curvatures are all different from each other :   $\Bigl(-3/ 4,
(-5+{2\over \sqrt5})/ 4, (-5-{2\over \sqrt5})/4 \Bigr)$  for type $IV$,
$\Bigl( 1-a^2, -(1+a)^2, -(1-a)^2 \Bigr)$  for type $VI_a$, but this
conjecture remains to be proved.  

\item From a geometrical point of view, two spacetimes solutions of
Einstein's equations are regarded as physically indistinguishable if they
are isometric. Ashtekar and Samuel \cite{Ash91} have however
emphasized that this may no more be the case in the
 the {\sl hamiltonian} formulation of general
relativity, in which the field equations are
derived from the Einstein action $I = \int R~\sqrt{-g} ~d^4x$.  In the
case of spatially homogeneous spacetimes, it was already known 
\cite{MaC71} that the
hamiltonian description was available for Bianchi class A and
Kantowski-Sachs space-times, but failed for Bianchi class B (for a
review, see \cite{Rya72,Rya75}). Ashtekar and Samuel \cite{Ash91}
have proved that the Lie groups underlying all class B spacetimes are
merely incompatible with a compact spatial topology, a result previously
pointed out in \cite{Ell74}.  This can be surprising since we have seen
that, for instance, locally hyperbolic 3--manifolds (corresponding to class
B types $V$ and $VII_a$) do admit compact topologies. But  the
hamiltonian picture of general relativity further constrains the
4--dimensional metrics (\ref{homo}), and thus imposes additional
restrictions on the topology of the spacelike sections.  This
result unveils an unexpected link between the metric and the topology
through the Einstein's field equations, which should play an essential role
in the minisuperspace approach to quantum cosmology \cite{Mis72}. 
\end{itemize}

\subsection {Example : a quasi-hyperbolic compact space}

 Fagundes \cite{Fag82,Fag85} presented a  ``quasi--hyperbolic" compact
space of the form $\Sigma = T _g \times \bbbs ^1$, with $\bbbh^2 \times
\bbbr$ as universal covering.  It is thus a homogeneous anisotropic
model. The g--torus $T _g$ with hyperbolic 2--metric is parametrized by
the coordinates $\rho$ and $\phi$, the circle $\bbbs ^1$ is parametrized by
the coordinate $\zeta$.

\begin{figure}[tb]
  \begin{center}
    \leavevmode
    \includegraphics{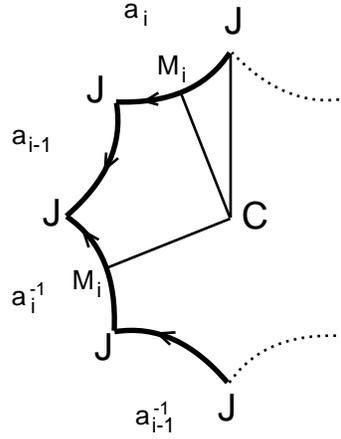}
\caption{\label{Mi}{\it The quasi-hyperbolic model of
Fagundes. The edges of the 4g-gon are identified by pairs, so that the two
$M_i$ are the same, and all the 4g points J are the same. Resolution of
the triangle $CJM_i$ gives $CM_i = a~cosh^{-1}(cot \pi/4g)$, $CJ =
a~cosh^{-1}(cot^2 \pi/4g)$.}}
  \end{center}
\end{figure}

For a better understanding the figure \ref{Mi} depicts a ``horizontal
section" $T_g$ of $\Sigma$. 
The edges of the 4g--gon have a length 
$\mid a_i \mid = 2a~cosh^{-1}(cot \pi/4g)$, and  
$\Sigma$ is described by the metric:
 
\begin{equation} \label{ksmetric}
d\sigma ^2 =  b^2 ~d \zeta ^2 +  a^2~  ( d\rho ^2+ sinh^2 \rho ~d\phi ^2),
~~a, b~ constants.
\end{equation}

The range of coordinates is
$-\lambda < \zeta < \lambda$ ($\lambda$ = const.), $0 < \phi < 2\pi$,
$0 < \rho < \rho_i$.

The volume of $\Sigma$ is $V = 8\pi g \lambda (g-1)a^2b$.
The anisotropy of $\Sigma$ is manifest in the horizontal section : from the
point of view of $M_i$, the points $C$ and $J$
have opposite images  at distance $\mid a_i\mid /2$, whereas  from the
point of view of $C$ only the $M_i$ provide opposite images at this
distance. 


\subsection {Spaces of constant curvature}
 
We emphasize that the preceding example was {\sl not} a space of
constant curvature.  Cosmology, however, focuses mainly on
locally homogeneous {\sl and} isotropic spaces, namely those admitting
one of the 3 geometries of constant curvature. 
Any compact 3--manifold $\cal{M}$ with constant curvature $k$ can be
expressed as the quotient   
${\cal M}~\equiv ~\widetilde{\cal M}/\Gamma$,
 where the Universal Covering space $\widetilde{\cal M}$ is either :
\begin{itemize}
\item  the Euclidean space $\bbbr ^3$ if $k = 0$
\item the 3-sphere $\bbbs ^3$ if $k > 0$
\item the hyperbolic 3-space $\bbbh ^3$ if $k < 0$.
\end{itemize}
and $\Gamma$ is a subgroup of isometries of $\widetilde{\cal M}$ acting
freely and discontinuously. The three following sections are devoted to a more
detailed description of such spaces.

 \newpage
\section {Three-dimensional Euclidean space forms}
\label{flatm}

The line element for the universal covering space $\bbbr^3$ may be written
as~:
\begin{equation}\label{e3m}
d\sigma ^2 = R^2 \{ d\chi ^2 + \chi^2 (d\theta ^2 + sin^2 \theta
d\phi ^2)\} 
\end{equation}

Its full isometry group is $G = ISO(3) \equiv \bbbr^3 \times SO(3)$, and
the generators of the possible holonomy groups $\Gamma$ (i.e., discrete
subgroups without fixed point) include
 the identity, the translations, the glide reflections and the helicoidal
motions
occurring in various combinations.  They generate  18 distinct types
of locally Euclidean spaces \cite{Wol84,Ell71,Art91}. The 17
multi--connected space forms are in correspondance with the 17
cristallographic groups discovered more than a century ago by
Fedorov \cite{Fed85}. Eight forms are open (non compact), ten are closed
(compact).

\subsection { Open models}

When $\Gamma$ does not include glide reflections, the space forms
$\bbbr^3/ \Gamma$ are orientable. They are four :

\begin{itemize} \item type $\cal E$. 

$\Gamma$ reduces to the identity, $\cal M$$ \equiv \bbbr ^3$.

\item type $\cal J$$_{\theta}$

 $\Gamma$ is generated by an helicoidal motion by an angle
$\theta$, $\cal M$ is the  topological product of a cylinder by $\bbbr$.

\item type $\cal T$$_1$ 

$\Gamma$ is generated by two independant translations, $\cal M$ is  the
product of a torus by $\bbbr$.

\item type $\cal K$$_1$ 

 $\Gamma$ includes a translation and an helicoidal motion of  angle
$\pi$ along a direction orthogonal to the translation.

\end{itemize}

When $\Gamma$ includes a glide reflection, the space forms $\bbbr^3/
\Gamma$ are not orientable.  We shall not describe them because of
their lack of interest for cosmology (cf. section 2)

\subsection {Closed models}

\begin{figure}
\centerline{\epsfig{file=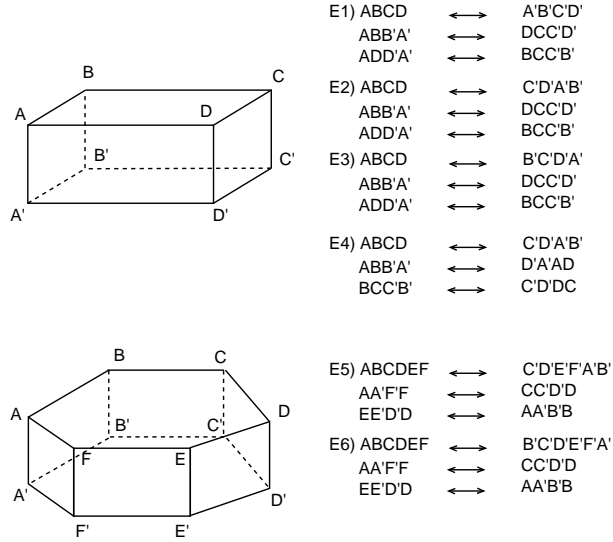, width=8cm}}
\caption{\label{euclid3}{\it The six locally Euclidean, closed, oriented
3-spaces}}
\end{figure}

The compact models can be better visualised by identifying
appropriate  faces of fundamental polyhedra. Six of them are
orientable (figure \ref{euclid3})

The fundamental polyhedron can be a parallelepiped. The possible 
identifications are then :

\begin{itemize} 
\item $E_1$ - opposite faces by translations. The hypertorus $T^3$
already mentioned in section 3.2.3, which is homeomorphic to
the topological product $\bbbs^1 \times \bbbs^1 \times \bbbs^1$, belongs to
this class  and, due to its simplicity, will provide a preferred field of
investigation in the second part of this article. 

\item $E_2$ - opposite faces, one pair being rotated by angle $\pi$

\item $E_3$ - opposite faces, one pair being rotated by $\pi / 2$

\item $E_4$ - opposite faces, all three pairs being rotated by $\pi$.
\end{itemize}

The fundamental polyhedron can also be the interior of an hexagonal
prism, with two possible identifications : 
\begin{itemize} 
\item $E_5$ - opposite faces, the top face being rotated by
an angle $2 \pi / 3$
 with respect to the bottom face

\item $E_6$ - opposite faces, the top face being rotated by an angle $\pi /
3$
 with respect to the bottom face.
 \end{itemize}

Finally, four spaces are not orientable and we shall not describe them
because of their lack of interest for cosmology (cf. section 2).

\newpage
\section {Three-dimensional spherical space
forms}\label{positivem}

 Three--manifolds of constant positive curvature were classified by Seifert
and Threlfall \cite{Sei30}. Their universal covering being the compact
$\bbbs^3$, they are necessarily compact. 

\subsection { The geometry of $\bbbs ^3$}\label{gs3}

The 3--sphere $\bbbs ^3$ of radius R is the set of all points in
4--dimensional Euclidean space $\bbbr ^4$ with coordinates $x^1, x^2,
x^3, x^4$ such that
\begin{equation}\label{S3E}
(x^1)^2 +  (x^2)^2 + (x^3)^2 + (x^4)^2 = R^2
\end{equation}

If we define angular coordinates $(\chi, \theta, \phi)$ by
$$x^1 = R~cos \chi, ~x^2 = R~sin \chi~cos \theta, ~ÊÊ
x^3 = R~sin \chi~sin \theta~cos \phi,~
x^4 = R~sin \chi~sin \theta~sin \phi $$

for $0 \le \chi \le \pi,~~Ê0 \le \theta \le \pi, ~~0 \le \phi \le 2\pi,$

then the metric $d\sigma ^2 \equiv (dx^1)^2 +  (dx^2)^2 +
(dx^3)^2 + (dx^4)^2$ on $\bbbs ^3$ may be written as

\begin{equation}\label{ms3}
d\sigma ^2  =
 R^2 \{ d\chi ^2 + sin^2\chi (d\theta ^2 + sin^2 \theta d\phi ^2)\}.
\end{equation}

The volume is 
\begin{equation}
vol(\bbbs ^3) =  \int_0^\pi {4 \pi R^2 sin^2\chi R d\chi} = 2 \pi ^2
R^3
\end{equation}

Another form of the metric, introduced by Robertson and Walker in the
Friedmann-Lema{\^\i}tre cosmological models (see, e.g., \cite{McV56}),
arises from the coordinate transformation $r = sin\chi$, which puts the metric
into the form :  
\begin{equation}\label{rws3}
d\sigma ^2 =  R^2~ \Bigl\{ ~\frac{dr ^2}{(1 - r^2)} + 
r^2 (d\theta ^2 + sin^2 \theta d\phi ^2)~\Bigr\}.
\end{equation}

\begin{figure}[tb]
  \begin{center}
    \leavevmode
    \includegraphics{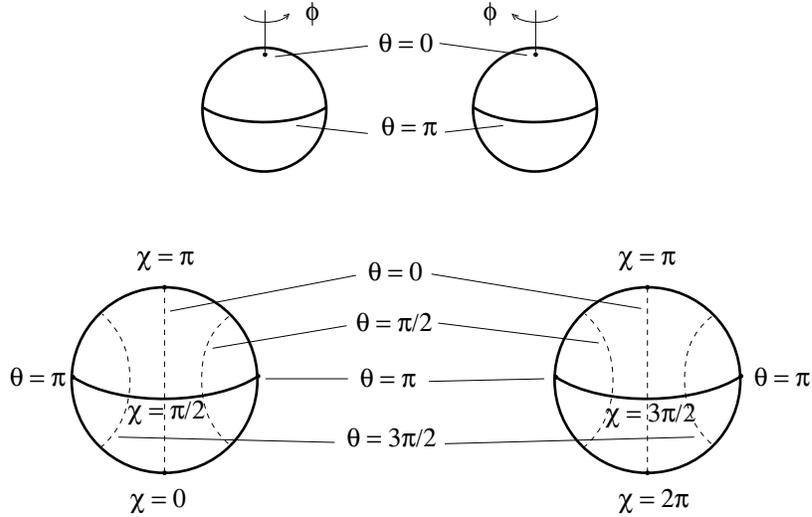}
\caption{\label{S3}{\it  Top : Representation of $\bbbs^3$ by two balls in
$\bbbr^3$ glued together.  Bottom : Behaviour of coordinates within
slices $sin \phi = 0$.}}
  \end{center}
\end{figure}

There are many ways to visualize the 3--sphere. One of them is to imagine
points of $\bbbs^3$ as those  of a family of 2--spheres which grow in
radius from $0$ to $R$, and then  shrink again to $0$ (in a manner quite
analogous to the 2--sphere which can be sliced by planes into circles).
Another convenient way (genially guessed in the Middle Ages by Dante in
his famous Divine Comedy, see \cite{Pet79}) is to consider $\bbbs^3$ as
composed of two solid balls in Euclidean space $\bbbr^3$, glued together
along their boundaries (figure \ref{S3}) : each point of the boundary of
one ball is the same as the corresponding point in the other ball. The result
has twice the volume of one of the balls.

\subsection{The holonomies of $\bbbs^3$}

The full isometry group of  $\bbbs^3$ is $SO(4)$.  A modern summary
by Wolf \cite{Wol84} gives an explicit description of each
admissible subgroup $\Gamma$ of $SO(4)$ without fixed point,
acting freely and discontinuosly on $\bbbs^3$~:

\begin{itemize} \item the cyclic group of order p, $Z_p$  ($p \ge 2$). 

A cyclic group just consists of the powers of a single element : $a, a^2,
\ldots, a^p = Id.$ (for instance the $p^{th}$ roots of unity
$exp(2\pi~mi/p),~m = 0, 1,\ldots, p-1$). In a more geometrical
representation, $Z_p$ can be seen as generated by the rotations by an
angle $2\pi/p$ about an arbitrary axis [$\theta$,$\phi$] of $\bbbr  ^3$.  

\item the dihedral group of order $2m$, $D_m$ ($m > 2$).

$D_m$ is generated by two elements $A$ and $S$ such that (in matrix
notation) $A^m = Id.$, $S^2 = Id.$, $SAS^{-1} = cA^{-1}$, where $c =
exp(2\pi~ki/m)$ is a $m^{th}$ root of unity. In a more geometrical
representation, $D_m$ can be viewed as generated by the rotations in the
plane by an angle $2\pi/m$ and a flip about a line through the origin. Such
symmetries preserve a regular  m-gon lying in the plane and centered on the
origin (figure \ref{dihed}). 

\begin{figure}[tb]
  \begin{center}
    \leavevmode
    \includegraphics{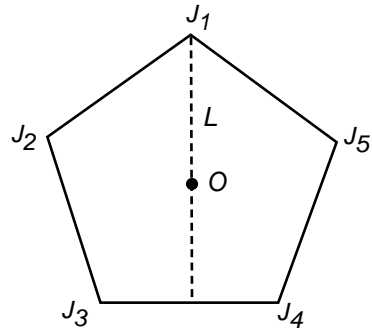}
\caption{\label{dihed}{\it The dihedral group $D_5$.
The pentagon $J_1 \ldots J_5$ is invariant by rotations of angle
$2\pi/5$ about the line L.}}
  \end{center}
\end{figure}

\item the polyhedral groups

They are the  symmetry groups of the regular polyhedra in  $\bbbr 
^3$ (figure \ref{poly}), namely~: 

\begin{itemize}

 \item the group $ T$  of the tetrahedron (4 vertices, 6 edges, 4
faces), of order 12;  

\item the group  $O$  of  the octahedron (6 vertices, 12 edges, 8
faces), of order 24 ;

 \item the group  $I$  of the isocahedron (12 vertices, 30 edges, 20
faces), of order 60. 
\end{itemize}
 \end{itemize}

\begin{figure}[tb]
  \begin{center}
    \leavevmode
    \includegraphics{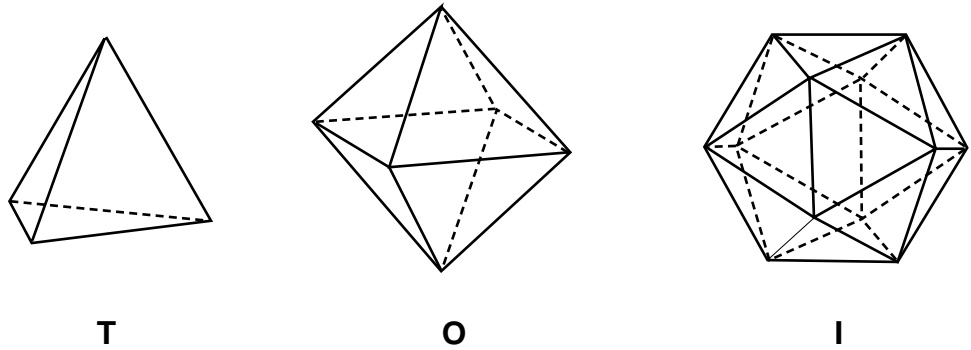}
\caption{\label{poly}{\it  The polyhedral groups.
\underline{Left:}~Tetrahedral group. \underline{Center:}~Octahedral
group. \underline{Right:}~Icosahedral group.}}
  \end{center}
\end{figure}

Note that there are two other regular polyhedra (historically known as the
Platonic solids)~: the hexahedron (cube) and the dodecahedron (12
faces). But the cube has the same symmetry group as the octahedron,
and the dodecahedron has the same symmetry group as the
icosahedron.

All
the homogeneous spaces of constant positive curvature are
obtained by quotienting $\bbbs^3$ with the groups described above.
 They are in infinite number due to parameters $p$ and $m$.

\subsection{The size of spherical 3-spaces}

The volume of  $\cal M$$ = \bbbs^3 /\Gamma$ is simply  
\begin{equation}\label{vol}
vol({\cal M}) = 2 \pi ^2~R^3 / \mid \Gamma \mid
\end{equation}
 where $\mid \Gamma \mid$ is the order of the group $\Gamma$.
For topologically complicated spherical 3--manifolds, 
$\mid \Gamma \mid$ becomes large and $vol(\cal{M})$ is small. There
is no lower bound since $\Gamma$ can have an arbitrarily large number
of elements. In contrast, the {\sl diameter}, i.e., the maximum distance
between two points in the space, is bounded below by $\frac{1}{2}
arccos((1/\sqrt3)~cotg\pi/5)~R \approx 0.326~R$, corresponding to a
dodecahedral space \cite{Ber80}.

\subsection {Examples}

\subsubsection{ The projective space}
 $\bbbp ^3 = \bbbs ^3/\bbbz _2$  is obtained by identifying on $\bbbs ^3$
diametrically opposite points : 

$(x^1, x^2, x^3, x^4) \equiv  (-x^1, -x^2, -x^3, -x^4)$ in (\ref{S3E}),
or,  equivalently,

$ (\chi, \theta, \phi) \equiv (\pi - \chi, \pi - \theta, \pi + \phi)$ in (\ref{ms3}).

\noindent
In contrast with its 2--dimensional analogue $\bbbp^2$, $\bbbp^3$ is
orientable. Its volume is  $ \pi ^2 R^3$. In $\bbbs^3$,  any two geodesics
starting from a point intersect also at the antipodal point, at a distance $\pi
R$ measured along any of these lines. In  $\bbbp^3$,
 two geodesics cannot have more than one point in common.

\subsubsection{A lens space} 
The spaces $ \bbbs ^3/\bbbz _p$ are called lens spaces, due to the
shape of their fundamental domain. Apart from the projective space, the
simplest lens  space  is  $ \bbbs ^3/\bbbz _3$, which divides
$\bbbs ^3$
 into  6 fundamental cells, each having a lens form.  The one  centered onto the
observer at  coordinates, say, ($\chi =0, ~\theta =0,~ \phi_0$) has for
boundaries~:
 the great circle ($\chi =\theta=\pi/2,~ 0 \le \phi \le 2\pi$);  the cone
of geodesics with summit ($\chi =\pi/6,Ê~ \thetaÊ=Ê0$) and
base the previous  great circle ; the symmetric cone with summit
($\chi =\pi/6,~ Ê\thetaÊ=Ê\pi$) and the same base.  When
this fundamental cell is translated, the points on the circle
($\chiÊ=\thetaÊ=Ê\pi/2, ~\phi$)  are transformed to
($\chiÊ=\thetaÊ=Ê\pi/2, ~\phi + Ê\pi/3)$.  Similarly  the points on the 
circle $\thetaÊ= 0$ have their value of $\chi$ increased by $\pi/3$. 
The maximum dimension of the fundamental lens is $ \pi R/2$. The
observer has 5 images of itself given by  ($\pi/3,~  0, ~
\phi_0$), ($2 \pi/3, ~ 0, ~ \phi_0$),  ($\pi , ~ 0, ~\phi_0$),  ($2
\pi/3,~  \pi, ~  \pi+\phi_0$),  ($ \pi/3,  ~ \pi, ~ \pi+\phi_0$).
 
\subsubsection{A dihedral space}
 The simplest ``dihedral" space is $\bbbs
^3/D_3$.   It divides the 3--sphere into 12 trihedral cells. The
observer at coordinates $(0,0,0)$~has  11  images of himself at
coordinates 
$(\pi/3,~  0,~ 0)$, $(2\pi /3,~   0, ~ 0)$, $(\pi , ~  0,  ~  0)$, $(2 \pi
/3, ~ \pi, ~  0)$, $(\pi /3,  ~ \pi,  ~  0)$, $(\pi /2, ~ \pi /2,  ~ 
0)$, $(\pi /2, ~  \pi /2, ~ \pi /3)$, $(\pi /2, ~ \pi /2,  ~  2 \pi /3)$, 
$(\pi /2, ~ \pi /2 ,  ~  \pi )$, $(\pi /2, ~  \pi /2, ~  4\pi /3)$, 
$(\pi /2, ~  \pi /2,   ~ 5\pi /3).$

\subsubsection{ The Poincar\'e dodecahedral space}
The Poincar\'e manifold \cite{Poi53} is an example of $\bbbs
^3/I$. The fundamental 
polyhedron is a regular dodecahedron whose faces are pentagons.
The compact  space is obtained in identifying the opposite faces after
rotating by  $1/10^{th}$ turn in the clockwise direction around the
axis orthogonal to the
 face (figure \ref{PSW}). This configuration involves 120 successive
operations and
 gives already some idea of the extreme complication of such
multi--connected topologies.

\newpage
\section {Three-dimensional hyperbolic space forms}
\label{negativem}
\subsection {The geometry of  $\bbbh ^3$ }

Locally hyperbolic manifolds are by far less well understood than the
other homogeneous spaces.  However, according to the
pionneering work of Thurston \cite{Thu79}, ``almost" all
3--manifolds can be endowed with a hyperbolic structure. Here we
present some elements of the theory;  for a
recent report, see \cite{Ben91}.

It is not easy to have an
intuitive representation  of $\bbbh^3$ because it cannot be imbedded in
$\bbbr ^4$. Instead, it can be seen as an hypersurface of equation  
$-(x^1)^2 +  (x^2)^2 + (x^3)^2 + (x^4)^2 = R^2$ in the Minkowski space 
of metric  $ds^2 = - (dx^1)^2 +  (dx^2)^2 + (dx^3)^2 + (dx^4)^2$. 
Hence the generators of the fundamental group $G$ of $\bbbh ^3$ 
are equivalent to  homogeneous Lorentz transformations \cite{Efi80}. 

If we introduce coordinates $(\chi, \theta, \phi)$ by
$$x^1 = R ~cosh \chi, ~
x^2 = R ~sinh \chi~cos \theta, ~
x^3 = R ~sinh \chi~sin \theta~cos \phi, ~
x^4 = R ~sinh \chi~Êsin \theta~sin \phi$$
 with
$0 \le \chi  < \infty$, $0 \le \theta \le \pi$, $0 \le \phi \le 2\pi$,
\noindent
the induced metric on $\bbbh ^3$ may be written as
\begin{equation} \label{h3m} 
d\sigma ^2 = 
R^2~ \Bigl\{~ d\chi ^2 + sinh^2 \chi ~(d\theta^2 + sin^2 \theta ~d\phi ^2)
\Bigr\}
 \end{equation}

The volume is infinite. The Robertson--Walker form of the metric
-- generally used in relativistic cosmology -- is obtained from the
coordinate change $r = sinh \chi$, which puts the metric into  :
  \begin{equation} \label{rwh3}
d\sigma ^2 =  R^2 \Bigl\{ \frac{dr ^2}{1 + r^2} + 
r^2 (d\theta ^2 + sin^2 \theta d\phi ^2) \Bigr\}
\end{equation}

Other forms of the metric are commonly used :

\begin{itemize}

\item In the upper-half space representation, $\bbbh^3$ is mapped onto 
$\bbbr^3_+ = \{(x,y,z) \in \bbbr^3  \mid z > 0 \}$ equipped with the
metric

\begin{equation} 
d\sigma^2 = \frac{dx^2 + dy^2 + dz^2 }{z^2}.
\end{equation}

The  lines and planes of $\bbbh^3$ become semi--circles and
semi--spheres of $\bbbr^3_+$, which orthogonally
intersect with the boundary.

\item In the Poincar\'e representation, $\bbbh^3$ is mapped into the unit
open ball $\{(x,y,z, \in \bbbr^3  \mid x^2 + y^2 + z^2 < 1\}$.
Hyperbolic lines and planes are semi-circles and semi-spheres
which orthogonally intersect the boundary $\bbbs^2$.

\item In the Klein model, $\bbbh^3$ is  mapped
 into the unit open ball in $\bbbr^3$, with Cartesian
coordinates $(x^i)$, with the correspondence~:  
\begin{equation} \label{kleincoord}
 x^1 = \tanh \chi~ \sin \theta ~ \cos \phi, ~
x^2 = \tanh \chi ~ \sin \theta~  \sin \phi, ~
x^3 = \tanh \chi ~ \cos \theta.
  \end{equation} 
 Then the distance between 2 points $\xx$ and $\yy$ writes~:   
\begin{equation} 
d(\xx,\yy) = cosh^{-1} ~\Bigl[\frac{1-\xx.\yy}{(1-\xx.\xx)~
(1-\yy.\yy)}\Bigr]^{1/2}. 
\end{equation}

The advantage of such a representation is that hyperbolic lines and planes
are mapped into their Euclidean counterparts.
\end{itemize}

\subsection{The holonomies of $\bbbh^3$}
The isometries of $\bbbh^3$ are most conveniently described in the
upper--half space model $\bbbr^3_+$. Their group is isomorphic to
$PSL(2,\bbbc)$, namely the group of
 fractional linear transformations acting on the complex\,\footnotemark[1] \footnotetext[1]{
Whereas the isometries of $\bbbh^2$ involved only real coefficients, cf.
  \secn{holoh2}.} plane~:
 $$z' ={az+b\over cz+d}, ~a,b,c,d \in \bbbc, ~ad - bc = 1.$$
This group operates also as the group of
conformal transformations of $\bbbr^3$ which leaves the upper
half space  invariant.  
Finite subgroups  are discussed in Beardon \cite{Bea83}. 

 \subsection {The size of compact hyperbolic manifolds}

 In hyperbolic geometry there is an essential difference between the
2--dimensional case  and higher dimensions. A {\sl surface} of genus $g
\ge 2$ supports uncountably many non equivalent hyperbolic metrics.
But for $n \ge 3$, a connected oriented n-smanifold supports {\it at most
one} hyperbolic metric. More precisely,  the {\sl rigidity theorem} 
proves that if two hyperbolic  manifolds, with dimension $n \ge 3$, have
isomorphic fundamental groups, they are necessarily isometric to each
other. This was proved by Mostow \cite{Mos73} in the compact case,
and by Prasad \cite{Pra73} in the non--compact case.
 It follows that, for $n \ge 3$, the volume of a manifold and the lengths of
its closed geodesics are topological invariants. 
  This suggested the idea of  using
the volumes to classify the topologies, which could have seemed, at a
first glance, contradictory with the very purpose of topology. 

Each  type of topology is  characterized by some lengthes. For compact
locally Euclidean  spaces, the fundamental polyhedron may
possess arbitrary volume, but no
more than eight faces. In the spherical case, the volume of $\bbbs ^3/
\Gamma$ is finite and  is an entire fraction of that of $\bbbs ^3$ (see eq.
(\ref{vol})),  the maximum possible value. By contrast,  it is possible to
tesselate $\bbbh ^3$  with polyhedra having an arbitrarily large number of
faces. This was already the case in dimension two, with for instance the
4g-gones whose angles are thinned down
by adjusting the surface on the hyperbolic plane. The role of the
volume in $\bbbh ^3$ generalizes that of the area in $\bbbh ^2$. 
Correspondingly, in  the three-dimensional hyperbolic case, the possible
values for the   volume of the $FP$ are bounded from below.   In
other words, {\sl there exists a hyperbolic 3--manifold with minimal
volume}. 

Particular interest has been taken by various authors in computing the
volumes of compact hyperbolic manifolds \cite{Neu85,Fuj90,Koj91}.
Little is known however about the set of all possible values of these 
volumes~: the minimal one $vol_{min}$ is not known, nor whether any
one is an irrational number\,\footnotemark[1] \footnotetext[1]{~The volumes of compact
hyperbolic manifolds are estimated by numerical computation.}
\cite{Thu79}. Thurston \cite {Thu82} proposed as a candidate for the
hyperbolic 3--manifold of minimum volume a space $\cal Q$$_1$ with
volume $vol({\cal Q}_1) = 0.98139 ~R^3$ (where $R$ is the curvature
radius of the universal covering space). The conjecture turned out to be
false when Weeks \cite{Wee85} and, independently, Matveev and
Fomenko \cite{Mat88} found a compact hyperbolic  $\cal Q$$_2$ such
that $vol({\cal Q}_2) = 0.94272 ~R^3$.  Since a ball of radius $R\chi$
has a  volume $\pi  R^3 [sinh(2\chi) -2\chi ]$,  this 
corresponds to a diameter $\approx 0.6~R$. 

However it is anyone's guess how the real minimal value
may be. Meyerhoff \cite{Mey86} has proved that $vol_{min}
> 0.00082 R^3$. The smallest $vol_{min}$, the more interesting the
corresponding manifold for cosmology (see next sections).

\subsection {Examples}\label{negativee}

Topologists have been able to sketch a classification of 
compact hyperbolic spaces in terms of volumes. A topology
is completely characterized by the number of faces of the fundamental 
polyhedron and by the various ways to identify them. This guarantees that
the  number of topological classes,  although infinite, is countable.

The full classification of three--dimensional
hyperbolic manifolds is far from  being fully understood today, although it
seems less unreachable than before.  Various
means are available to build an infinite number of
hyperbolic spaces. Thurston \cite{Thu79} has given a procedure for
effectively constructing hyperbolic structures by gluing together ideal
polyhedra. The idea goes back to Poincar\'e, see e.g. \cite{Mas71}. However
the construction of {\sl closed} manifolds is far more complicated
than that of non--compact ones \cite{Gut79,Apa87}.  Many
authors use the Dehn's Surgery method, which consists in removing
certain ``regular" pieces of a manifold and gluing them back with a
specific twist. 
We give below some well-known examples.

\subsubsection{Non-compact models}

\begin{itemize}
\item
It is  possible to construct a 3--space of
constant  negative curvature having the metric
\begin{equation}
\label{}
d\sigma^2 = d\chi^2 +  \cosh^2\chi~ d{\sigma'}^2,
\end{equation} 
where $d{\sigma'}^2$  is the metric (\ref{h2m}) of a locally hyperbolic
surface $\bbbh^2 / \Gamma$.
Since there is an infinite number of 
topologies on $\bbbh^2 / \Gamma$, this offers a way of 
building an infinite number  of topologies for locally hyperbolic
spaces. These space are not compact in the direction
orthogonal to
 $\bbbh ^2/\Gamma$ ($ -\infty < \chi < \infty$).  
  
\item \label{horn}
Sokoloff and Starobinskii \cite{Sok75} have considered
multi--connected hyperbolic spaces whose fundamental polyhedron
is the non compact domain  comprised between two ``parallel" (that
is, non intersecting) planes.  With the coordinate transformation 

$$x = - \ln[\cosh \chi - \sinh \chi~\cos \theta ] $$ 

$$y = \frac{\sin \theta ~\cos \phi}{\mbox{Argth}
\chi - \cos \theta}$$ 
$$z = \frac{\sin \theta ~\sin \phi}{\mbox{Argth} \chi - \cos \theta}$$
the
metric (\ref{h3m}) takes the form~: 

\begin{equation} \label{xh3}
d\sigma ^2Ê= Êdx^2Ê+ e^{-2x} (dy^2 + dz^2). 
\end{equation}

The boundaries of the fundamental domains are
defined by the relation $a ~e^{-x} = \Lambda$. Each domain (in particular
the FP)  represents the interior of a  ``cylindrical horn".
Holonomies occur via the identifications $y \rightarrow y + ma$,
where $a$ is the circonference of the cylinder and $m$ an integer. 

\end{itemize}

\subsubsection{Compact models}

\indent
{\bf The Seifert-Weber Space}.

\begin{figure}[tb]
  \begin{center}
    \leavevmode
    \includegraphics{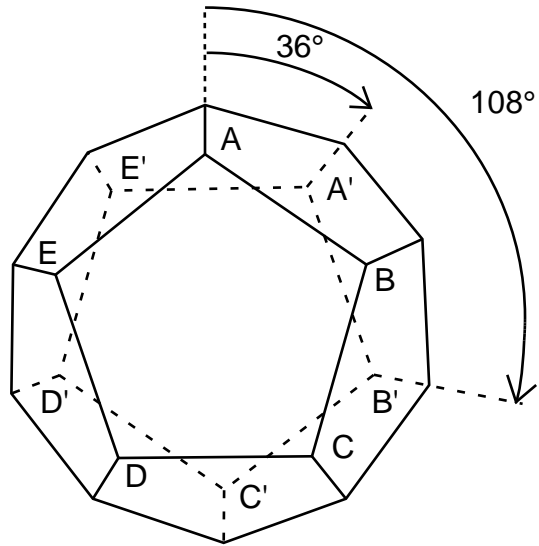}
\caption{\label{PSW}{\it The  spherical Poincar\'e space and the closed
hyperbolic Seifert-Weber space are respectively obtained by
identifying opposite faces of a regular dodecahedron after rotation by
$\pi/5$ and $3\pi/5$.}}
  \end{center}
\end{figure}

Seifert and Weber \cite{Sei33} have obtained a compact hyperbolic
manifold whose  fundamental polyhedron is a
dodecahedron, with opposite pentagonal faces fitted together after
twisting  by 108 degrees (figure \ref{PSW}).

{\bf The L\"{o}bell Space}. \label{lobell}

\begin{figure}[tb]
  \begin{center}
    \leavevmode
    \includegraphics{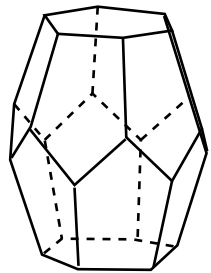}
\caption{\label{Lobell}{\it  The FP for the L\"{o}bell
topology is made of 8 such 14--hedra pasted together. All the angles in the
figure are right angles in $\bbbh^3$.}}
  \end{center}
\end{figure}

 L\"{o}bell \cite{Lob31} has constructed a compact hyperbolic manifold,
later on studied by Gott \cite{Got80} in a cosmological context.  The FP is a
14 faces polyhedron, two faces of which are regular rectangular
hexagones and the 12 others rectangular regular pentagones (figure
\ref{Lobell}). The formulae of hyperbolic  trigonometry permit to
estimate the surfaces of the faces from their angular deficits :    the
area of each pentagon is $(\pi/2) R^2$, that of each hexagon is $\pi
R^2$, while the edges have a length $1.32~ R$. Around each
vertex, 8 polyhedra can be placed and glued together to tesselate
$\bbbh^3$.   In fact, an infinite number of compact  hyperbolic
3-spaces can be build by pasting together various numbers of
these 14--hedra, and suitably identifying the unattached faces.

{\bf The Best Spaces}. \label{Bests}

\begin{figure}[tb]
  \begin{center}
    \leavevmode
    \includegraphics{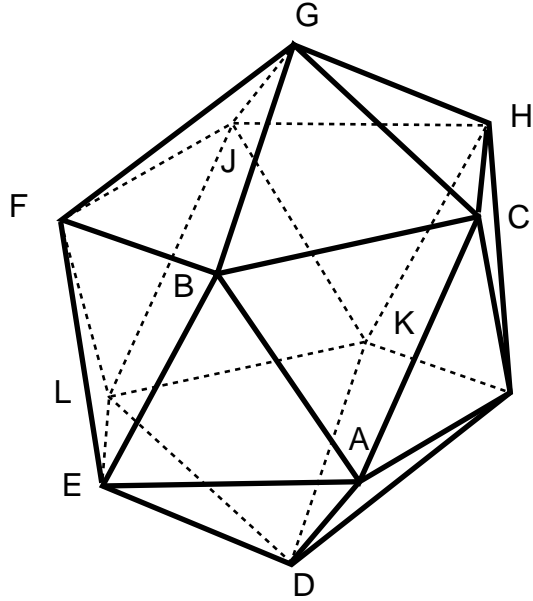}
\caption{\label{Best}{\it  The closed  hyperbolic Best space.
The pairwise identifications of faces are $ADI
\leftrightarrow BGC$,  $ICA \leftrightarrow ABC$, 
$EFL \leftrightarrow DIK$, 
$DEL \leftrightarrow KLD$, 
$GFB \leftrightarrow HJK$, 
$FGJ \leftrightarrow GJH$, 
$ABE \leftrightarrow LJF$, 
$FEB \leftrightarrow CGH$, 
$AED \leftrightarrow KIH$, 
$JKL \leftrightarrow CHI$ (where the order of vertices in the faces
in maintained in the identification).}}
  \end{center}
\end{figure}

 Best \cite{Bes71} has constructed 
several compact hyperbolic manifolds whose FP is a regular
icosahedron.  One of them was studied in details by  Fagundes
\cite{Fag86,Fag89} in a cosmological context. Its outer structure
is represented in figure (\ref{Best}). 
The corresponding generators of the holonomy group are
expressible in terms of $4 \times 4$ matrices corresponding to 
homogeneous Lorentz transformations; for details, see Appendix A
of \cite{Fag89}. The manifold  is avantageously described in the 
Klein coordinates (\ref{kleincoord}).

{\bf The Weeks Space}\label{Weeks}

\begin{figure}[tb]
  \begin{center}
    \leavevmode
    \includegraphics{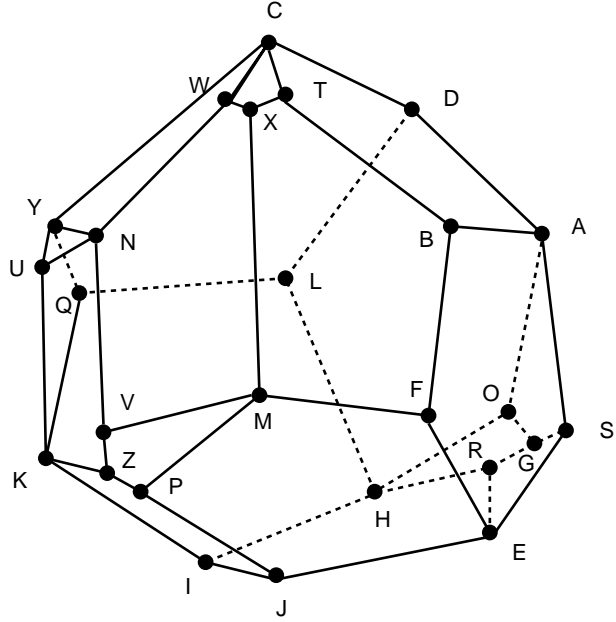}
\caption{\label{Wee}{\it The closed  hyperbolic Weeks space.
The pairwise identifications of faces are $ABTCD
\leftrightarrow HLDAO$,  $SEFBA \leftrightarrow FMPJE$, 
$RHIJE \leftrightarrow IKQLH$, 
$FMXTB \leftrightarrow UKZVN$, 
$WNUYC \leftrightarrow IJPZK$, 
$CYQLD \leftrightarrow MXWNV$, 
$CTXW \leftrightarrow HOGR$, 
$KQYU \leftrightarrow ERGS$, 
$MPZV \leftrightarrow ASGO$.}}
  \end{center}
\end{figure}

The Weeks manifold \cite{Wee85} is a polyhedron with 26
vertices and 18 faces, among which 12 are pentagons and 6 are
tetragons. Its outer structure
is represented in figure (\ref{Wee}). It has the peculiarity to be the
smallest compact hyperbolic manifold presently known. Given the fact
that, in quantum cosmology, the probability for spontaneous creation of a
compact universe is bigger for a small one than for a large one, the Weeks
space was studied by Fagundes \cite{Fag93} in a cosmological context.
Fagundes provides also numerically the coordinates of the vertices and
the 18 generators of the holonomy group (also expressible in terms of  $4
\times 4$ matrices corresponding to Lorentz transformations). 

\section{Multi--connected cosmological models} \label{Cosmic models}

\subsection{Simply and multi--connected models}

The metric of spacetime specifies its local  geometry but not its topology.  Such a
metric, solution of the Einstein's equations for   a given form of the cosmic
stress--energy tensor,    may correspond to different models of the Universe with
different spatial topologies. On the other hand, the  local property of being a
solution of  the Einstein's equations does not automatically guarantee  that the
boundary conditions are satisfied. Thus,  solutions corresponding to  the
same metric but to  distinct topologies may have different status.    

Among the   cosmological models,  the most usually considered  in the litterature
are those with  simply--connected space. We will refer  generically to them 
as the Simply--Connected Models, hereafter SCM's. However, the assumption of
simple--connectedness  is  arbitrary and can be dropped   out.  Our
interest in this paper    focusses on   the   multi--connected cosmological models
(hereafter MCM's), defined as those having   multi--connected (oriented) spatial
sections\footnotemark[1] \footnotetext[1]{They have been also called ``glued--together" or ``spliced" universes
\cite{Sok75}}.   All of them obey  the Einstein's equations.

The most celebrated cosmological solutions of Einstein's equations  are the
homogeneous and isotropic \frl ~ (FL) models, which obey the cosmological principe,
\ie, where spatial sections have constant curvature. Beside  the usual  ``big--bang"
solutions, the FL models also include   the de Sitter solution, as well as  those
incorporating a cosmological constant,  or a non standard  equation of state. From a
spatial point of view,  the FL models fall into 3 general classes, according to
the sign of their spatial curvature $k=-1$, 0, or~$+1$. The
spacetime manifold is advantageously described by the \rw ~metric 

\begin{equation} \label{rwmetric} 
ds ^2 = c^2~dt^2 - R^2(t) ~d\sigma ^2, 
\end{equation} 

where  $$  d\sigma ^2 = d\chi ^2 + S_k ^2 (\chi ) (d\theta ^2 +
sin^2\theta ~d\phi ^2) $$  is the metric of a 3--dimensional homogeneous manifold,
flat ($k=0$, see eq. \ref{e3m}) or with curvature  ($k=\pm 1$, see eq.
\ref{rws3}, \ref{h3m}). We have defined the function  \[ \left\{  \begin{array}{ll}
  S_k(\chi)Ê= \sinh(\chi)   & \mbox{if  $k=-1$}\\

  S_k(\chi)Ê= \chi & \mbox{if  $k=0$}\\

  S_k(\chi)Ê= \sin(\chi)  & \mbox{if  $k=1$}
\end{array}
\right. \]
and $R(t)$  is the scale factor, chosen equal to the spatial curvature radius for non
 flat models, so that  $k/ R^2$  is the  spatial curvature.   The
quadratic form  $R^2(t) d\sigma ^2$  denotes the metric of space at the cosmic time
$t$, which remains homothetic during the cosmic evolution.  The coordinates $\chi$,
$\theta$ and $\phi$ are {\sl comoving}~: they remain the same for a cosmic object (a
galaxy)  in free fall, \ie,  which follows the cosmic expansion.  The proper distance
(from the observer) to  any object, evaluated at the present time
$t_0$, is simply $d_{proper}=\chi ~R(t_0)$. It remains constant if the object   
exactly follows  the expansion law  and  is called the  {\it proper comoving 
distance}. The metric of the comoving space is $R_0 ^2~d\sigma ^2$,  $R_0 = R(t_0)$
being the present value of the scale factor. 

Quite usually  an other radial coordinate, the ``circumference" coordinate  $r =
S_k(\chi)$, is used instead of $\chi$ in the metric.   For simply--connected models,
the whole space is described when the longitude (or right ascension) $\phi$ varies 
from $0$ to $2\pi$; the latitude (or declination)  $\theta$ varies  from $-\pi/2$ to
$\pi/2$ ; the radial   coordinate   $\chi$  goes from  $0$ to $\infty$ if $k = 0$ or $
-1$,  or from $0$ to $\pi$ if $k=1$.   If $r$ is used instead of $\chi$, it goes from 
$0$ to $\infty$  if $k = 0$ or $ -1$,  or from 0  to 1, and then back from 1 to 0  if
$k=1$. For multi--connected models, the space is smaller and the variation range of the
coordinates $(\chi, \theta, \phi)$ is reduced. 

To any MCM is associated an unique SCM sharing exactly the
same  kinematics and dynamics. The universal covering  of the spatial  sections of a MCM 
may be identified  to the  spatial sections of the corresponding SCM ($\bbbr ^3$,
$\bbbh  ^3$ or $\bbbs  ^3$ for FL models ) at the same time of its evolution.  In
particular, the scale factors $R(t)$  are exactly identical. In fact, most characteristics of the
\frl ~models are preserved when we turn to the MCM's, and this is a precious
guide for their study.  

It is presently believed that our Universe is correctly described by a \frl
~model. But the values of the cosmic parameters are not known accurately enough to
decide the sign of the curvature.
It is worth to remark that the  multi--connectedness of space would provide an
independant information on the cosmic parameters.  For instance, the type of
non--trivial topology observed  would dictate the sign of the spatial
curvature, since the structures of the holonomy groups are completely different for the
cases $k = 1,~0,~ -1$.  
Thus, beside its own specific  interest, multi--connectedness would offer a very efficient  
tool of investigation  in observational cosmology.

\subsection {Properties of the Friedmann--Lema\^\i tre models} \label{Friedmann models}

The FL models are specified by the sign of the curvature ($k=0,-1$ or 1) and
the scale factor $R(t)$, as it appears in the metric (\ref{rwmetric}). As usual, we
define the Hubble constant $H_0 = \frac{\dot{R} }{ R }\mid _0$, and the deceleration
parameter $q_0 =  \frac{R \ddot{R} }{ \dot{R} ^2 }\mid _0$, where the subscript
$0$ means that the quantity is evaluated at the present time. Thus $H_0$ is also
the constant appearing in the Hubble law $c~z = H_0 ~D$ for galaxies.  The
dynamics of the Universe, for the matter--dominated era,, \ie, in the last ten billion years, obeys the Friedmann
equations in the following form : 
 
\begin{equation}
\frac{\ddot{x}}{H_0^2} = \frac{\Omega}{2x^2} + \lambda ~x
\end{equation}

and 

\begin{equation}
\frac{\dot{x} ^2}{H_0^2} = \frac{\Omega}{x} + \lambda ~x^2 + 1 - \Omega
-\lambda, 
\end{equation}

where 

\begin{equation} \label{redshift}
x(t) \equiv  \frac{1}{1+z} = R(t)/R_0
\end{equation} 

and the dot denotes
time differentiation.  Note that, in this latter equation, the redshift $z=z(t)$
is regarded as a timelike coordinate, as usual in the \frl ~models. This is not
contradictory with the use of $z$ as a spatial coordinate also, through the fact
that the null geodesics establish a relation between look--back time and distance
from the observer.  As usual, $\Omega = 8\pi~G\rho_0/3~H_0^2$ is the
(present) density parameter of the universe and $\lambda = \Lambda/3~H_0^2$
is the ``reduced" cosmological constant.  These equations allow us to evaluate the function
$R(t)$, to estimate the age of the universe (since the duration of the radiation era was
negligible compared to that of the matter era), etc. We will often refer to $h$ as the Hubble
constant $H_0$ in units of 100~km~s$^{-1}$~Mpc$^{-1}$, so that distances estimated
from redshifts are expressed in units of $\hmpc$. This defines the Hubble time $H_0^{-1} =
9.78~10^9~h^{-1}$~years, and the Hubble length  $c~H_0^{-1} = 3000 ~\hmpc$. Note
that the function $R(t)$  defines a correspondence between
redshift and cosmic time through the relation (\ref{redshift}) which holds in MCM's
as well as in SCM's. 

\subsection {Homogeneity, Isotropy and   Finiteness} \label{Homogeneity}

The most commonly studied universe models are based onto the cosmological
principle which implies   spatial homogeneity. Also, spatial  isotropy
is assumed, in accordance   with the observed  distribution of cosmic
objects, and with observations of the CMB.  This  implies, from Schur's theorem,  that
space has constant curvature.  The existing  litterature on multi--connected
cosmological models (with some exceptions, \eg ~\cite{Sok75,Fag83b})  almost
exclusively consider this case. 

The isometry group $G/\Gamma$  of a manifold is smaller than $G$, that of
its universal covering. As a consequence, the isotropy of space is 
broken in multi--connected models, excepted for the projective space (see  \S 5.2).
This breaking of symmetry may be  apparent through the presence of some    
principal directions. In a  cylinder  $\bbbr \times S^2$ for instance, compact in 2
dimensions and infinite  in the other, the metric tensor is exactly the same  at
every point~: it keeps  local  homogeneity. However,  it is not globally
isotropic and has not the maximal symmetry. It is worthy to note that globally
anisotropic models do not contradict observations, since the homogeneity of space
and the local isotropy ensure the complete isotropy of the \cmb, and the
statistical isotropy of the distribution of discrete sources \cite{Ell71,Sok74,Fag82}.
However, as we shall see in \S 12, global anisotropy  can influence the spectrum of
density fluctuations. 

A major interest of the MCM's  come from the fact that the
compact (finite) or non compact character of space is not linked to the sign of
the curvature, unlike for the  simply--connected ones. Multi--connected models
with zero or  negative curvature can be compact in some, or all their
dimensions.  For instance a toroidal universe, despite its zero  spatial
curvature, has a finite circumference and a finite volume which may in
principle  be measured.  It contains a finite amount of matter. But a
cylindrical universe (in the sense that the spatial sections are cylinders) 
is compact in 1  dimension only and has an infinite volume, although   a
finite circumference in the principal direction. 

It is well known that, in    the  simply--connected \bb~ models,  the
homogeneity of space cannot be {\it explained} but is assumed in
initial conditions. These initial conditions Êare often said, in this
regard,   to be       very ``special'',  in the sense that  curvature  and all
physical  properties  are identical in   regions  of space which 
have never been in causal contact~: the values of the metric at 
different positions in  space, although they  may be seen as independent initial
conditions, are indistinguishable, thus requiring a ``fine tuning".  This so--called 
``homogeneity problem'' has been particularly emphasized in relation
with the \cmb ~observations, which confirm the fundamental prediction of the \bb
~models that  the Universe was homogeneous better than to $10^{-5}$ at the
recombination period, when this radiation was scattered
for the last time.  

It has been  suggested that  a past inflationary era could   be an explanation  for the
homogeneity of the Universe.  But convincing arguments in favor of inflation only
exist  in models where space was already homogeneous before inflation \cite {Gol92},
so that the homogeneity problem is only pushed back in time. Moreover, no satisfactory
model for inflation exists. In any case,    there is no guarantee that inflation, which  involves
quantum effects  interacting with   gravity, may be treated in   classical (\ie, non quantum)
cosmology. No theory for quantum gravity or quantum cosmology is presently  available.
But  it is certain  that, if this happens,  topological questions will have a very important role
to play   (see   \secn{Quantum cosmology}).

In order to   explain the observed homogeneity of space,  it would be tempting to
invoke a causal process making homogeneous    an initially
heterogeneous universe. But, because of the  causality
constraints, such a process had no time to act before \rec, at a scale sufficiently
large to account for the isotropy of the \cmb. This is true in models without
inflation. In models with inflation, on the other hand, causal processes had no time
to homogeneize the universe before the occurrence of inflation, in order to allow
this latter to start. Thus the  past attempts to propose
chaotic models which homogeneize with time  \cite{Ree72}  failed because of  the
observed isotropy of the \cmb.  Such models could however be reconciled with observations
if the Universe is multi--connected  \cite{Ell71,Ell79,Got80}.
In that case a {\it small} Universe could have become totally causally connected
before the \rec ~period.  We discuss this possibility in \secn{Homogeneization}.

\subsection{Quantum cosmology and the early Universe}

\subsubsection{Quantum cosmology} \label{Quantum cosmology}

Since the \gr~ theory provides no prescription concerning the spatial topology,
the questions of connectedness  and homogeneity of the Universe may well
finally  relate  to   quantum cosmology. In   classical (non quantum) 
relativity, a theorem due to Geroch \cite {Ger67} states that no
topological change may occur in a non singular spacetime. However it
is believed \cite{Gib90,Hor91} that a theory of quantum gravity, if any, would
allow    changes of the topology of space.  Although this
field is   presently not fully developed,  various approaches have been
proposed to address the question of a quantum origin of the universe,
or of quantum transitions in its very primordial state. Such studies
are aimed  to answer the questions concerning the origin of the
geometry of the universe, the value of the cosmological constant,  the
material content, and the origin of cosmic fluctuations. In these
approaches the connectedness of space plays a very important role since
it is related to its compactness. 

Here we do not intend to fully review these approaches but only briefly mention
the role that topology can play. This problem has been addressed by various authors,
(\eg, \cite{Atk82,Gon89,Gur86}).   Gurzadyan  \& Kocharyan \cite{Gur86} considered
for instance  the framework developed by Hawking \&
Hartle  \cite{Har83,Haw84}. Very shortly, a wave--function of the universe $\Psi$
is defined, which obeys the Wheeler--de~Witt equation, an analog of the Schr\"odinger
equation for quantum cosmology.  $\mid \Psi \mid ^2$  defines the amplitude of
probability associated to  the corresponding universe. 

The wave--function $\Psi$ only depends on the characteristics of space and of
its content,  (but \cite{Gur86} consider only the case without matter,
although with a cosmological constant), \ie, a 3--manifold $S$ with Riemaniann   
metric $h_{ij}$ and matter field configuration $\Phi$ on $S$. According to the
``quantum--geometrodynamical" formalism, the wave function $\Psi$ may be calculated,
in the quasi--classical approximation, as an integral  over all 4--dimensional
manifolds ${\cal M}$ which admit $S$ (with its metric) as a boundary. This
defines the field of action   for the Wheeler--de~Witt equation :     the
infinite--dimensional space of all 3--dimensional Riemaniann metrics  $h_{ij}$, called
the superspace.     Usually only homogeneous and isotropic closed spaces are
considered, so that superspace is reduced to ``minisuperspace''. In this framework,
\cite{Gur86} have compared the probabilities for creation of universe
with closed space, of constant positive ($\bbbs^3$ or a multi--connected
space having  $\bbbs^3$  as universal covering), zero ($T^3$ or a
multi--connected closed space having  $\bbbr^3$  as universal covering), or
negative (a multi--connected space having  $\bbbh^3$  as universal
covering) curvature. They  conclude that the creation probability of a
spherical \frl ~universe is larger than a toroidal one, which is itself
larger than one with negative curvature. For inflationary universe
(defined as created with a very high Hubble parameter), the 3
probabilities become equal.  They also calculated the probability that a
transition with a change in the topology of the universe, allowed in
quantum (although not in classical) cosmology, occurs. For a toy--model,
they found that the transition from a sphere to a torus (with equal
volumes) is practically impossible, for instance much  less probable than
a transition from a sphere to an other one with different radius.  

We do not want to insist too much on the details of these  results, in which 
matter  and matter creation have been neglected.  Moreover, quantum
cosmology is only very tentative and lacks an admitted interpretation.
But we want to emphasize  that :

- Connectedness is to  play in quantum cosmology  a role as much important, and
probably more,  than curvature. 

- Multi--connected models are at least as  probable than simply--connected ones. 

\subsubsection{Quantum effects in the early universe}
   
By the play of expansion, the spatial dimensions of a MCM  were  very small in the
primordial universe. In the first moments they may have been comparable to  the scales
of microphysics and quantum physics, thus allowing quantum and other peculiar 
effects. This is related to the thermal story of the primordial Universe,   the
generation of primordial fluctuations, the matter--antimatter asymmetry, \etc

Quantum field theory plays an important role in the description of the early
universe~: the distributions and   properties of the matter, radiation, and
energy contents  rely on quantum physics and statistics. It
has also been realized in the   recent years that quantum physics could also play a
role  through   the fundamental state -- the vacuum --  of some quantum fields. The
popular inflation idea,  for instance, results from the hypothesis that the dynamics
of the Universe was  dominated by the vacuum energy in a distant past, 
also in relation with  the question of the cosmological constant. 

There is no wide consensus over the concept of vacuum energy  which lies at
the interface between quantum physics and general relativity, not
compatible  in this regard. On the other hand, the Casimir effect \cite{Cas48},
observed in the laboratory, has been interpreted in terms of vacuum energy, \ie, the
energy of a  fundamental state of a quantum field. Quantization in flat space is
usually made with boundary conditions at   infinity. Some constraints (like Ê the
interposition of conducting plates for an electromagnetic field) may
impose   different boundary conditions at finite distance. The
consequence is the suppression of some modes for the vacuum state, with a
different associated energy. The observed Casimir effect is
interpreted as the (dynamical) consequence of the energy difference
between the 2 different  vacuum energies, with and without the
plates. 

This has led to the idea that vacuum energy, and vacuum energy {\sl differences} can
play a role in cosmology. We must however recall that the concept of an absolute
vacuum energy remains controversial. Even the interpretation of the vacuum energy
  differences  associated to the Casimir effect are not very clear. 
Despite these difficulties,  such  considerations have been extended to curved
spacetime. This is the case of the   Unruh (quantum vacuum effect in Rindler
space, \ie, in the space seen by an accelerated observer) and Hawking (quantum
effect in the vicinity of a black hole)  effects, which   present analogies with 
the Casimir effect. But there is no consensus  either about the  generalization of
quantum theory, not speaking about vacuum effects, to curved  spacetime. 
All these effects are induced by a modification of the boundary
conditions imposed to quantum fields. Since  the multi--connectedness of
space would  also modify those (by closing space), similar consequences
may be  expected.  

Bytsenko \& Goncharov (\cite{Byt91,Gon91})
have   evaluated  the {\it topogical Casimir effect} of a massless real scalar
field on multi--connected spacetimes. Already for a fixed spacetime model,
with a given topology (characterized by its holonomy group $\Gamma$),  they remarked
that    different {\it topologically inequivalent configurations}  
  of the scalar field do  exist. To each such configuration $C$ is associated a 
different vacuum energy, called ``Casimir energy" $E$.  It is  
defined, as usual in quantum field theory,  as  the vacuum expectation value of the
corresponding hamiltonian $H$
 $$E(\Gamma,C) = <0 \mid H \mid 0>,  $$ which depends both on $\Gamma$ and
$C$.  Considering    hyperbolic spacetimes  $\bbbr \times \bbbh ^n /
\Gamma$, of dimension $n+1$, where $\bbbr $ is the time line, $\bbbh ^n$ is the
$n$--dimensional Lobachevsky space and $\Gamma$ a  discrete group of isometries of 
$\bbbh ^n$  without fixed points, they were able to calculate
$E(\Gamma,C)$.

In their first paper \cite{Byt91}, they  consider   the
3--dimensional   spacetime $\bbbr \times \bbbh ^2 / \Gamma$~: the spatial part 
$\bbbh ^2 / \Gamma$ is a compact surface of genus $g > 1$. They
evaluate the  number of different topologically   configurations to be $2^{2g}$ 
and  calculate the corresponding  topological Casimir effect, for each of them.
Their following paper \cite{Gon91} turns to the more
realistic case of  a 4--dimensional hyperbolic spacetime 
$\bbbr \times \bbbh ^3 / \Gamma$, where $\bbbh ^3/\Gamma$ is compact. A first
evaluation of the Casimir energy leads to a formula which contains an explicit
infinity. Since  it does not depend on the characteristics of $\Gamma$,
they interpret it as the full (infinite) energy of the spacetime $\bbbr
\times \bbbh ^3$. Thus   they throw it away and obtain, after this renormalization, 
the desired Casimir effect as the finite shift between the (infinite) energies
associated to  $\bbbr \times \bbbh ^3$ and  $\bbbr \times \bbbh ^3 /
\Gamma.$      A vacuum energy density is then calculated by dividing by
the volume. One part of this energy comes from the topology of space
itself. The other comes from the peculiar topological configuration of
the field under consideration.    

The authors do not  provide a tractable   formula, and they do not  try to interpret
the possible physical or cosmological consequences of their calculations.
Considering the uncertainties associated to quantum theory in curved space time 
and to the concept of vacuum energy, this would probably have been premature. But
their work illustrates well how the topology of space must be taken into account
for such effects. In particular it follows that no discussion concerning the
cosmological constant or vacuum effects in cosmology can avoid to address the
question of the connectedness of space. 

Quantum field  theory in compactified spacetime may also play a role in other
contexts. For instance, Elizalde \& Kirsten \cite{Eli94} mention, beside the Casimir
effect itself,  the influence of the topology on the effective mass of a quantum
field, or on  particle creation. They   consider in more details the possibility of a
topological symmetry breaking    generating mass for a quantum field. 
Despite the lack of a well defined framework in which to study all these  
topics,  the various existing publications show that quantum effects in the early
Universe cannot be studied without reference to   topology. 

\newpage
\section{Observing a multi--connected Universe}\label{sec10}
\subsection {The  universal covering space as the observer's world} \label{universal
covering}
 
Multi--connected spaces,   regardless of their spatial curvature, are compact in
one spatial direction at least, and possibly in the 3.     They can have a finite
volume even if the curvature is negative or zero,  for instance when the cosmological
constant $\Lambda=0$ and the density parameter  $\Omega \leq 1$. Such models with
compact   spatial sections  are called  generically  ``small universes"
\cite{Ell86}.

The aim of   cosmic topology  is to select, among the models having    
$\bbbh^3$, $\bbbr^3$, or $\bbbs^3$ as universal   covering  space, those compatible
with the present observational data, and  to propose observational tests of
multi--connectedness.  Part of this task has been undertaken  by various authors and
will be reviewed there.

Celestial objects lie in real space where they can be characterized by 
3 spacelike coordinates~: in general   2 angular coordinates,  labelled $\theta$
and $\phi$ (right ascension and declination), and a  distance (for instance the
proper distance $d_{proper}$), not  directly measurable in cosmology, and  
usually represented by   the  comoving  coordinate  $r$ or $ \chi$ (see eq.
(\ref{rwmetric})). Comoving objects (like for instance the galaxies which follow
the cosmic expansion) keep fixed values of comoving coordinates $r$, $\theta$ and
$\phi$. Events occur in spacetime, and are defined by a spatial position
\{$\theta$, $\phi$, $r$ \} and one time coordinate $t$. The 2 angular
coordinates  $\theta$ and $\phi$ are observable. But, in general, only one more
coordinate is observable, say  the redshift  $z$ which has  a mixed (both spatial
and temporal) nature. In addition, or as an alternative, other types of  
distance can be observed~: the luminosity--distance $d_L$, the
angular--diameter--distance $d_{AD}$,  or  similar quantities.   We will  
generically refer as $d_{obs}$  to such observable  quantities   like $z$,   $d_L$
or $d_{AD}$   Ê  as $d_{obs}$, as opposed  to the proper distance $d_{proper}$.
ÊThus   $\{ \theta, \phi, d_{obs} \}$  form a set $S_{obs}$ of {\it observed}
coordinates.

 All cosmic  information comes through light--rays,   along null geodesics of
spacetime. The  observer lies at one end;   an  event (emission of radiation from the
source) at the  other. In a SCM, there is in general one and only one null
geodesic relating a given     object at a given position in space  to
the observer (see however counter examples below). This creates a one--to--one
correspondance between an object in space (at distance $d_{proper}$) and an event in
spacetime (characterized by a set of  observable coordinates $S_{obs}$). 
However, the set $S_{obs}$ of measurable quantities  characterizes the geodesics,
and not directly the cosmic object. The mentionned correspondence has for
consequence that the  redshift inceases monotonically with the distance to   the
emitting object. 

\begin{figure}[tb]
  \begin{center}
    \leavevmode
    \includegraphics{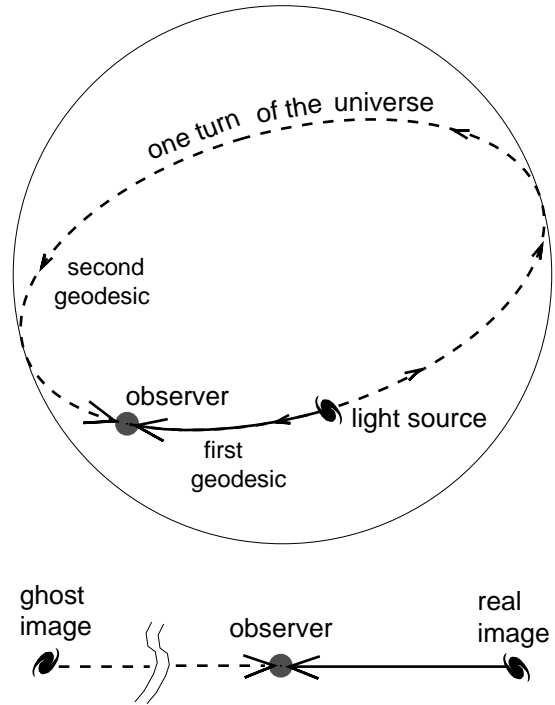}
\caption{\label{fig10_1}{\it Ghosts of the second kind in a spherical
(simply--connected) Universe.}}
  \end{center}
\end{figure}

Even for the SCM's,  there is one case where this correspondence may not  not hold~: 
in the \frl ~models with positive constant spatial curvature,   space  $\bbbs^3$  is
compact.  Light--rays   may in principle make more than one turn around the Universe
before reaching the observer. In such a case, they would generate different ``ghost
images" of a unique  source (figure \ref{fig10_1}).  Ghosts of this type are  called
``ghosts of the second kind". In most \frl ~models  the universe is not old enough and  such
light--rays had no time to perform one single  turn, so that  ghosts
cannot be observed. In some models  with non zero cosmological constant 
however such ghosts may be expected \cite{Pet67,Kar67}. In this case, they would be
observed with very different redshifts, although they concern the same source. Because of
a gravitational lensing argument \cite{Got89}, the nearest ghost image would be at a
redshift $z > 3.28$. In such models, the antipode must have a redshift larger than that of
a particular multiply lensed quasar at $z = 3.28$, because just beyond the antipode there
would be an overfocused lensing case which would typically not produce lens images.
Also, just before the antipodal redshift, lensing cross sections would blow up, giving an
excess of lenses in a narrow redshift range with large separations, an effect which is not
observed. 

\begin{figure}[tb]
  \begin{center}
    \leavevmode
    \includegraphics{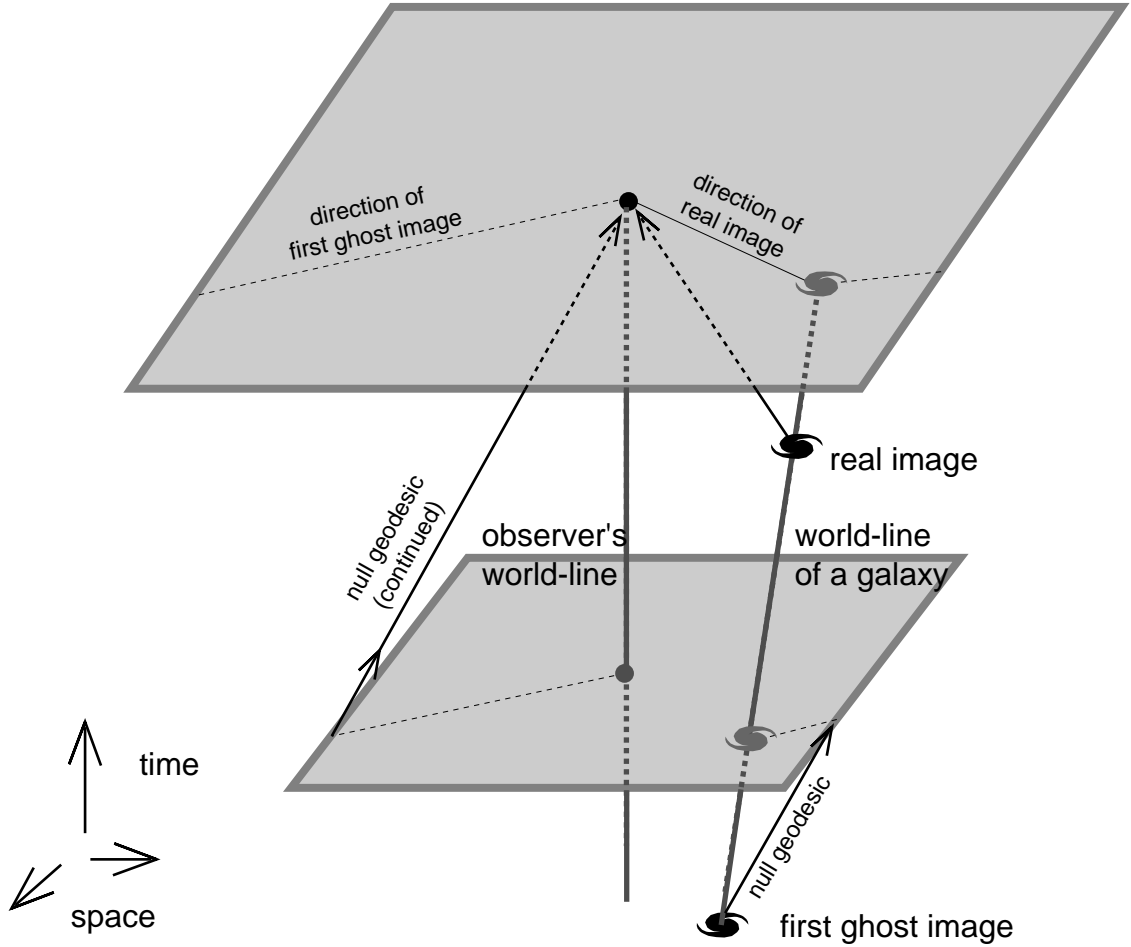}
\caption{\label{fig10_2}{\it Real image and first ghost in a multi--connected
universe.}}
  \end{center}
\end{figure}

In a MCM, there is in general no  correspondence  between  $S_{obs}$ and  a
position in real space.   But the  correspondence   between  $S_{obs}$  and an 
event in {\sl spacetime} does remain. Thus no one--to--one 
relation exists  between $z$ (or $ d_{obs}$) of an image and the distance of the
corresponding emitting  source  in real space. To such  a  (unique) source, at a
given  position in space,   are in general associated many images with different
redshifts.  This is due to  the multi--connectedness of space,
which implies   that  many spacetime geodesics link a spatial position to the observer
(at present time).The nearest image  is called  ``real" and the others are
called ``ghosts" (figure \ref{fig10_2}). Each image corresponds to a different null
  geodesic linking the source  to us. To each of them is associated a set
$S_{obs}$ of observable quantities.

\begin{figure}[tb]
  \begin{center}
    \leavevmode
    \includegraphics{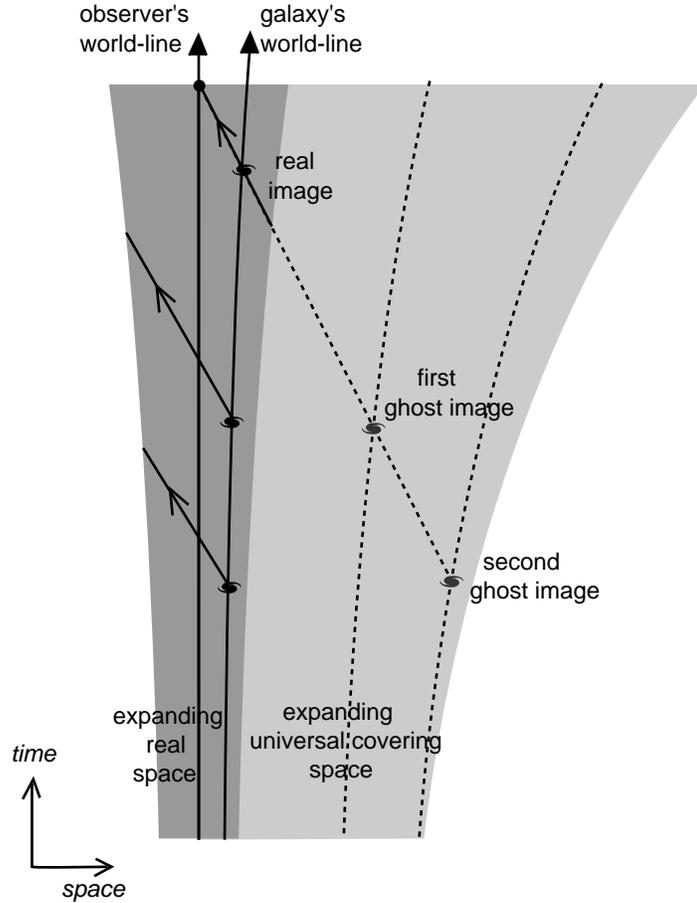}
\caption{\label{fig10_3}{\it Different ghosts of the same galaxy  in a
multi--connected universe.}}
  \end{center}
\end{figure}

To deal with this   situation, it is convenient toÊwork in   the
UC space~: the  one--to--one correspondence holds between $S_{obs}$ and a spatial
position  {\it in the UC},  which can be called the ``observer's world". But 
the   correspondence  between these positions in the UC and those in real space
is not univocal.  In the UC, a different  position is associated to each one of
the bunch  of geodesics linking   an  object in real  space  to the  observer. 
Each of them   corresponds to   a ghost image   of the real object,  uniquely
related to a geodesic  and to the corresponding  set $S_{obs}$ (figure
\ref{fig10_3}). The unique   image of the object which lies inside  the fundamental
cell  and thus  coincides with the original object, is called ``real". All 
observable properties  of a ghost in the UC space identify  to the properties of an
object  at the same position in the real space of the associated  SCM (and linked
by  the same geodesic).  All relations  between   $z$, the observable distance  $
d_{obs}$ and the look--back time    hold   exactly as  in the corresponding  SCM~:
to each redshift $z$ is associated an   instant $t$ of emission, by the same formula
$1 +z = \frac{R_0}{R(t)}$ than in a SCM (which involves the curvature of
spacetime), where $R_0=R(t_0)$ is the present value of the scale factor.  Most of
the usual  cosmological formulae may still be used, although  they take their sense
and validity in  the UC, and not in the real  space.

\subsection {Comoving space and real space}

In  multi--connected models  like in SCM's,  the geometry of space expands 
proportionally to the scale factor $R(t)$, in accordance to the Friedmann 
equations. Space at any time $t$ is exactly  homothetic to space
at any  other time, for instance  at the present time  $t_0$,  with the ratio
$R(t)/R_0 =  1/1+z$. 
To any position   $P$ in space at  time $t$  corresponds, by the same
homothety,  a  comoving position  $P_0$ in the present space. To avoid  
comparisons of spatial properties at different times,  all spatial   positions   
of cosmic objects are   considered in  the 
comoving space. This is allowed by the fact that all  spatial structures and
relations, as well as the topology,  are preserved by the homothety. For instance,
rather than using the proper distance $d_{proper}(t)$ between 2 celestial objects  at a
time $t$, it is convenient to  refer to their  comoving proper distance $d_{proper}(t_0) =
(1+z) ~d_{proper}(t)$.

 This  is still possible   in the   MCM's. The geometry of space
expands homothetically and all distances (including the  dimensions of space
itself) vary so.  The   homothety of ratio $\frac{1}{1+z}$   also preserves
all topological properties. The UC at time $t$ is similarly imaged to the comoving
UC. The relation between space and its \uc ~is exactly identical to that between the
comoving  space and its comoving \uc, and the  comoving \uc ~may be 
identified with  the comoving space of the associated  SCM.  Thus the search   for the
topology of spacetime, already  reduced  to that of space,  reduces further  to that 
of the  comoving space. Throughout this paper, if not specified
otherwise, space will  refer to comoving space, the \uc ~will refer to the comoving
\uc, \etc  ~Like in standard cosmology, all distances,  volumes, densities,..
 will be comoving quantities. 

\begin{figure}[tb]
  \begin{center}
    \leavevmode
    \includegraphics{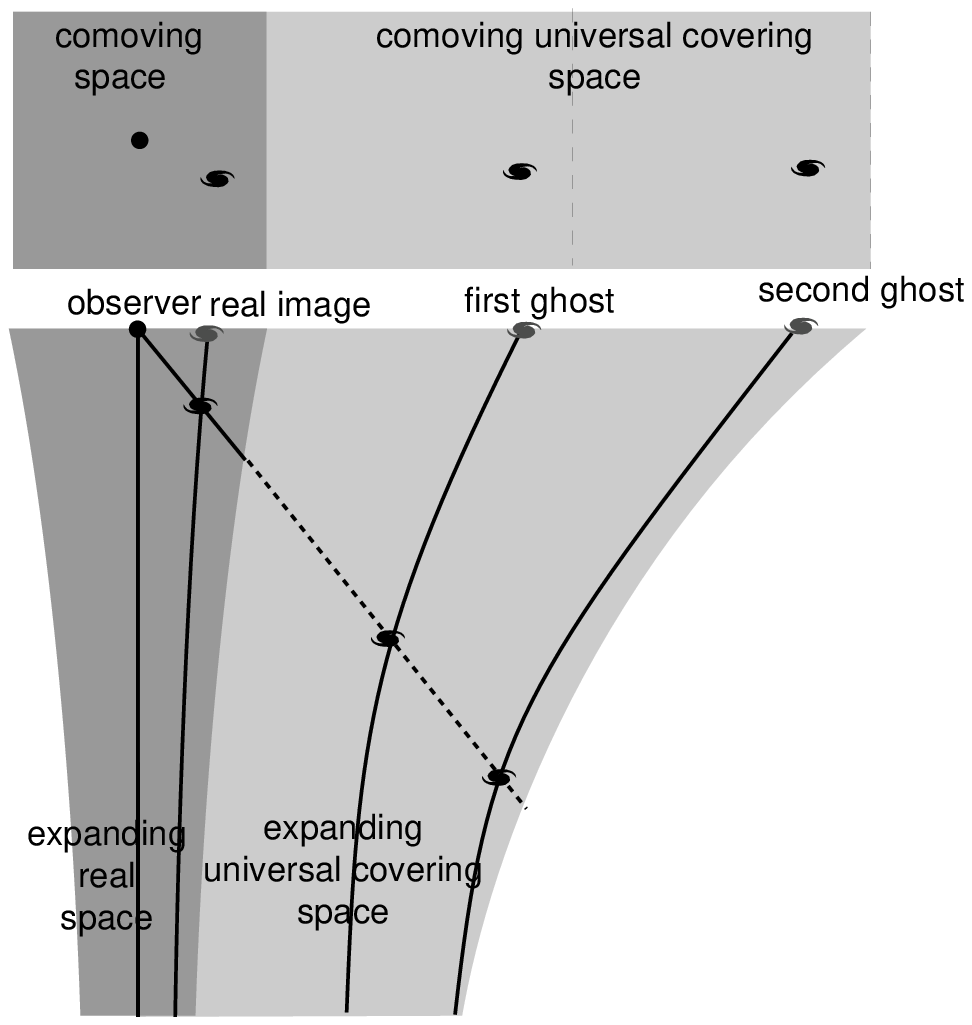}
\caption{\label{fig10_4}{\it Ghosts in spacetime and in the \uc.}}
  \end{center}
\end{figure}

\subsection {Spatial scales} \label{spatialscales}

 \subsubsection{Hubble length}

In SCM's, the only scale related to comoving space is its 
curvature radius,   identified to   the present value of the scale factor  $R(t_0) =
R_0$ (there is  no scale at all in the flat case). Thus $R_0$  is the natural length unit in
comoving space,  and   in  the UC  for a MCM. For instance, in the   spherical
simply--conected models,  space has a finite  volume $2\pi ^2 R_0^3$. 
 The Friedmann equations imply the relation
  \begin{equation}
\Omega + \lambda Ê-Ê1= \frac{k~c^2~Ê}{~ H_0^2~ R_0^2}.
\end{equation}

Unfortunately, the real value of $R_0$ remains unknown.
The only cosmological length to which we
have a direct observational access is  the Hubble length $$L_{Hubble} = c
H_0^{-1}=3\,000~  h^{-1}~\mbox{Mpc}.$$  If we define  $f=\sqrt{ \mid
\Omega+\lambda-1\mid }$, we have for a non--flat universe \footnotemark[1] \footnotetext[1]{for a flat universe,
the value of $R_0$ remains arbitrary}~:  
\begin{equation}
R_0 = L_{Hubble}~ f^{-1} = 3\,000 ~(f~h)^{-1}\mbox{Mpc}.
\end{equation}
Current observations imply $R_0 > .5 ~L_{Hubble}$ and $R_0 >  
L_{Hubble}$ if $\Lambda =0$ \cite{Ber80}. 

Another  natural cosmological length is associated to
the cosmological constant:
   $$L_{\Lambda} = \sqrt{c^2/\Lambda} =
1.17~10^{28}~\lambda^{-1/2}~h^{-1}~cm =  400 /\sqrt{\lambda}~\hmpc$$.     

 \subsubsection{Particle Horizon}  \label{horizon}

The (present) particle horizon is the distance corresponding to an infinite
value of the redshift~: 
\begin{equation} \label{horizone}
L_{(z=\infty)} =  R_0~\chi_{(z=\infty)} = c~R_0 \int^{t=t_0}_{t=0}
\frac{dt }{R(t)}.
\end{equation}
 It  depends on the dynamics only. In standard cosmology (without inflation), the
fact that the expansion law  does not differ too much from a power law $R(t) \propto
t^{\gamma}$ (with $\gamma \approx 2/3$ for matter dominated models) implies that  
\begin{equation}
L_{(z=\infty)} = \chi_{(z=\infty)}~
R_0  \approx \frac{c t_0}{(1-\gamma)}, 
\end{equation}
where  $t_0 = u ~H_0^{-1}$ is
the present age of the universe and $H_0^{-1}$  the   Hubble time. In FL models,  
$u \approx \int ^1_0 dx~F(x),$ with $F^{-2}(x)~\equiv~   \frac{\Omega }{x} +  x^2~ \lambda  -
\Omega -\lambda.$ In the Einstein--de Sitter solution ($\Omega=1$, $\Lambda=0$),
$u=2/3$. More generally,  $u< 1$ if  $\Lambda=0$  and  remains  of the order of unity for
admissible models. It results that $L_{(z=\infty)}= R_0~\chi_{(z=\infty)} =
\frac{u}{(1-\gamma) } ~L_{Hubble}$, where $ \frac{u}{(1-\gamma )} $ is of the
order of unity (this is not the case, however, in models with inflation since the
expansion  does not follow a power law).

In any case,  since the universe Êwas opaque before the moment $t_{rec}$ of the 
recombination,   the comoving length $L_{rec} = R_0~\chi_{rec}$ can  be 
considered as a {\it physical}  horizon, where  
\begin{equation}
 \chi_{rec} =c\int^{t=t_0}_{t=t_{rec}}  \frac{dt}{R(t)}.
\end{equation}
 In models without
inflation, $L_{(z=\infty)}$ and $L_{rec}$ almost coincide. In the case with
inflation, $L_{rec}$ corresponds more than $L_{(z=\infty)}$ to the intuitive
notion of horizon and  represents with a good precision the (comoving) radius of
the {\it observable} universe. Thus, to avoid any ambiguity,  we define in the
following  $L_{h} =L_{rec}$ and  $\chi _{horizon}=\chi _{rec}$. It results 
that in all cases (with or without inflation),  $$\chi _{horizon} \approx
\frac{u~f}{(1-\gamma )} $$ and  $L_{h} \approx
\frac{u}{(1-\gamma )} ~L_{Hubble}
 = 3\,000~ \frac{u}{(1-\gamma )}  ~h^{-1}~\mbox{Mpc}, $
where $u$, $f$ , $h$ and $1-\gamma$ are of the order of unity.

The concept of horizon keeps its exact validity in  the MCM's, but must
be applied to  the \uc ~space~:  an  image is potentially visible iff  its 
(comoving) distance is smaller than  $L_{h}$ in the \uc. This sets {\it
a priori} no constraint about the position of the real object. 
A particle horizon can be similarly defined at any epoch $t$, from   \eqn
{horizone} where the upper bound of integration $t_0$  is replaced by $t$.

\begin{figure}[tb]
  \begin{center}
    \leavevmode
    \includegraphics{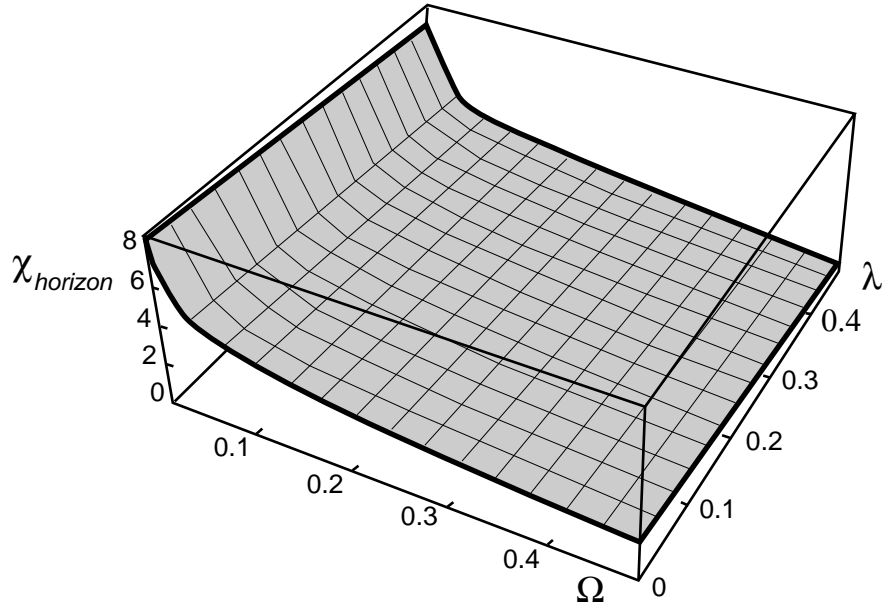}
\caption{\label{fighorizon}{\it The horizon length in curvature radius units.}}
  \end{center}
\end{figure}

In the non flat case,   topological constraints impose precise relations between the
dimensions of space compared to the radius of curvature. On the other hand,
the potential visibility of ghosts images is linked to their situation with respect to
the horizon. Thus  $\chi_{horizon}$,   the ratio of the horizon length
over the curvature radius has to be maximum if
observable effects are expected.   Figure  \ref{fighorizon}  displays $\chi_{z =
\infty}$ (for a standard \frl ~model) as a function of the cosmic parameters $\Omega$
anf $\lambda$. 

\subsubsection{Spatial scales associated to multi--connectedness}
   
In a MCM,   additional spatial scales are  associated with the topology, those
of  the \fp. The geometry  suggests to   compare them with  $R_0$ but
it is often more convenient,  for  observations, to compare them to   
 $L_{Hubble}$ or  $L_{h}$, or to evaluate them in~Mpc or $\hmpc$.
Observable effects linked to the multi--connectedness will only occur if these
scales are smaller than  the size of the observable   universe, \ie, the horizon
length.  We  call $\alpha$ the smallest  length associated with  the
\fp. As already mentionned, the ratio $\alpha/ R_0$ can only take specific
values in a non flat space~: when $k>0$ (resp. $k<0$), the geometry imposes  a
maximum (resp. minimum) value for  it. It remains arbitrary in flat space.
An other scale   is involved, the  maximum length $\beta$ inscriptible in the
Ê\fp,   which is also the diameter   of the  sphere inscribing it. $\beta$ is also the
maximum distance between 2 images of the same object belonging to adjacent cells.
$\alpha/2$ and $\beta/2$ are the minimum and the maximum distance of the observer,
assumed at the center of the \fp, to its boundaries.

\begin{figure}[tb]
  \begin{center}
    \leavevmode
    \includegraphics{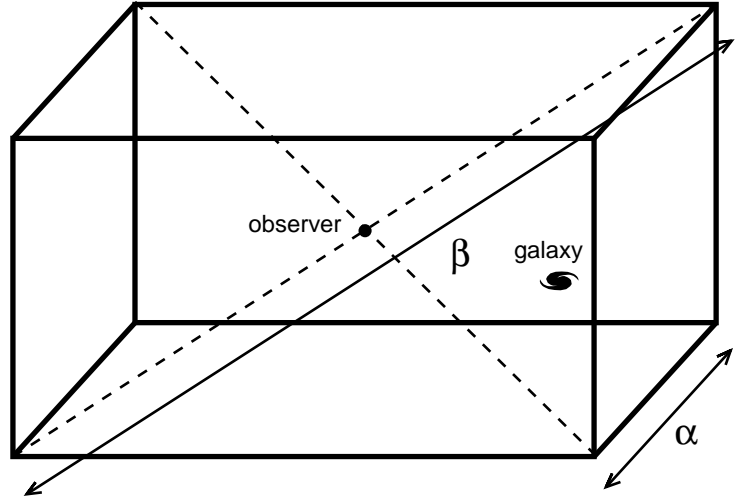}
\caption{\label{fig10_5}{\it Characteristic lengths of a toroidal universe.}}
  \end{center}
\end{figure}

For instance, the geometry of the 3--torus (figure \ref{fig10_5}) is defined by the 3
lengths $\alpha_x$, $\alpha_y$ and $\alpha_z$, with $\alpha = min (\alpha_x,
\alpha_y, \alpha_z)$, the length of the smallest side,  and
$\beta= \sqrt{(\alpha_x)^2+(\alpha_y)^2+( \alpha_z)^2}$, the length of the diagonal.
If real space has the topology of such a 3--torus, immediate observations
 impose that $\alpha$  cannot be too small,  for instance smaller than the size
of the \mw. The observations detailed in the next sections    allow to
increase this lower limit. 

\subsection {Multi--connected universes with   flat spatial sections}

For simply or multi--connected models, space is Euclidean if $\Omega+\lambda=1$.
The \uc ~is   $\bbbr $$^3$ and  space has no characteristic scale~: the present
value $R_0$ of the scale factor may be chosen arbitrarily and thus cannot be used
as an intrinsic   length unit.
A convenient metric is (\ref{rwmetric}) Êwith $S_0(\chi) = \chi$.   The variety of
topological spaces with flat curvature has been detailed in \secn{flatm}. Among the non
compact models (where space is infinite in at least one direction, so that $\beta= \infty$),    
three  types, beside  $\bbbr $$^3$  itself,  are orientable and thus are
candidates for   MCM's. Since they are non compact, they have been less discussed
in the litterature. In many aspects, their properties are intermediary between
those of SCM's and those of, for instance, the  torus. 
 Thus, for the cosmological purpose it is especially interesting to consider  
the  6 kinds of orientable closed spaces with constant flat curvature described in
\secn{flatm}.

The 3 dimensional hypertorus $T^3$ is the simplest case. Its fundamental
cell is a  parallepiped with 3  arbitrary dimensions $\alpha_x, \alpha_y$, and
$\alpha_z$. Its principal directions define 3 preferred orientations in space,
thus breaking the global isotropy. This is a general property of multi--connected
models.  For this space, there is no mathematical constraint on the values of
$\alpha$, the smallest side of the FP, and $\beta$, its diagonal. 
Multi--connected universe models with toroidal space have been widely studied. For
instance all the images of an object lying onto one of the principal directions are
distributed with a periodicity in comoving proper distance. This 
has motivated observational     searches for periodicities.
  On the other hand, the hypertorus
produces some images of the observer which could be potentially visible in opposite
directions (if they are within the horizon). This has also motivated specific searches that we
present in  \secn{associations}. 

Cosmological models where the fundamental polyhedron is paralellepipedic, but
where faces are twisted before identifications would appear,  in some aspects, 
alike the hypertorus. It should be mentionned that, in the case where the faces are
rotated by $\pi /2$, the arbitrariness is reduced because of the additional condition
$\alpha _x = \alpha _y$ necessary to allow for the identification.  Models based on a
fundamental cell with  hexagonal faces have not been   studied to our knowledge.
 
\subsection {Multi--connected universes  with  positive spatial curvature}

In this case, the UC --  the 3--sphere $\bbbs $$^3$ --   is compact,  so that all
models also have compact spatial sections.  The present radius of the universe
$R_0$ is chosen to coincide with the curvature radius of (comoving) space.  
   $\bbbs $$^3$  is advantageously described with  the  comoving  metric
(\ref{rwmetric})Ê with $S_1(\chi) = sinh(\chi)$.  Its whole extent     is described by 
$\chi$ from $0$  to $\pi$ , $\phi$ from  $0$ to $2 \pi$ , and $\theta$ from  $0$ to
$\pi$. 

\subsubsection{The elliptical space}\label{elliptical}

 As soon as 1917, de Sitter  \cite{deS17a,deS17b}
distinguished the spherical space $\bbbs $$^3$ and the projective space $\bbbp $$^3$
that he called the elliptical space. Then he compared the properties of cosmological
models based on these two spatial parts. For instance they have for respective
(comoving) volumes $2 \pi ^2~R_0^3$ and $ \pi ^2~R_0^3$. The maximum distance between 2
points is $ \pi R_0$ in $\bbbs $$^3$ and $ \pi R_0 /2$ in $\bbbp $$^3$. But the
main difference to his eyes was due to the presence, in $\bbbs $$^3$, of an
antipodal point associated to any point, and in particular to the observer, at a
distance of  $ \pi ~R_0$ precisely. To avoid this undesirable fact, he claimed
that cosmological models with  $\bbbp $$^3$ are much preferable than those with
$\bbbs $$^3$ and he mentionned that Einstein had the same opinion. Since both
spaces are, for a given value of $R_0$, of approximatively the same size (only a
factor 2), it would be very difficult to decide from observations in which we
live, if it appeared that $\Omega + \lambda > 1$. In this case, if we follow
Einstein and de Sitter, $\bbbp $$^3$ would be a better assumption than $\bbbs
$$^3$.   

However, de Sitter correctly remarked that, for admissible values of $\Omega$ and
$ \lambda  $ it is probable that the most remote points lie beyond the horizon,
so that the antipodal point, if any, would remain unobservable. Eddington
\cite{Edd23} also referred to elliptical space as an alternative  as much as attractive
as $\bbbs $$^3$. His discussion for comparing both models led him to very
interesting discussions about the nature of space. He pointed out for instance that
many conceptual difficulties arose from the fact that space is  often, and
erroneously, considered as a ``continuum in which objects are located". 
He suggested then to abandon this misconception and to consider space as a
network of intervals. In this case, the structure of the whole space appears
only as the lattice--structure of all cross--connexions between points. 
We will not here go deeper into these discussions but only point out again that
there is no simple argument according to which space should be
simply--connected. Friedmann \cite{Fri24} and Lema\^\i tre \cite{Lem29} also
discussed these possibility and preferred the elliptical form of closed space for
aesthetical reasons. 

Narlikar \& Seshadri \cite{Nar85} examined the conditions in which ghost  images of
celestial  objects may be  visible in a \frl ~model with
elliptical space. They suggested that, in such a model the apparent cutoff in the 
redshift distribution of quasars may be due to this effect (see \secn{quasars}). 
  
\subsubsection{Lens spaces and similar MCM's }

 None of these authors mentionned  other multi--connected spaces
in a cosmological context. This was done by Gott \cite{Got80}. For instance, the rotations of
the  cyclic group   $Z_n$  ($n>2$) (see \secn{positivem})     generate MCM's where  
space is $\bbbs$$^3~/~Z_n$. The larger the value of $n$, the smaller   the  
dimensions of space.  The simplest, and  largest,    case    is  the lens space
$M=\bbbs$$^3/Z_3$, which tesselates $\bbbs$$^3$  
into  6 replicae of the fundamental cell   having  a lens form.  In the
corresponding MCM, an observer (at  $\chi =0$) has 5 images of itself, whose
coordinates are given in \secn{positivem}. The minimum distance between images is  
$\alphaÊ= \pi R_0 /3$ and  the maximum dimension of the fundamental lens is $\betaÊ=
\pi R_0 /2$.  Gott   remarked that in a standard \frl ~model, this is
larger than  the size of the  horizon,  so that   no observable effects would  take
place in such MCM's. Interesting cases for observational cosmology  involve
fundamental  cells of smaller sizes, and thus higher values of $n$. 

 The dihedral 
group  $D_m$ ($m>2$) gives the spaces $\bbbs$$^3/D_m$.    For instance, $D_3$ 
generates  11  images of the observer  (coordinates in \secn{positivem}). MCM's can
also be constructed from the polyhedral groups  : $T$ divides $\bbbs
$$^3$ into 24 tetrahedral  quotient spaces, with observer's nearest  image   at a
distance   $\alpha = \pi R_0/3$.  $O$ divides  $\bbbs $$^3$ into 48   octahedral
quotient spaces, with observer's nearest  image  at $\alpha =  \pi R_0/4$.
 $I$ divides  $\bbbs $$^3$ into  120 dodecahedral  quotient  spaces, with observer's
nearest image   at $\alpha =  \pi R_0 /5$. For this latter case, Gott  showed that
the fundamental  cell may be inscribed in a sphere of radius $ \beta/2 =
0.338~R_0$.

The spaces  $\bbbs $$^3/Z_n$ and  $\bbbs $$^3/D_m$ with  small values of $n$ or
$m$ have    dimensions   comparable  to $R_0$   or to the
horizon length.  This is also true     for $\bbbs $$^3/T$, $\bbbs $$^3/O$ and
$\bbbs $$^3/I$. In any case   $\beta$ remains always greater than $0.326~R_0$,
the value for $\bbbs ^3 / I$.  In these cases, where  ghost  images would be at 
distances comparable to the horizon,   very few observable effects can be
expected. On the other side, very large values of $m$ or $n$ would lead to
identification lengths so small that they are excluded by  the present observations
(see below). The most interesting cases would therefore lie in the intermediate
range.  

\subsection {Multi--connected universes with negative spatial curvature}
\label{negativemodels}

When the covering space is   $\bbbh $$^3$, 
the scale factor  $R_0$ may be chosen to coincide with the radius of curvature,
and  may be used as   a  unit length. The metric  is usually written  under
the comoving   form    (\ref{rwmetric})
with $S_{-1}(\chi) Ê= \sinh \chi$. The coordinate  $\chi$  varies from $0$ to
infinity, $\theta$ and $\phi$ from 0 to $2\pi$.   An  infinite number of MCM's with
negative spatial curvature may be build by various means
as depicted in  \secn{negativem}.

Symmetrically to the case of constant {\sl positive}
curvature, where   the volume of a manifold  is an entire fraction of   $2
\pi^2~R_0^3$ which is thus an upper bound,  there is a minimal value $vol_{min}$  for
the volume of a manifold with constant {\sl negative}Ê curvature (see
\secn{negativem}). Its value remains however unknown, and it is not known either
if there is a minimum value for $\alpha$. We decribe in \secn{minimum} models
where space could have a volume equal to   $vol_{min}$.
 Given this constraint, the most interesting models for cosmology   
are  those with the maximum ratio $L_{h}/R_0$, \ie,  low values of $\Omega$ and,
to a lesser extent, of $\lambda$, as it appears in figure  \ref{fighorizon}.  

\subsubsection{A toy spacetime in 3--dimensions} \label{toy2}

A 3--dimensional universe model  introduced by Fagundes
\cite{Fag82,Fag83a,Fag85}   presents a pedagogical   interest. Its  
(two--dimensional) spatial part is the  2--$torus$   $T_2$, whose \fp~ is a regular
octogon as described  in \secn{2torus} and  \secn{flatsexamples}.  

This 3--dimensional spacetime admits the metric     
\begin{equation}
ds^2 = a(\eta)^2 (d \eta^2 - d\sigma ^2) , 
\end{equation}
  a restriction of the \ks ~metric
(\ref{ksmetric}).
In this formula,  $a(\eta)=R(t)$ is the scale factor, with present
value $a_0=R_0$  and   $\eta$ is the  conformal time defined by $a(\eta)~d\eta = c
dt$ which becomes, after integration, $c t = a_0~ (\sinh \eta -\eta)$. The
spatial part of the metric
\begin{equation} \label{h2metric} 
R^2~d\sigma ^2 =R^2~ (  dr^2 +
\sinh^2r~ d\phi ^2 )
\end{equation} 
has a   negative curvature $-1/R^2(t)$.   

The comoving space, with a curvature radius $R_0$,  is described by the metric
$R_0^2~ d\sigma ^2$. Although the  model does not intend  to represent the  real
universe,  Fagundes \cite{Fag85} derived  the coordinate of the 
horizon     $\eta _{horizon} \approx 6$, corresponding to a distance  
$\approx 6~R_0$.
  The redshift of a source at a 
coordinate distance $r_1$ is given by : 
$1 + z_1 =  \frac{ a_0 }{ a ( \eta _1)}$, with $\eta _1 =\eta _0 - r_1$. 
The distance between 2 adjacent images,  equivalent to the length of the smallest
geodesic loop in $T_2$ is $\alpha \approx 3~R_0$.  The nearest image of  a
source appears about halfway to the horizon, thus very far away.
This illustrates  that the fundamental cell cannot be made very small  in a MCM's with
negative curvature.  Fagundes emphasized this by calculating  that a source at
redshift  $z=0.7$   would have its first image at $z=$42.  The purpose of his paper
was mainly pedagogical, to illustrate how multiple images and counter images of
quasars could occur, with the idea that this could solve the quasar redshift
controversy. We present   the 4--dimensional generalisation of  his
model   \cite{Fag89} in \secn{toy3}, and we discuss the implications
of multi--connectedness onto the     quasar distribution  in~\secn{quasars}.  

\subsubsection {Non--compact   models}
\label{Non compact}

Some examples of non compact   multi--connected hyperbolic spaces were presented
in  \secn{negativee}.  There is a general procedure which allows to build
such a space $\bbbh ^3 / \Gamma$ from a multi--connected surface $\bbbh ^2 /
\Gamma$~: it is non compact in the direction orthogonal to  $\bbbh ^2 / \Gamma$. 

 Gott \cite{Got80}   applied this procedure, tesselating  $\bbbh^2$ with    regular
n--gones. According to a theorem of  hyperbolic geometry, their area is   $R_0^2$
times the angle deficit [$\pi(n-2)$~-~the~sum~of~all~angles~at~vertices]. 
For instance the octogon in  $\bbbh^2$ has  all angles $=\pi/4$, and  a surface 
$4\pi R_0^2$. Then the identification of the sides 1 with 3, 2 with 4, 5 with 7  and 6
with 8  gives to each octogon the topology of a sphere with 2 handles. 

This may be generalized to  $n$--gons ($n=4g$), which have  the
topology  of a sphere with $g$ handles. Thus    $\bbbh$$^2$ can be tesselated
with cells topologically equivalent to  spheres with $n$ handles.    The one
considered by Gott has its \fp ~formed from 2 regular tetrahedra with vertices  at
infinity, which are glued together. It is non compact.

Another MCM with non compact hyperbolic space was considered by Sokoloff
and Starobinskii \cite{Sok75}. Its spatial  part is  the ``cylindrical horn'' described
in \secn{horn}. In the frame of this model they studied the structure and the growth 
of the density--perturbations which could have led to the formation of galaxies,
clusters, etc. Those are discussed in \secn{gf}. 

\subsubsection {Compact  models} \label{quasi--hyperbolic} 
\label{toy3}

The topologies of $\bbbh^3$ which give observable effects are those 
having Êthe smallest \fp ~as possible   (in analogy  to the high $n$ or $ m$ modes
in the spherical case).  We list below the main MCM's  constructed from one of
the   compact manifolds   $\bbbh$$^3/\Gamma$

\begin{itemize} 
\item 
Gott \cite{Got80} proposed a model whose spatial part is the    L\"{o}bell
space \cite{Lob31},   described in \secn {lobell}.  He
calculated $\alpha = 2.64 ~R_0$ and  showed that  the present horizon 
contains no more than about 10 replicae of the \fp.     The L\"{o}bell
space is thus 
too large to give  interesting observable effects. Considering  the
time evolution, he showed that  the entire space entered   the cosmological  
horizon at a redshift $z\approx6$  (this is defined as the first moment where  
the whole space,   whose proper volume  increases $\propto R^3(t)$, is contained
within the horizon). 

\item 
Fagundes \cite{Fag83b} Êstudied  a ``quasi--hyperbolic" model where space
has the topology of $T_g~ *~ S^1$. Here $S^1$
is the ordinary circle, parametrized by the coordinate $\zeta$ and $T_g$ is the
 $g$--torus, described by the 2 coordinates $\rho$ and
$\phi$. It should be emphasized that the  \uc ~space is not $\bbbh ^3$ but  the product
$\bbbh ^2 \times \bbbr$. This space, being anisotropic, has therefore no constant
spatial curvature. 
This model 
generalizes the 2--dimensional study   of \secn{toy2}. 

This MCM may be described by the Kantowski--Sachs  metric (\ref {ksmetric})
\begin{equation} 
   ds ^2 = c^2 dt^2 - R^2(t)~  ( d\rho ^2 + sinh^2 \rho ~d\phi ^2 ) - b^2(t) ~d
\zeta ^2.
\end{equation}
It can be also  written in the conformal \rw ~form~:

\begin{equation}
ds^2 = a(\eta )^2~ (d \eta^2 - d\sigma ^2) = c^2~dt^2 - R^2(t)~d\sigma ^2,
\end{equation}

where $a(\eta )=R(t)$ is the scale factor, with present value $R_0$,    $\eta$ the
conformal time defined by $a(\eta )~ d\eta = c~dt$,   and $R~d\sigma $ the spatial
metric. 
 
Solving the Einstein equations for a dust--filled Universe, Fagundes derived the
parametrized solution~:  
$$ ct = R_0~( \sinh  \eta - \eta),$$
$$a(t) = R_0~(cosh \eta -1),$$
$$b(t) = 3R_0 ~(\eta~cosh \frac{\eta}{2} - 2).$$
Restricting  his discussion to the  2--torus ($g=2$), Fagundes explored some
properties of the Cosmic Microwave Background, that we describe in see \secn{cmb}.

\item
Fagundes \cite{Fag89,Fag86} also studied    models where space is
a Best model \cite{Bes71}. The Best spaces (see    \secn{Bests}) have  one of the    
topologies    $\bbbh^3/I$,  where $I$ is  the regular   icosahedron. 

Fagundes \cite{Fag89}   investigated  the dynamical properties with  the
parameters $\Omega =0.1$ and $\Lambda=0$.     The redshift $z_1$ of a source
at a coordinate distance $\chi_1$ is given by :  $1 + z_1 = \frac{R_0}{ a
( \eta _1)}$, with $\eta _1=  \eta _0 -\chi_1$. Fagundes calculated the 20 generators
of the holonomy groups which transform  any source, in the \fp,   into 20  images 
in the adjacent cells, and more beyond.   He remarked that  images of the  center of
the \fp ~are given, in  the \uc ~$\bbbh^3$,  by reflexions on the 20 faces of the \fp.
 
Fagundes was interested in the existence of multiple images of a given source,
with a main focus towards our own Galaxy, the Milky Way. He calculated the
positions of ghost images  with a method  similar to the ray--tracing procedure used
to   synthetize realistic  images  by computer.  He emphasized the existence
of   conjunctions (or oppositions and  other associations)  of images onto the sky.  He 
discussed  in particular a case where the original source has a redshift of
0.124,  the conjunct of  4.263, and associated images of 3.14, 2.34, and
1.94. In his opinion,  they could be good candidates for the discordant  quasar
associations. We discuss these topics in    sections  \ref{galaxy} and \ref{quasars}.

\end{itemize}
 
\subsubsection{The minimum volume model} \label{minimum}

As we mentionned in \secn{negativem},  the possible values of the volume of a space
with constant negative curvature are bounded from below.  But the limiting value
$vol_{min}$ remains unknown. Weeks  \cite{Wee85} and Matveev \& Fomenko \cite{Mat88}
found the manifold with the minimum volume presently known, \ie, 0.94~$R_0^3$, which is
described in \secn{Weeks}. On the other hand Meyerhoff  \cite{Mey86}Êdemonstrated that
$vol_{min} > V_{Meyerhoff} =  0.00082~R_0^3$.  Recently, Hayward \& Twamley
\cite{Hay90} have studied multi--connected models based onto the
Weeks--Matveev--Fomenko manifold. They considered, in this framework, the possible
isotropization of the \cmb ~(see \secn{cmb}) and tried to explain the apparent
periodicity in the galaxy distribution found by Broadhurst \etal ~\cite{Bro90}. 
Although there is no known manifold corresponding to the  volume $V_{Meyerhoff}$, they
made the hypothesis that there could exist one and guessed the possible properties of
a universe having it as a spatial part.  For these 2 manifolds respectively, they
considered a (hyperbolic) sphere having the volume of the whole space and considered
its diameter $L_{max}$ as an upper bound for  $\alpha$. For the 2 models they found
respectively the values 

\begin{equation}\label{VWMF}
L_{max,WMF}=\frac{3552}{\sqrt{1-\Omega}}~\hmpc
\end{equation}

and 

\begin{equation}\label{VMey}
L_{max,Meyerhoff}=\frac{348}{\sqrt{1-\Omega}}~\hmpc .
\end{equation}

Comparing
these expressions to the present observational limits derived from the observations of
galaxy clusters (see \secn{clusters}), they concluded  that both   manifolds are
presently   admissible.

\subsubsection{Barrel models} \label{barrel}

Most observational  tests could concern only local parts of the universe, \ie, at
distances much smaller than the curvature radius of space. At these scales  curvature
effects play no role. Thus   Sokoloff \cite{Sok75,Ruz75} remarked that all
compact MCM's, with negative spatial curvature, have the same asymptotic
structure. To    describe this he employed   the Euclidean metric,  
approximately valid for scales  not too large, in polar coordinates $x$, $r$, $\phi$.
Then the considered MCM's  correspond to the ``barrel structure" defined by the
  identifications~:
$$x = x + n~h$$ and $$ \phi = \phi / n ~a,  $$ where $n=$0, 1, 2,... $h$ is a
constant identification length and $a$ an identification angle ($a=0$
corresponds to cylindrical space). A MCM  with negative spatial curvature would look
very alike such a barrel space. Such models are of course anisotropic with
the $x$--axis as a preferred direction (or $a=0$). They were studied
\cite{Sok75,Ruz75} in relation with cosmic magnetic fields (see
\secn{magnetic}).

\section {Ghosts in multi--connected Universes}\label{sec11}

The hypothesis that the real universe is multi--connected  has various implications.
On one hand, if the identification scale is of the order of the horizon  or larger,
no directly observable effect is expected, although
 primordial quantum effects could have generated interesting effects. On the other
hand, if the  topological scales  $\alpha$ and $\beta$ are significantly smaller 
than the present  horizon,  observable effects will be manifest, among which the
most interesting is the existence of ghost images. We will concentrate now on the 
``small universes", defined as those where $\alpha < L_{h}$,    of the order
of a few  100 or 1~000~Mpc. Our knowledge of the universe gets worse when the
spatial  scale increases,   with the horizon  as a   limit. Thus, the smaller the
basic cell (\ie, the \fp, which identifies with the real space) the   easier are  
topological effects to  observe.  How do  the present observational data  constrain
the possible multi--connectedness  of the  Universe~? And, more generally, what
kinds  of tests are conceivable~?  These  questions  are developed in this  section
and the following ones. We will   also discuss the  suggestions that
multi--connectedness of the Univers could  explain    some   observational results 
not presently understood, like discordant redshift associations,  redshift
periodicity, distribution of   the gamma ray bursts,\ldots
 
\subsection {Geodesics and ghosts}

The topology of a given \frl ~model  is characterized byÊ the shape and the
dimensions of the \fp. The simplest case of the torus  $T^3$, for instance,
corresponds to    a parallelepiped   with  3 identification lengths 
$\alpha_x,\alpha_y$ and $\alpha_z$.  We   refer   as $\alpha$ and $\alpha_{max}$
respectively to the smallest and the  largest of these lengths (see
\secn{spatialscales}). \sos ~have defined   the characteristic radii  $R_1$  and 
$R_2$~: by definition,   there is no ghost image of a source nearest  than $R_1$ to
the observer, so that     $R_1 = \alpha/2$ (figure  \ref{fig10_5}). Also, by
definition,   no single original source lies beyond   $R_2$, which appears therefore
as  the radius of the smallest sphere containing entirely  the fundamental cell~: 
$R_2 =\beta /2$, where  $\beta= (\alpha_x^2Ê+Ê\alpha_y^2Ê+Ê\alpha_z^2)^{1/2}$ is the
diagonal. The real image of any celestial object is
always closer than  $\beta/2$ to the Earth.  The torus considered by Gott
\cite{Got80}   has  all fundamental lengths equal to $\alpha= 28.5~\hmpc$, so that 
$\beta = 50~\hmpc $ (note that he calls $R_H$   the quantity  $ \beta /2$). In the
more general case, the  relation between $\beta$ and   $\alpha$ depends on the
geometry.
 
\begin{figure}[tb]
  \begin{center}
    \leavevmode
    \includegraphics{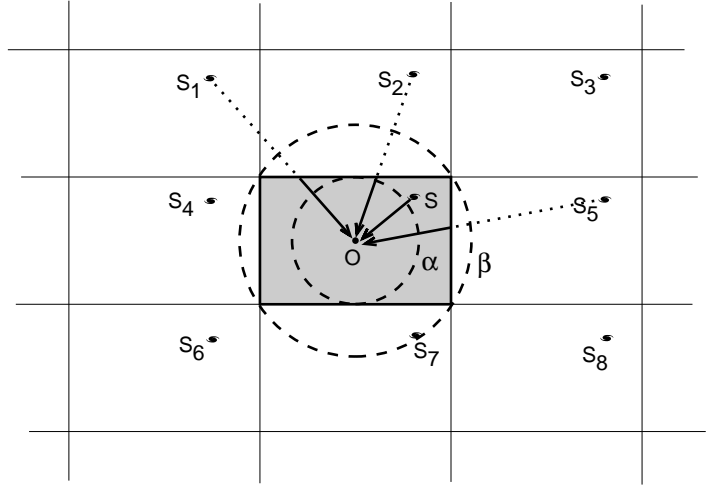}
\caption{\label{theghosts}{\it Multiplication of images
in the \uc ~space of a toroidal universe.}}
  \end{center}
\end{figure}

In a MCM, many null geodesics start from the   position of an object (e.g., a light
source)  in real  space, to reach the present observer (figure  \ref{theghosts}).
They   represent  light rays which have been emitted at different epochs. To each of
them corresponds a set $S$ of observable quantities (see \secn{universal covering}),
and an image of the real object.  These images,  replicae  of the original object in
the \uc, are called ``ghosts'' and can be labelled by an index $i$.  All these
images have the same status but  a region  of the \uc ~is arbitrarily chosen to
match the (multi--connected) real  space.  This ``real part'' of the \uc, the basic
cell, is     generally chosen to coincide  with the \fp ~centered on the observer, 
although this choice is for pure convenience in homogeneous models. Then, the unique
image situated in this real part of the \uc  ~is called the real object and  all
other will be called ghosts.  In the following we will assume with no loss of
generality (when space is homogeneous)    that the observer lies at the  center of
the \fp. 

The ghosts are images of a real object, so that  their associated proper  
(comoving) distances, in space,  are rigorously identical to   $\hat{d}$, that of   
the original  (not taking into account  its possible proper motion during the
light--crossing--time). But  instead of the proper  distance, measurements provide 
other observable distances $d_{obs}$ (the luminosity--distance,  the 
angular--diameter distance, the parallax--distance, \etc) of the ghosts. In general 
the values of $d_{obs}$ differ  from a ghost to another of the same object.   As we
mentioned earlier, $d_{obs}$ is to be  interpreted as a distance  in the \uc ~space.
Work in the \uc ~rather than in real space thus allows us to  conserve the  usual
cosmological interpretation of   all distances. In the \uc, it is also   possible
to  assign a proper (comoving) distance $d_i >\alpha /2$  to each ghost (only the
original image has $\hat{d} =d<\alpha /2$; conversely, only ghosts are observed at
$d>\beta/2$). The ghosts are in general   observed with  different redshifts $z_i$. 
The holonomies which define the topology are isometries of the comoving \uc ~space,
and thus concern the proper comoving distances  $d_i$.

The most natural way to prove that our universe is multi--connected would be to
show, ponctually or statistically, the existence of ghost images  of objects or of 
specific configurations. In the following we  present different methods which have
been, or which can be applied in this purpose. The number of potential  ghosts of an
object equals the ratio of the volume of the \uc ~space  to that of the \fp, and
thus  depends on    the cosmic parameters and  on the topology   considered~: it is
finite in the case of positive constant spatial curvature, infinite otherwise.  
Only   ghosts   nearer than the horizon can be seen, so that their number  is
reduced to the number  of  cells in the {\sl observable} universe, \ie, about
$L_{h} ^3 / (\alpha \beta)^{3/2}$, where $(\alpha \beta)^{3/2}$ is a rough
estimate of the volume of the \fp.

The distinct null geodesics associated to different ghosts of a same object
correspond to different look--back times (defined as the temporal delay between
emission and reception of the radiation) and redshifts. This would allow to
distinguish  ghosts of the same object,  appearing  on very close lines of sight (or
on the same one), from   multiple images created by gravitational lensing, which
have equal redshifts and look--back times.  For a ghost, all usual cosmological
relations hold when they involve distances  or quantities estimated in the \uc. In
particular, the larger this distance, the longer   the look--back time and the
redshift (both larger than for   the original image, the nearest  in the \uc). The
more distant   a ghost in the \uc, the younger appears the corresponding object. For
instance, a ghost of the Sun   could unveil its aspect    millions or billions  
years ago. 

The differences in  the look--back times for ghosts of a same  object correspond to
the time necessary for   light  to   made  different numbers of turns around the
universe. Thus they appear approximatively as  multiples or combinations of the   $ 
\alpha_i /c$. But    cosmic objects have a finite lifetime. For an object with an
intrinsic   lifetime $T$, the number of visible ghosts   is limited to~$ \approx
(cT)^3 / (\alpha \beta)^{3/2}$.  Too distant (resp. near)  ghosts would correspond
to  a time of emission when the object was  not yet  (resp. no more) existing. This  
limits    the possible observation of ghosts to objects with   lifetime $T \ge 
\alpha /c$. This is crucial for quasars, that we discuss in \secn {quasars}.

Moreover, even if an object does exist   during  a long    period of time, this does
not guarantee that it will be easily recognizable. For that, it would be   also
necessary   that  its  appearance has  not changed too much during its lifetime.
Thus the tests must be performed on objects which live  long enough, and which keep
a sufficiently constant    aspect   during their life. 

\subsection{Searching for  ghosts} \label{Searching for  ghosts}

The usual cosmological relations  hold in the \uc ~: the more distant   a ghost, the
weaker  its luminosity, the smaller its angular size (the possible reconvergence of  
geodesics due to  the spatial  curvature is avery large scale effect which    plays
no role in current  observations).  Very  distant ghosts cannot be observed    for
the same reasons than distant objects in  the SCM's. Observing techniques    define 
a  practically usable portion of space,   inside which a ghost remains observable
{\it and recognizable}, depending on the type of objects (or configurations)
considered, and of the observation procedure. This limits again   the  number of
observable ghosts.  In addition some peculiar reason may forbid to observe   a
ghost    visible in principle~:   an  high obscuration     region, or an other
object ahead on the \los ~can mask it.  It results that no test which  
depends        on the observation of a small number of ghosts only can  be used to
rule out a MCM, since  the non--observation of an expected ghost  could be
attributed to many different reasons.

Additional effects may also perturb the observation of ghosts.  Because the Universe
is not exactly homogeneous, the null  geodesics  are not exactly those of the
spatially homogeneous spacetime. They  are deformed by the density inhomogenities,
leading to the various consequences of  gravitational lensing~:  deformation,
amplification, multiplications of images...  A ghost so  amplified, or distorted,  
may   become hard to recognize. The expected effect is however rather weak in
typical conditions~: the maximum angular displacement of a distant source  can be
estimated to be  less than $1'$ for lensing by a cluster of mass~$\approx
2~10^{13}~M\sol$ and dimensions $\approx ~ $3~Mpc. In a MCM, however, there is also
a non zero probability that the light from a ghost is gravitationnally distorted by 
the object itself (this is expected when the object lies in a principal direction). 
Such     configurations have not been considered up to now at our knowledge,
although they could give raise to curious   effects.   In general, however, the
expected influence of gravitational lensing is      weak enough to be neglected
\cite{Dem87}.

Finally, Êproper velocities, which reach  typically   $V_p \approx 500$~km~s$^{-1}$
or more  for cosmic objects, must be taken into account.  During a time $\approx
\alpha /c$, when  a  light ray  turns around a small universe, the real object moves
by~$\approx V_p~\alpha /c$. Thus the position of the next ghost, in the UC, is
 shifted by the corresponding angle, typically a few arcminutes. For this reason,
any search for ghosts must be performed with a finite spatial resolution of this
order. 

The simplest conceivable   test is the search for possible  ghosts of a specific
class of objects~:  individual  galaxies (possibly with some peculiar
characterisitics);  clusters,   superclusters  or peculiar associations of
galaxies;  quasars, active galactic nuclei, or radiosources; and more generally  any
type of recognizable systems, like for instance peculiar  associations of   objects
(chains, rings, voids,...).  The   first attempts were devoted to the  search    
for  ghost(s)  of an   individual peculiar  object~:  the \mw ~or another typical
galaxy,  the Virgo and  Coma clusters, the Local Supercluster, \etc ~Further
studies, that we also describe in the next sections, have been devoted to the search
of peculiar configurations (coincidences, oppositions, or other) or of periodicities
of images on the sky. 
 
\subsection {Ghost images  of individual galaxies} \label{galaxy}

A fascinating possibility,  in a sufficiently small universe,  lies in the fact that
ghost  images of our own Galaxy should be visible (thus, from outside). Their
number,    distances and orientations depend on the topology.  The nearest  must be
at a distance  $\alpha/2$. So,   not seeing any image of our Galaxy up to a distance
$d$  would allow us to  exclude topologies with $\alpha < 2d$. A sphere of radius
$R$  in the \uc ~contains  between $(\frac{2 R}{\alpha })^k$ and $(\frac{2
R}{\beta})^k$ images, where $k$  is the number of  principal directions. The maximum
distance $d$ up to  which we would be able to recognize our galaxy  has been
discussed by  Sokolov and Shvartsman \cite{Sok74}. They deduce that $d > 7.5~ 
h^{-1}$~ÊMpc, implying $\alpha >15 ~h^{-1}$~Mpc (see also   \cite{Fag86,Fag87}). 
However,    many galaxies are visible in the sky and there is no simple way to decide if such
observed galaxy may be or not a ghost image of the Milky Way. It would thus be very
difficult      to decide  in this way if we do, or do not, live in a MCM.

The main  interest offered by  the \mw  ~comes from its  situation  at the node
between the principal directions  (in a homogoneous MCM), so that it lies
automatically on each of them. This implies an exact quantization, in  proper
comoving distance,  of its  ghost images~:  series of  images of our Galaxy  should
appear  in the principal  directions, having for proper comoving distances   entire
multiples of   $\alpha _i$, the distance to the first ghost in this direction.  This
equality must be considered with a   spatial resolution  depending  on our  proper
motion, see \secn{Searching for  ghosts}. ÊAlso,  most  MCM's  predict that, if a
ghost is present in one direction, an other  must be present in the opposite
direction at similar distance and redshift. Some MCM's also predict ghost images  in
perpendicular (or in other well defined) directions. Thus the observation of similar
images at identical redshifts, and  in related (preferentially opposite, or
perpendicular) directions,  is expected in a  MCM.  This would  also be the case if
images are observed in  a common direction with  quantized    comoving distances. To
apply this latter test, it remains to decide what kind of image must be searched
for, or in other words, what was the appearance of the \mw ~some millions years
ago.  Fagundes \& Wichoski \cite{Fag87}, following an ealier suggestion by
Lynden--Bell, proposed  that our galaxy was a quasar a long time ago. In this case,
any observed quasar could \apriori ~be a ghost image of the \mw. We present in 
\secn{quasars}  their  search for quasar--images in opposite or related   directions
on the sky.  

Fagundes \cite{Fag89}   examined the occurence  of    images of the \mw ~in the MCM
decribed in \secn{toy3},  where space is the Best model. The large number of
principal directions makes this peculiar   geometry      especially favorable.  For
this model,  Fagundes was  able to show that several  conjunctions between source
and  images should be expected,  in specific regions of the sky  distributed along
an equatorial band. The situation would however not be so favorable for other 
models.  Thus, although he expressed the hope that  discordant redshifts could be
accounted for in this way (see \secn{discordant}), no firm conclusion may be drawn
from these studies. 

Presently, no source has been recognized as an image of our Galaxy and this search 
has failed to provide  interesting limits~: the derived constraints are much weaker
than  those derived from other types of objects. It remains thus very few  hope to
use the \mw ~to test in any way the hypothesis of multi--connectedness.

Multi--connectedness would also imply the presence of several   ghost images of any
individual galaxy.  Demian\'nski \& Lapucha searched for instance, without success,
opposite images of galaxies in the sky. However, as  stated by \sos,  it would be
very difficult to recognize that different galaxies  are indeed  images of  an
unique  one.  Moreover,   the spatial coverage of galaxies does not extend very  far
away, so that, compared to   the other possibilities, galaxies do not appear as good
candidates to test   the multi--connectedness of the universe. 

\subsection {Ghost images of clusters and superclusters} \label{clusters}

The brighter -- and the more recognizable -- a type of  objects, the greater 
interest  it offers for   testing multi--connectedness.   \sos ~examined clusters of
galaxies. Contrarily to, \eg, quasars, the lifetimes of clusters seem   sufficiently
long to guarantee an  appearance almost constant during the time necessary for 
light to cross a small Universe. Although the Virgo cluster -- the nearest of the
Abell clusters -- could appear \apriori as an interesting candidate,  it is  not
very rich and not easily   recognizable. The Coma cluster -- the more  prominent in
our neighborhood, and  also the best studied  -- appears therefore more promising.

\subsubsection {The Coma cluster}

Pointing out that Coma    could  be hopefully recognized   from other clusters,  
Gott \cite{Got80}   used it as a candidate for the search of ghost images. It has  
an optical luminosity of $2~ 10^{13} L_{\odot}$ and an elliptical shape. It is 
dominated by the 2 giant galaxies NGC~4874 and NGC~4889, although most of  other
condensed clusters have only one central dominant galaxy. Moreover,  NGC~4889 does
not have many compagnons whereas NGC~4874 has a dozen  satellites. Of course   a
ghost image   would reveal an object older by a look back time $ T \approx \alpha
/c$ taken by   the light  rays   to  turn once around a small Universe (Coma, as we 
known it,   could also   be a ghost itself, in which case  the search could unveil
the real image, younger  by Ê$T$).  How would appear  Coma,   younger or older   by a
time $ T$~?  Using the argument that Coma seems to have  been stable over  billions
of years, Gott  concludes that its appearance would not be too much  different, if
$T \approx $~some~10$^8$~years.  

Coma is about $70~ \hmpc$ from us, in the direction of the north galactic pole.
Estimating that a ghost image of it could not have escaped detection at a  distance
lower than $140~\hmpc$ , \ie,  in a sphere of radius $70~\hmpc$ centered on the
cluster itself, Gott  deduced  $ \beta > 140~\hmpc$,  and $\alpha \geq 60 ~\hmpc$.
Since  Coma is much  richer  than the other close Abell clusters,  we do not expect 
stronger constraints from other {\sl  individual} clusters.

\subsubsection {Other clusters}
 
\sos ~tried to establish constrains from the study of catalogs of clusters.  They
considered the Abell catalog,  which contains 2~712 rich clusters,   and the  Zwicky
catalog, which  contains  9~730 clusters of all types. Both are limited to 
redshifts $\approx 0.2$, corresponding to   about $600~\hmpc$, covering  only the
northern half of space around us.   They concluded that rich clusters detected up to
this distance must be originals since the closer clusters, which  are poorer, 
cannot be identical  objects in a more recent stage of evolution. Applying this
constraint to a toroidal universe, they  concluded  that $\beta > 600~\hmpc$. 
Demia\'nski \& Lapucha~\cite{Dem87} searched, without success, opposite pairs in a
catalog 1889 clusters.

  Gott \cite{Got80} constructed simulations of a $T^3$ universe~:    real galaxies
were disposed in a  cubic  cell  ($\alpha_x = \alpha_y=\alpha_z =  27.5$~Mpc, so
that $\beta = 50$~Mpc).  A numerical code was used to provide a pattern of clustering
and a correlation function in agreement with  observations in the nearby universe.
Ghost images were then calculated in the \uc,   to simulate the appearance of the
universe as it would be seen from a randomly selected point.  He concluded that the
multiple images of  rich clusters could not have escaped  detection and that   a
survey up to magnitude $14$  is able to exclude  values $\beta<$~25~Mpc (for the
case of a torus). Scaling argument allowed him to conclude   that  the corresponding
limit for the Shane--Wirtanen count survey is $\beta \approx $330~Mpc.

Lehoucq \etal \cite{Leh94} devised a different test which is able to detect any type
of multi--connectedness in a catalog of objects (see \secn{lelalu}).Applied it   to
recently compiled   catalogs of  clusters, they   concluded to similar limits. On
the other hand,  Fetisova \etal ~\cite{Fet93}~reported a peak about 125~$\hmpc$ in
the correlation function of rich clusters. According to \cite{Leh94}, this could be
a sign of multi--connectedness with that identification length. But this
characteristic scale, present  in   a catalog of {\sl rich} clusters  ($R \ge 2$)
including 70 objects only,  does not appear for a larger catalog  including less rich
clusters. This evidence is thus  not sufficient in favour of  multi--connectedness. 

\subsubsection {Superclusters} \label{superclusters}

 Beyond clusters, superclusters constitute  the next step in the hierarchy of cosmic
scales.  Gott \cite{Got80}  remarked that the  Serpens--Virgo region, containing
several rich clusters,  constitutes the most prominent large structure,   at a mean 
distance of  $280~\hmpc$.  Arguing that there is no image of this structure nearer 
to us, at least  in the  directions covered by the Shane--Wirtanen survey, he
pointed that there is no image closer than  $200~\hmpc$ from the source itself. He
deduced that $\beta > 400~\hmpc$, close to the limits derived from galaxy
clusters.   In the future,  it will be interesting to search for images of our Local
Supercluster, or of other  recognized superclusters (see \secn{quantiz}). But our  
view of the large scale matter distribution,  faraway from us,  is    presently too
imprecise.  

\subsection {Quasars as ghosts} \label{quasars}

\subsubsection {Quasar associations or oppositions} \label{associations}

Quasars occupy   a large volume of the universe. Their strong optical  luminosities 
--   typically $L\approx 10^{46} $~erg ~s$^{-1}$ -- make them interesting potential
candidates as ghost images of closer galaxies or quasars, since they can be observed
at very  large distance. On the other hand,     the quasar phenomenon is probably 
short--lived compared  to the expected time necessary for a photon to turn around a
small  Universe.   Energetic considerations suggest that their lifetime
$T_{quasar}$  is   shorter than  $10^9$~years. Thus quasars   would notÊ allow to
recognize identification lengths $\alpha$ larger than  $\approx c~ T_{quasar}\leq$
2~000~Mpc. Given the  limits about a few~100~Mpc    already  established for
$\alpha$, quasars would offer an interest for topology only if their lifetimes are
larger than $10^8$~yrs.  These limits could however be trespassed, as remarked by  
\sos, if  their activity takes a  recurrent form.    Also Pa\'al \cite{Paa71}
remarked that, although individual quasars may have a short lifetime, they may be
members of larger associations which survive much longer, and thus could reveal a
possible multi--connectedness of space.

Quasars may be and have been observed   very far away. For some peculiar MCM's,
ghost images    could  appear as quasar associations at large distances (not
necessary with  counterparts at closer distances). A sign of multi--connectedness
could be for instance the  observation of a chain of nearby quasars,   with
progressive redshifts and similar characteristics : they would be successive
snapshots  of a  same object at different moments of its evolution (with  slightly
different  positions   because of   the  proper motions).  Although we may hope the
presence of   such effects,  which would offer positive arguments for
multi--connectedness,   no model allows to predict them with certainty.  Thus, given
our bad knowledge of  the quasar phenomenon and  our ignorance of the   topology,   
quasars cannot  be used to disqualify a possible MCM~: the absence of an expected
effect  may always be attributed, for instance,  to their too short lifetimes. 

The situation is similar with radiosources since their peculiar shapes are probably
strongly modified  during the time necessary for light to cross  the universe.  Thus
there is no hope
  to recognize different images of a same source by a morphological criterion.  
   
Fagundes \& Wichoski \cite{Fag87} examined the possibility that, in a toroidal
universe,  some  quasars  could be past (ghost) images of our own Galaxy (see
\secn{galaxy}). They considered  the {\sl Revised Optical Catalog  of Quasi  Stellar
Objects}Ê \cite{Hew81},  completed by about $1\,500$ sources, in which they searched
for equidistant and oppositely lying QSO's. To take into account   possible 
errors,    proper velocities and   gravitational lensing, they adopted a tolerance 
$\frac{\Delta z}{z}Ê\leq 5\%$ and $\Delta \theta \leq 2\deg$.  They found 32 such
pairs, representing 0.0028 \% of their sample. Monte Carlo simulations of their
model led to the conclusion that this percentage is not significant  enough to allow
any conclusion.  Searching  also for orthogonal images,  they found 2 cases  (with 
the directions [$\alpha=$13~h~7mn,$\delta = -65\deg~ 7']$ and 
[$\alpha=$23~h~7mn,$\delta = +48\deg~ 9']$  for rotation axes). If significant,
these cases would imply $\beta \approx 4\,000 \hmpc$. They also remarked that these
images are not necessarily the closest ones. 
 Demia\'nski \& Lapucha \cite{Dem87} searched opposite pairs of quasars in {\sl The
Catalog of High Redshifts} compiled by Triay~\cite{Tri82}. They found 12 candidates,
a number that thy did not estimate sufficiently significant to conclude.

Narlikar \& Seshadri \cite{Nar85} examined the case of elliptical space $\bbbp
^3$.   They calculated that, in such models,  there is a maximal redshift for
directly observable objects and suggested that this could give an account for the
apparent cutoff in the redshifts of quasars. However, a cutoff at a value $z \approx
4$ would require  an excessively high value $q_0 > 4$. Moreover these authors did
not discussed the fact that, in such a model, ghost images could populate the high
redshift region. 

This leads to the conclusion that  such methods are not very efficient. They can
allow neither to reject nor to support any MCM.  

\subsubsection {The question of discordant redshifts} \label{discordant}

The cosmological interpretation for the redshifts of galaxies, quasars  and distant
objects, is presently widely accepted. There have been  however (and still are a
few)   isolated  claims that some observed  associations of cosmic objects could not
be explained in the framework of the  big--bang cosmology. Such  associations are
defined as  statistically significative reunions of cosmic objects -- galaxies
and/or quasars -- in the same {\sl projected}~region of the sky, although with
different redshifts. Evidence for such associations were for instance presented by
Arp and Hazard \cite{Arp80}  and where reviewed in \cite{Bur80}. It has been
argued     that this could be the sign of  non cosmological redshifts, the objects
being  {\sl physically}   associated despite their different redshifts.  After so
many years,  no convincing explanation    has been however proposed  for the origin
of  non cosmological redshifts and this hypothesis  is usually rejected. Moreover,
the observational  evidence in favour of the reality of such associations is very
poor and controversial. 

 Some authors have remarked that multi--connectedness of the Universe   could offer
an alternative explanation. In simply--connected   \frl~ models,  such
configurations are  highly improbable. But Fagundes  \cite{Fag85,Fag89}  emphasized 
that such situations are naturally   expected   in  some MCM's, without abandoning
the cosmological interpretation of redshifts. This  comes from the fact that 
multi--connectedness modifies  the relation between redshift and distance~: apparent
associations would be due to an accumulation of different ghost   images of a same
physical  source in a given direction of the sky.  Although such effects may be
expected in many   MCM's,   Fagundes restricted  his discussion to the cases
with      negative  spatial curvature, as suggested by the present astrophysical
evidences.  This is however,     unfortunately, the less favorable case, because
of   the  necessarily large  values of the topological scales     (see
\secn{negativemodels}). In addition, the geometry is more complicated, and the
calculations more difficult in this case. Exploration of the same effects in  
MCM's  with zero or  positive spatial curvature, and with a cosmological
constant,    has not  been done at our knowledge.
 
Rather than presenting a complete quantitative discussion,  Fagundes  illustrated
in  this peculiar model the possibility ofÊ images  concentrations,  which could
appear as  discordant associations.  He \cite{Fag85} firstly illustrates this
effect  in the case of the 3--dimensional ``toy--spacetime"  presented in 
\secn{toy2}, where space is the  2--dimensional double torus  $T_2$, a simple example of
compact topology with constant negative curvature.  In this model, the distance
between 2 adjacent images,  \ie,    the length of the smallest geodesic loop in
$T_2$, is $  \approx 3.06~R_0$, compared to  the horizon size $   \approx 6~R_0$. A
source at redshift  $0.7$ would have its first image at redshift $42$.     Fagundes
(following \cite{Mag74}) calculates the elements of the  holonomy group which
transforms the \fp ~into its images and generates a tesselation  of  $\bbbh^2$. This
group also Ê  transforms any  source in  the FP into the ensemble of its  ghost
images.  

In   general  the sources and its images do not coincide into the sky. But Fagundes
remarked that   many geodesics cross  themselves. This gives  rise to images  in the
same, or in opposite directions of the sky.  He concluded  that this model is 
potentially able to generate discordant associations.  But  he calculated also  the
time delay corresponding to the light travel around the  universe,   $\delta t
\approx 2.87~R_0/c$, larger than the Hubble  time. It seems doubtful that   any
cosmic  object could maintain its  nature and appearance over such a long delay. 

In a subsequent paper, Fagundes \cite{Fag89} considered  a 4--dimensional spacetime
whose spatial sections are described by  the Best model of \secn{toy3}. In this
complicated geometry, he was able to  calculate the position of {\sl potential}
associations or conjunctions~: such effects  are expected if a cosmic light--source
lies precisely at one of  the few positions that he found. His main interest
concerned  images of our own Galaxy,  but he   also emphasized that  conjunctions of
quasar images  can  be interpreted as discordant associations. As he pointed out, 
such a situation would provide us with different images of the same quasar at
different periods of its evolution, and thus offer very valuable information about 
the evolution of quasars. From the calculations in his prototype model, and for a
peculiar choice of the geometry,  he derived  favourable positions  for conjunctions
inside a narrow band on the sky. But he remarked also that this is not the case in
general. Thus his conclusion remains qualitative and tentative~: there may be 
discordant associations  due to this effect but there is no convincing evidence,
neither theoretically nor observationally. 

Considering the   low probability that a quasar lies exactly at the right position
to generate this effect,  and given that, even when allowed by the geometry,  the
effect may  be unobservable because of   the chronological constraints (derived from
the study of the toy--model), it seems  that the explanation for the  discordant
associations must be searched elsewhere.  On the other hand it is now believed that 
many apparent  associations   can be explained by   gravitational lensing  effects
which had been previously widely  underestimated. The very rare remaining cases would
be  pure  coincidental projections,  expected in any distribution. 

\subsection{Periodicities in the   distribution   of cosmic  objects}
\label{quantiz}
 
\subsubsection{A large scale periodicity in the galaxy distribution ?}

Most 3--dimensional galaxy--surveys are either shallow ($z<0.03$), gathering a few
thousands objects in a wide solid angle, or narrow, covering a very small solid
angle  up to a large redshift.  The shallow surveys have not shown any sign of
periodicity. Broadhurst \etal \cite{Bro90} have however  performed a very deep
``pencil beam" survey extending to $z \approx 0.5$, in a solid angle smaller than 1
square degree. The  geometrical characteristics of the  resulting 
galaxy--distribution    depends on the spatial curvature of the Universe, since this
is the case for the correspondance between the observed redshifts and the 
distances. Interestingly they remarked that, if $q_0 = \Omega /2 - \lambda = 0.5$,
this distribution shows an apparent  periodicity~: galaxies lie in discrete peaks
separated by   128~$\hmpc$. Subsequent reports (mentionned for instance in
\cite{Hay90})  have been made for periodicities  of  109~$\hmpc$ and 125~$\hmpc$  
in 2 other directions. The original periodicity was revised to the value of
135~$\hmpc$, with a ``best choice" $\Omega =0.1$. No convincing interpretation has
been given yet of this result, whose significance is hard to establish. Thus it is
interesting   to ask if it could  be a  sign of  multi--connectedness of the
Universe.  The characteristic length appears however  shorter than  the present
limits  $\beta > 600 \hmpc$ imposed by clusters, and this makes    such an
interpretation not easy. It also appears very much smaller than the limit derived
from   the \cmb ~ anisotropies (\secn{cmb}) so that it may work only if these limits
may be   reconsidered.  

Hayward \& Twamley \cite{Hay90} have however examined this possibility  in the
framework of the MCM's with negative curvature and minimum volume (see
\secn{minimum}). Since a MCM does not predict periodicity along an arbitrary \los, 
it is extremely unlikely -- as they remarked --  that  galaxies   belonging to one
peak  are ghost images of galaxies of another peak.   But Hayward \& Twamley
suggested  that the results of \cite{Bro90} could be explained by a peculiar model,
where real space is quite devoid of galaxies almost  everywhere, excepted in a large
system which identifies more or less to  the observed ``Great Wall", with a scale of
100--200~Mpc.  Thus the structure observed in  \cite{Bro90} would be the  collection
of the ghosts of the Great Wall. No strong argument is however given to support  
this idea. Moreover, the cluster distribution does not show this characteristic
length (excepted a marginally significative  excess in the correlation function of
rich clusters at a comparable scale, see \cite{Fet93}).  

It remains  thus difficult to account for   the observed quasi--periodicity in the
galaxy distribution in terms of  multi--connectedness.  Given that it is not so
improbable that  an arbitrary \los , when  cutting a ``normal" (non periodic)
distribution of points,  generates a quasi--periodic distribution, a peculiar
explanation may be not absolutely required \cite{Par91}. 

\subsubsection{Periodic redshifts of quasars }

Various authors have reported  an observed periodicity in the distribution of quasar
redshifts~: references can be found for instance in  \cite{Fan90},  beginning with a
paper of Burbidge \cite{Bur68}. Fang \etal \cite{Fan82} claimed an observed
periodicity in the quantity $w = log(1+z)$. Chu and Zhu \cite{Chu89} also reported a
periodicity  in the redshift distribution of the $Ly~\alpha$ absorbing clouds.
Beside a non cosmological interpretaion  for the redshifts  or selection effects, a
multi--connected geometry for the Universe has been invoked as a possible explanation
  \cite{Fan83,Fan87,Fan90}.

In a small Universe, an original object gives rise to a large number of  ghosts
which, in general,  lie onto different lines of sight not directly related to   that
of the original, and without 
 periodicity in their redshift or distance distribution.   

In the very peculiar case of an  object lying onto (or very close to) one of the
principal directions,      ghosts are expected in the same (or related) \los,
with     proper (comoving) distances   periodically distributed (the period
 is $\alpha _i$,    the identification length corresponding to this principal
direction).  
 But only a    very small percentage of objects   (our own position may be the only
case)    lies very near a principal direction. On the other hand,  for any  
observed  ghost, other ghosts are  also expected (although not necessarly
observable) on the same \los,   at proper (comoving) distances which are entire
multiples of the first one. Thus  a periodicity is  associated to any \los, but they
all differ. Thus,  no global periodicity is   expected in the distribution of 
ghosts.

Is there any chance to observe the periodicity   for  those objectsÊ close to a
principal direction~?  Since  the redshift--distance relation is non linear, no
redshift periodicity is expected   (excepted for ghosts   sufficiently close   in
the \uc, so that the linear approximation applies). In addition,  the
periodicities     $\alpha _i$   differ in general  from one    principal direction 
to another, excepted for the   peculiar models where all identification lengths are
equal.  Thus no observable    periodicity is expected, as concluded     for
instance   by Ellis \& Schreiber \cite{Ell86} and   Ellis  \cite{Ell87}. However
\flz~ readressed the question by considering a   small Universe  where space is a
torus  $T^3$.  He claimed that, in this case,  a resonant peak should appear in the
power spectrum of the redshift distribution.    He considered universe models with a
number of cells  inside the horizon $N_{cell} < 500$,   numerated $(l,m,n)$, the FP
being $(0,0,0)$. From an object at the real position $(x_1,x_2,x_3)$, there is a
ghost  in each cell $(l,m,n)$, at a redshift $z_{l,m,n}(x_1,x_2,x_3)$, that he
calculated for a matter--dominated universe. The    dynamical  clustering of objects
and their proper velocities  were  taken into account by a   modification in the
positions of the ghosts,  by a    random quantity  in the range $[0,d]$, in each
dimension. He constructed a synthetic universe with    a cubic \fp ~(for which  the
maximum periodicity is expected) and  $\alpha_1=\alpha_2=\alpha_3=\alpha = 480~\hmpc
=  L_{h} /2.5$. Only 10 original objects were distributed in the FP and their
ghosts  assumed to be visible up to $z=3$. With $d=0$, he generated  a distribution
of 659 ghosts.  As expected from the arguments above,  no periodicity in $z$ could
be found. 

He considered   then another model   with   $d = L_{h}/20$ and a source  now
present  at the observer's position.  Some peaks   appeared in the redshift
distribution of ghosts, in particular  one at $z\approx 2$, that he claimed to be
similar to those really observed. He stated  that the conclusion \cite{Ell87}    
according to which  no $z$--periodicity is expected   applies only when no object is
present nearby the observer.  To express his result,  he   introduced a Fourier
transform and presented the $z$--distribution under the form of a power spectrum. 

Thus,   as Fang pointed himself, a periodicity (or, equivalently, a  peak in the
Fourier distribution) only appears if one, or some, objects are present  nearby the
observer. Although this is the case for his simulation,  it is easy to understand 
that  the whole  signal comes in fact from the ghosts of   that   peculiar object at
the origin~: it is recognizable in his  simulation  only because    the dilution
effect (1 object  over 10) is    artificially very small. For instance,   the
observer sees 6 images of himself at the distance $\alpha$, and 12 images of himself
at a distance $\alpha \sqrt{2}$ ; this is sufficient to explain the observed peak. His model
is thus very special since it invokes a very improbable position of the earth. 

In conclusion, no  global redshift or distance  periodicity   can be expected in the realistic 
MCM's, and there are very few hopes to recognize the   periodicities   which apply to
the ghosts of the rare objects onto  the principal directions. We have discussed in  
\secn{galaxy}  the search for ghosts of an object located at the origin. But if we
assume a density $n$ of real sources in   real space  and a spatial resolution $d$,
the number of objects in the cell is $n~a^3$, compared to  $n~d^3$  in our
neighborhood.  The signal--to--noise ratio of the expected peak is therefore $
N_{cell}~n~d^3 / \sqrt{N_{cell}~n~a^3}$. Since $N_{cell} \le ( 2~L_{h} /
a)^3$, it follows that it is unrealistic to expect an important enough signal to
noise ratio. Recent simulations of a toroidal universe \cite{Leh94} have confirmed
that no observable periodicity in redshifts or distances appears. 

\subsection {An universal statistical method to test   the MCM's} \label{lelalu}

The   tests mentionned up to now  have not given, and are not able to give, very
convincing results, because they suffer from various limitations.  Searches for
images of a peculiar object (the \mw, a peculiar cluster) are limited by  the fact
that these images coulds remain hidden for a peculiar reason not linked to topology
(obscuration, impossibility to recognize the object at  a different age, and seen
under a different orientation).   Such tests   only use a very small part of the
available information on the structure of the Universe.  On the other hand, 
searches for periodicities, or associations of images in related directions, are
also limited~: they  concern only peculiar models,  and they involve   a very small
percentage  of observable  objects.  For this reason, Lehoucq \etal \cite{Leh94}
proposed  a  test with more general validity. Ideally, such a test should be able to
answer to the question of multi--connectedness of the Universe (at a given scale)
independently of the type of topological model assumed (\ie, of the   holonomy of
space).  Rather to try to detect a small population of ghosts  with  peculiar
properties, it is more advantageous to use all the  information distributed among a
whole population of images. This is the philosophy   subtending this direct
holonomy--searching method.

There is a common property  shared by all MCM's~:  in the \uc, a ghost is obtained
from the original   object -- or from the nearest ghost -- by one of the holonomies.
These holonomies are  isometries  analogous  to the translations in $\bbbr ^3$.  In
the case of the torus, for instance, they are the translations by the    3 vectors
$\alpha_i ~\ee_i$ ($i=$1,~2,~3 for $x,y,z$), where the  $\ee_i $  are the 3
unit--vectors in the 3 principal directions, and by the compositions of them. Let us
consider a small universe of  volume $V  \approx (\alpha \beta)^{3/2}$. The test
consists in an histogram of all spatial separations (exactly comoving proper
distances) in a catalog of observed objects, like clusters of galaxies. If the
universe is multi--connected at an appropriate scale, many objects are in fact
ghosts, related to their original by one of these translations. The result is that
peaks corresponding to the translation lengths must appear in the histogram. The
presence of such peaks is the signature of multi--connectedness. 

A complete catalog of observed objects,  having a volume $V_{catalog}$ in the \uc,
must contain a percentage $ V/ V_{catalog}$ of original images, the remaining being
ghosts. Every original object must have,  in the catalog,  a number $\approx
V_{catalog} / V  -1$ of ghost images. Thus, among the $N^2$ pairs of objects in the
catalog, approximatively $\approx V_{catalog} / V  -1$   concern objects related by
one of the 3 basic translations mentionned. If we consider that $N^2 (V  /
V_{catalog})^2 (1- V  / V_{catalog}) $ pairs have separation smaller than
$\sqrt(\alpha \beta) $ and thus do not need to be examined, it is easy to understand
that a high signal is expected, in the sense that a significant number of pairs
should show a vectorial separation $\alpha_i ~\ee_i$. This remains valid for a MCM
with {\sl any topology}.

\begin{figure}
 \centerline{\epsfig{file=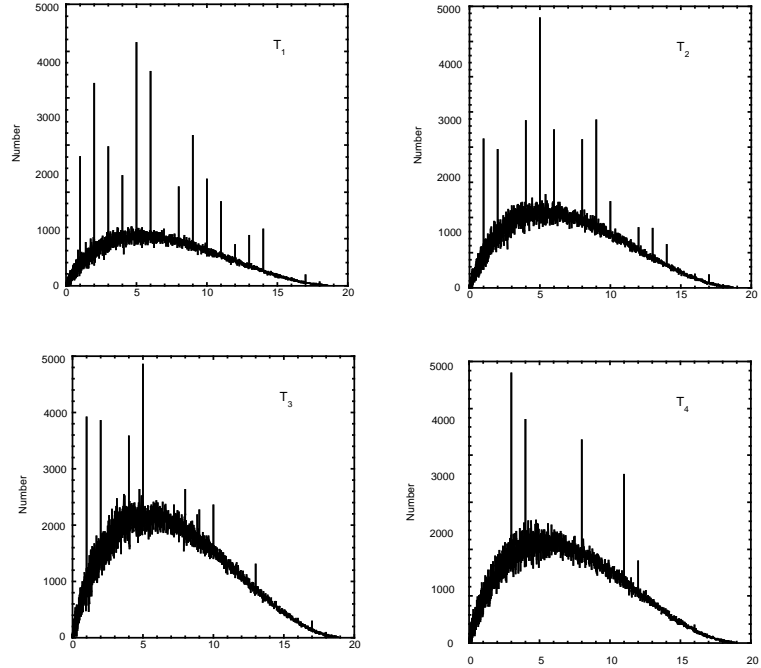, width=10cm}}
  \caption{\label{figlelalu}{\it The histogram of pair
separations (in comoving proper distances) in the galaxy distribution, for a
computer generated toroidal Einstein--de Sitter universe. 50 galaxies are randomly
distributed in the cubic  \fp,  of size 2~500~Mpc. The images are assumed to be visible
up to a redshift $z = 4$. The peaks reveal the  holonomies.   }}
\end{figure}

The basis of the test is just to search for peaks in the histogram. A peak indicates
the  presence of an holonomy with the corresponding scale   $\alpha_i $,
corresponding to a vector $\alpha_i ~\ee_i$. This signature is completely
independent of the type of holonomy.  Numerical simulations of a toroidal universe
have shown effectively that this signature appears very clearly (see figure
\ref{figlelalu}), in conditions where absolutely no periodicity in distance (see
\secn{quantiz}) is recognizable. Thus a negative result will be sufficient to
exclude non trivial topology of any type, at the   scale of the catalog under
examination. In case of a positive result,  immediate further tests would provide
easily the principal directions, the identification lengths,   the type of topology,
and also the curvature radius of space (see \secn{Cosmic models}). The application
of this test to the Abell catalog of clusters     confirms the conclusions of
\secn{clusters} : the universe cannot be multi--connected with a scale $\beta <
600~\hmpc$, whatever its topology. 
 
\subsection {The distribution of gamma--ray bursts} \label{grb}

Gamma--ray bursts were discovered already 20 years ago but their nature and
mechanism remain still unknown.  In particular we have no idea of  their distances
and there is presently no consensus about their nature, galactic or extragalactic.
Although first observations were consistent with a galactic distribution, the recent
results by Meegan \etal ~\cite{Mee92} from  BATSE (Burst And Transient Source
Experiment) suggest a cosmological distribution. However Quashnock \& Lamb
\cite{Qua94} analyzed the angular distribution of detected bursts and concluded that
they appear significantly clustered on angular scales $\approx 4\deg$. From this
they suggested that repeated bursts from the same object occured, what  would 
challenge the possible models for cosmological gamma--ray bursts. The burst
distribution was later analysed by Narayan \& Piran \cite{Nar93} who found also an
excess of antipodal pairs. They concluded that both effects were probably due to
some selection effects. In this spirit,  Maoz \cite{Mao93}  proposed an explanation
as a ``ring bias" due to the satellite localization procedure. 

Alternatively, Biesada \cite{Bie93} suggested that these correlated   bursts could
be due to the multi--connectedness of the Universe, if   their sources  lie on the
principal directions of a small universe. Since the gamma--rays from opposite bursts
reach us at approximatively the same time (compared to cosmological time scales),
this would imply that the considered bursts all lie at   distances $\approx \alpha_i
/2$, $\alpha_i$ being  the identification length along a principal direction. This
restricts strongly the number of sources which may give rise to such effects.
Biesada also correctly remarked that the radiation from a source   near a principal
direction,  and  that from its ghost image, should appear as one bright and one
faint  burst in very close directions.  The two corresponding bursts should have
been emitted with a time delay $\alpha_i /c \approx 10^9~yrs$.  He did not discuss
the probability that the same source should experience two bursts separated by a
time exactly $\alpha_i /c$. Since  these correlated bursts would imply   a very
improbable coincidence, it is unlikely that multi--connectedness is the correct
explanation for the angular correlation of bursts. 

\section{Backgrounds and fields in multi--connected  universes}\label{cmb}

Two different  kinds of studies  have been made concerning the \cmb ~(hereafter
CMB) in relation with the possible  multi--connectedness of the Universe.

 The first
is  concerned with  the  homogeneity of  the Universe, especially as it  appears
through the high isotropy of the  CMB. Although this homogeneity is a {\it
postulated property} of the \frl ~models, the question of its origin has been often
raised, given that it applies to different parts of the Universe which have never
been in  causal contact. Tentative answers   were considered in the framework of the
MCM's. 

The second point of interest concerns   the weak angular fluctuations of the CMB, 
which have now been detected. Such fluctuations are predicted by
the current  models of galaxy and structure  formation, and also by some additional  
processes. It is thus interesting to ask how the situation is modified in a MCM.
Interest toward this question  was recently  renewed thanks to the last results
of the \cobe ~satellite, and recent papers have examined the resulting constraints
for the MCM's. 

\subsection{An early homogeneization of the Universe} \label{Homogeneization}

Since  homogeneity and isotropy are the funding  assumptions
of the \frl ~models, it is a logical evidence that they  cannot find an
{\sl explanation} in the framework of these models themselves, even if their causal
structure has been modified by an early  inflationary phase (see
\secn{Homogeneity}). It is often claimed for
instance  that the initial conditions of the \frl ~models are  ``special" or
``improbable''. But there is no defined  framework  in which these words can be given
a precise meaning. Quantum cosmology, for instance, which suggests a process 
distributing the  initial conditions, is not presently
operative enough to draw firm conclusions. 
 
Concerning inflation, often   suggested as a possible explanation for
homogeneity, the current  models  assume that it occurs in a universe whose
dynamics is already described by a  \frl ~phase,   \ie, with
initial conditions  also   ``special''~: if  the 
pre--inflationary  universe was not already sufficiently homogeneous,  it
has been shown  \cite {Gol92} that  inflation would not have developed  and  led to 
a presently homogeneous universe. In any case,   if the  CMB     homogeneity 
results  from  inflation (although no satisfactory candidate has been proposed for
the ``inflaton"), it   only   reflects  the homogeneity  already present in the
pre--inflationary universe,   at  smaller scales. 

Thus the causality problem is not solved by inflation and this
motivates the search for a causal process  which could have homogeneized  an
initially chaotic (in the sense ``highly inhomogeneous")  universe.   Rees
\cite{Ree72} proposed for instance a   model with chaotic initial conditions, in the sense
that density fluctuations are always of the order of unity at the scale of the
horizon, before the homogeneizing process was active. But this process makes the
universe to  become more homogeneous with time,  over spatial scales comparable to
the (ever increasing) horizon~\footnotemark[1] \footnotetext[1]{This assumes that the dynamics of this
chaotic universe is not too far from that of a \frl ~model with comparable density,
so that the concept of horizon keeps the same meaning}.  This process is however
``violent", so that  a lot of  energy is  generated   when regions of different
densities merge. This energy would be at the origin of   the photons which are now
present in the Universe. But, in such a model, the process of homogeneization  is
not expected to  have  stopped over the cosmic history; in particular,    strong
anisotropies   should have been imprinted over the CMB at the moment of the
\rec , so that this  model cannot account for its observed isotropy.
Rees \cite{Ree72} suggested  that   a transition from   chaotic state to a more
uniform situation could have   occured   at $t \approx t_{eq}$, near the equivalence
between    the matter and radiation densities,  (\ie, before the \rec ~according to
the usual cosmic chronology). 

Ellis \cite{Ell71,Ell79}     pointed out  that such a model would be compatible
with observations   if the Universe is multiconnected~:  a {\it small} Universe
could have become totally causally connected before the \rec ~period. Following
these first ideas,   Gott \cite{Got80} examined the possibility that the Universe
was homogeneized before the \rec~ by some causal process~: as he stated, ``a
multi--connected Universe produces the special initial conditions required by the
Rees  chaotic model." The solution comes from the fact that the spatial volume of
a compact MCM  has a finite and ``small" value $V$ (and all dimensions are  
$< \beta$). Thus after a time   $ \approx \beta /c$, where $c$ is the velocity
of light \footnotemark[1] \footnotetext[1]{The exact relation depends on the dynamics of the model}, the
whole space is contained within  the  horizon, \ie, has become causally connected.
Therefore  a multi--connected universe is  and remains completely   causally
connected from an age  $t \approx \beta /c$. For a sufficiently small universe,    
this occurred before the \rec, so that the  CMB isotropy  corresponds to
the homogeneity of the universe at the \rec,   generated by causal effects.   Gott
\cite{Got80} and \sos ~discussed in more details this possibility.

Gott examined first the simple case of a torus $T^3$ ($\Omega = 1$). The    3
lengths $\alpha _x$,  $\alpha _y$,  and $\alpha_z$  are of the same order
$\alpha$, to allow a sharp transition between chaotic and smooth Universe
\cite{Ree72}. The requirement that the transition  occurs near $t_{eq}$  imposes  
$\alpha /1.15 \approx \beta /2 \approx R_{horizon}(t_{eq})$, the horizon radius at
this period.  In the case where the FP is an hexagonal prism, the sharp transition
condition suggests a side of the hexagon $\approx 1/ \sqrt{3} $ the height
of the prism, with $\alpha /1.31 \approx \beta /2 \approx  R_{horizon}(t_{eq})$. 
He applied  the limits on  $\alpha$ derived from observations of clusters or
superclusters (\secn{clusters}).  The conclusion was  that,   for a toroidal  
space,  thermalization of the CMB is marginally  excluded by observations, but also
marginally admissible.  He also stated that, in other types of  MCM (with non--flat
space),   the geometrical constraints and the value of $R_0$ imposed by observations
do not allow for  thermalization. 
 \sos ~ also examined the question~: with the constraint $\beta \ge
600~\hmpc$, they estimated that   a   \frl ~model   (without inflation) can
only have been homogeneized  at a redshift $z \approx$~10--100 (note
that this redshift varies like $1/\beta$), later than the \rec.  

Hayward \& Twamley \cite{Hay90}~adressed the same question in the framework of the
MCM's with negative spatial curvature having the minimum spatial volume (see
\secn{minimum}). They considered a model based on the 
Weeks--Matveev--Fomenko    manifold with a volume $V_{WMF} = 0.94~R_0^3$
and a maximum value $L_{max}$ for $\alpha$ given by eq. (\ref{VWMF}). They considered
also a model based on an hypothetic manifold of volume $V_{Meyerhoff}=
0.00082~R_0^3$  and  $L_{max}$ given by eq. (\ref{VMey}). 
They stated that homogeneization could have been efficient at the \rec ~if $V <
V_{rec}$, the volume of a causally connected region at the \rec. They conclude that
a model based on the Meyerhoff arguments may do the job if $\Omega < 0.54$,
although the one based on the Weeks--Matveev--Fomenko   manifold requires $\Omega <
0.011$, out of the permitted range. 

These conclusions are based on the hypothesis of   a standard \rec ~at $z_{rec}
\approx  1\,400$. They
would not remain valid   if  the cosmic matter  had suffered a late
reionization. In such a case,    the \cmb ~would originate from, or would have been
 modified at an epoch    much later than  $z_{rec}$. This was emphasized by  Hayward
\& Twamley who concluded optimistically, but without really convincing argument, 
that isotropy of the CMB could be explained in the framework of their models. On
the other hand, Gott remarked that the late 
thermalization of the CMB photons,   required in this case,   is difficult  to
include in the model.  He thus concluded that this is marginally  possible  only if
$\Omega >1$.

Thus, although multi--connectedness indeed modifies  the causal structure of the
Universe, calculations show that it is unlikely  to explain the
spatial homogeneity. This latter must therefore be either admitted as an observed
fact without explanation, or  accounted for by  quantum effects in the early
Universe.  
 
\subsection{The temperature anisotropies of the CMB}

\subsubsection{Temperature fluctuations}

The search for anisotropies in the CMB is an old story. The dipole
anisotropy, detected in 1977, is now well interpreted as the Doppler shift due to
the motion of the Earth with respect to the \lss.  Its existence, magnitude,
and interpretation   remain exactly identical in a MCM.  Beside the
dipole,   anisotropy has been detected by  the DMR instrument on the {\it COBE}
satellite,  at a very low level $\dtt \approx 10^{-5}$, and at large angular scales
$\theta \geq 7 \deg$ \cite{Smoot92}. In particular the quadrupole
component, corresponding to an angle $\theta = 90 \deg$,  appears very low.  It is
generally admitted that  this anisotropy is already present at   the \rec
~epoch. Its level and characteristics may be compared to the  predictions of   the
cosmogonic models, the prototype beeing the cold dark matter  model (hereafter cdm).  In
this frame, an anisotropy  at a given angular scale $\theta$ is related 
to a  fluctuation of
a spatial  scale $L = \frac{\theta}{0.95 \deg}~100~\hmpc$  at the  \rec ~time.
This relation remains valid, in the \uc, for a multi--connected universe, since it 
concerns  observational quantities. But, in this  case,  $L$ may be  greater than
the dimensions of (real)  space, so that no physical fluctuation  exists at this
scale. Various papers have explored the consequences of this situation. 

A tentative detection of the quadrupole anisotropy in the CMB radiation was 
announced in 1981  \cite{Fab81} and  Fagundes \cite{Fag83b} suggested that it could
be due to    multi--connectedness. He estimated the expected effect  in the
framework  of his ``quasi--hyperbolic" model (\secn{quasi--hyperbolic}). The latter 
presents a fundamental anisotropy, described by a parameter $\varepsilon$, which 
induces a quadrupole component on the CMB~: 
\begin{equation} 
	Q =2.7~\frac{\varepsilon}{4}~\mbox{K}. 
\end{equation} 

Fitting   the claimed observational result, he obtained $\varepsilon \approx
1.3~10^{-3}$, from what he predicted  an  anisotropy
in the cosmic expansion, unfortunately too weak to be detected. 
The quadrupole result of 1981 \cite{Fab81} has now been
rejected but it remains  that multi--connected models may break the symmetry of
space, with   interesting consequences for the CMB.
 
In particular,  multi--connectedness   modifies
the relation between angular anisotropies of the CMB and the spatial 
fluctuations
present at the  \rec. This is especially the case for values of the topological
scale    $\alpha$ smaller than the  horizon length at the \rec.  The    \lss ~is  a
spherical surface  of radius $\approx L_{h}$ (the present horizon length),
centered on the observer, in the \uc. Observing  directions
separated by an angle $\theta$ is equivalent to observe a comoving  length $L =
\frac{\theta}{0.95 \deg}~100~\hmpc$ on this surface (thus in the \uc). In 
a MCM,   large values of $\theta$ may correspond to values of   $L$   comparable to,
or   greater than $\alpha$.

The temperature fluctuations $\dtt$   of the CMB are interpreted as the effect of
inhomogeneities  at the \rec ~time. 
They   are usually  developed into spherical harmonics through the formula
\begin{equation} \label{harmonique}  \frac{\Delta T}{T}
({\bf \hat{q}}) =  \sum _{l=2}^{\infty}~  \sum _{m=-1}^{m=l} ~a_{lm}~ Y_l^m({\bf
\hat{q}}),
 \end{equation}
 where   the coefficients  $a_l^m = \int_{4\pi}~d\Omega ~\dtt({\bf \hat{q}})~ Y_l^m({\bf
\hat{q}})$
 characterise the intensity associated to the 
 harmonic $Y_l^m({\bf \hat{q}})$. The vector     ${\bf \hat{q}}$ 
denotes a given line of sight on the sky. Homogeneity of space implies a
global  isotropy of the CMB, so that the moments are rotation--invariant. It is
thus sufficient to consider the terms 

\begin{equation}
		a_l^2 = <\mid a_l^m \mid ^2> =\frac{1}{2l+1}Ê\sum _{m=-l}^l  
~|a_l^m|^2,
\end{equation}

defining the angular power spectrum (the two first components define the   dipole
and  quadrupole).
 
\subsubsection{Density perturbations } \label{gf}

According to the widely admitted gravitational instability scenario, galaxies and
other cosmic structures   result from the collapse     of initially
small density fluctuations. These fluctuations may be treated as small
perturbations superimposed onto    the  strictly  homogeneous \frl ~models. They are
generally expressed by their expansion into eigenfunctions of the 3--dimensional
covariant Laplace operator, usually in Fourier modes. In some respect, a MCM is
equivalent to a SCM with additional periodic boundary  conditions. As a result,
some modes in the expansion are
suppressed, those which do not satisfy these conditions.  Sokoloff
and Starobinski \cite{Sok75} considered these  missing modes  and defined
``G--domains" as the resulting singularities in the distributions of galaxies or
clusters, appearing as  dark and light spots or bands in the sky. It is important to
emphasize that such features are completely different from the 
ghosts previously discussed. 
In the present case,  a special pattern resulting from the multi--connectedness
would be present   {\sl inside} the \fp, and thus be in
principle recognizable even in the case where no ghost is observable.  The authors
concluded, optimistically, that such  features could be detected, or their absence
demonstrated. But their result strongly depends on their particular model, with a
complicated geometry. These effects deserve to be explored in more details, because
they offer the possibility of   observable consequences even if the
multi--connectedness  scale is of the order of the present horizon.  

\subsubsection{Origin of temperature anisotropies}

Temperature anisotropies of the CMB may be  of two kinds. ``Secondary'' anisotropies
are    imprinted on the CMB   later than the \rec. We will not consider
them here, with the idea that multi--connectedness does not modify their
characteristics.  On the other hand, ``intrinsic'' anisotropies are   imprinted by
the fluctuations in   the matter  density at the  \rec. Although three types
of effects (due to the fluctuations in  density, velocities and potential)  
simultaneously contribute,  only the fluctuations in the gravitational potential $\delta
\phi$  (Sachs--Wolfe effect) matter at  angular  scales  beyond $\ 1
\deg$,   leading to temperature anisotropies 
\begin{equation} \label{sw}
 \dtt \approx \delta \phi /3.
\end{equation}
Following the gravitational instability scenario,  the fluctuations $\delta \phi$ of
the gravitational potential   are related to those $\delta \rho $ of the  mass
density  through the Poisson equation. Thus, finally, the statistics of  $\dtt$  
derives  from that of   $\delta \rho$, or from the density contrast  $\delta =
\frac{\delta \rho }{ <\rho>}$.

Usually, the statistical properties of the scalar field  $\delta (\xx) $ (in real
space)  are expressed through its spatial    Fourier  modes   $\delta_k $. 
 In the idealized case  of a  gaussian    statistics,
 the power  spectrum $P(k) = < \delta_k ^2>$  contains all the information concerning
the statistics.    Given a model  for large scale structure 
formation, the coefficients $a_l^2$ may be estimated from the predicted Ê power
spectrum $P(k)$. Most models predict, at least at the scales under concern,  a
power law spectrum $P(k) = <\mid \delta _k \mid ^2> \propto k^n$ (that we assume now
in the spatial range under study). This corresponds to \rms ~average  density and
mass contrasts, at the scale $L$
 $$< \mid \delta \mid ^2>^{1/2} \propto \mid 
\frac{\delta M}{M} \mid ^2> ^{1/2} \propto L^{-\frac{3+n}{2}},$$  and to an average
potential fluctuation $ < \mid  \delta \phi \mid  ^2 > ^{1/2} \propto L
^{\frac{1-n}{2}}$. The corresponding power spectrum for the Fourier transform of the
potential is $ < \mid \phi _k \mid ^2> \propto k^{n-4}$. The formula (\ref{sw})  
allows us to relate the modes $a_l ^2$ to the index $n$. The classical formula
reduces to 

\begin{equation} 
	< a_l ^2> = 16~\pi \sum _k \frac{\mid \delta _k \mid ^2
~j_l ^2 (ky) }{(ky)^4}, 
\end{equation}

where $j_l$ is the $l^{\mbox{th}}$ order spherical Bessel function, and $y$ is
the radius of the \lss, well approximated by the present horizon scale (see
\secn{horizon}). The summation extends over all the Fourier modes denoted by
$\kk$. Multi--connectedness would limit the possible modes.   

\subsection{Influence of multi--connectedness}

Stevens \etal ~\cite{Ste93} and Starobinskii \cite{Sta94} have evaluated this
effect, and compared it to the results of the \cobe ~satellite. In a SCM, the
sum, extending over all values of $k$, may be estimated through an integral and
leads to the classical result 

\begin{equation} \label{thesum}
<a_l^2> = <a_2^2> \frac{\Gamma [(2\ell + n -1)/2]}{\Gamma [(2\ell + 5 -n)/2]}
\frac{\Gamma [(9- n )/2]}{\Gamma [(3+n)/2]}, 
\end{equation}

where $n$ is the slope of the power--law spectrum, and $\Gamma$ the gamma
function. For instance,   the \hz ~value $n=1$
(relevant for the  cdm models) leads to  $$<a_l^2>  \propto
\frac{1}{l(l+1)}.$$

However multi--connectedness modifies  the situation.   When space is compact and 
finite, at least in some directions,   only a restricted collection of
wavevectors $\kk$ are allowed. In the   case of a torus with sides $\alpha
_x, \alpha_y, \alpha_z$, for instance, the allowed vectors have components $k_x
=\frac{2 \pi n_x }{ \alpha _x}, k_y =\frac{2 \pi n_y }{\alpha_y}$, and $k_z =\frac{2
\pi n_z }{ \alpha_z}$,  where $n_x,n_y,n_z$ takes entire values. The sum in
(\ref{thesum}) is thus restricted to these discrete values. This modifies the
spectrum of temperature anisotropies in two respects. First the ratio of the
temperature fluctuations level   at large angular scales over that at smaller scales is
decreased,   because there is no direct source (fluctuations of gravitational
potential) at the larger scales. Second, the dependence on $\theta$ around the
large scales is also modified. Both effects have been considered.  
  
Stevens \etal ~\cite{Ste93} considered the simplest case   $\alpha _x=
\alpha_y= \alpha_z = \alpha$. An  \hz ~spectrum  of fluctuations ($n=1$) is assumed, as
suggested  by scale invariance arguments, like those resulting   from the idea of  
inflation   (in a MCM, inflation could occur in the same conditions than in a SCM).
 This spectrum is  normalized with    the value observed by \cobe ~at the
angular scale $\theta = 18 \deg $.
From this fluctuation spectrum,  they  estimated the statistics of the temperature
anisotropies as a function  of $\alpha$. Their result includes the values $\alpha =
$500, 2~700, 33~000, and 70~000~$   \hmpc$ , corresponding respectively
 to 0.15, 0.8, 1 and 20 times the horizon length at  \rec. They concluded that the 
  \cobe ~observations  could fit their  MCM    only for $\alpha > 2~400~ \hmpc$, 
compared to an horizon size of 3~000 $\hmpc$ for their model. For   other MCM's with
also zero  spatial curvature,  a cubic \fp ~with  identifications after 1   or 3
rotations of 180$\deg$,    they  obtained limits of 1~600 and
$2~900~ \hmpc$ respectively. 

Starobinskii \cite{Sta94} remarked that, for any MCM with dimension much
smaller than the horizon, the power at large angular  scale becomes  much  weaker
than the values observed by \cobe. Moreover the dependence on   $\theta$ must be 
such that   $a_l^2$ remains almost constant, independent of  $l$. This is in
contrast   to the dependance $a_l^2 \propto \frac{1}{l(l+1)}$ predicted by the 
$n=1$ spectrum and  in accordance with the \cobe ~observations. More precisely he
estimated the expected fluctuations to follow

\begin{equation}
a_l ^2 = \frac{2 \pi}{9} \sum _k <\delta \Phi ^2>~(\frac{1}{k L_{rec}})^2,
\end{equation}

where both $k$ and $L_{rec}$, the length of horizon at the \rec, are in
comoving units. From this he concluded, with the 
 same reasoning than Stevens \etal, that   \cobe ~results exclude a  very small
Universe.  ÊFor an identification length much smaller than the present horizon
length, this result is independent   of the peculiar topology, and of the slope of
the  power spectrum. This is due to the fact that the fluctuations   at large angular
scales are created as some ``queue--effect'' of the spatial fluctuations at much smaller
scales (the only existing). In  this case, the constancy of  $a_l$ for large $l$
does not depend on  what  happens at much smaller scales.  

Starobinskii defined another criterion to compare MCM's with the  \cmb ~observations.
He defined the   mode  $l_m$ as that having the largest multipole
amplitude $(\dtt)_l = \sqrt{\frac{2l +1}{4 \pi}}~a_l$.  
\cobe ~results imply $l_m \le 6$. Applying this constraint to toroidal universes
 with three different identification lengths he concluded that the smallest of these
values, $\alpha$, must be larger than $0.75 L_{h} \approx 9~000~$Mpc.  

Starobinskii also considered cylindrical models,  with only one or two
 compact dimensions. In these two cases, he concluded that the 
identification lengths (1 or 2)  must also obey the previous  constraint. But he 
also  remarked that some symmetry (planar or axial, respectively) must be present
in the CMB fluctuations.  Further observations with improved precision may
be able to exclude (or recognize) such symmetries and thus to improve the
constraints on MCM's.

Both papers consider the case of toroidal topology (including the
degenerate cases where only 1 or 2 identification lengths are present). Their result
rely onto the following hypotheses~: \begin{itemize}

\item the Universe is spatially flat with $\Omega=1$.

\item    there is no strong  reionisation after \rec ~with a high optical depth,  so
that the \cmb ~originates from  $z_{\rec} \approx 1400$. 

\item    the \cobe ~anisotropies are intrinsic and due to the \sw ~effect only, with
a   negligible  variance.

\item      the density fluctuations at \rec ~have a gaussian statistics.

\item Starobinskii makes no hypothesis concerning the shape of the fluctuations spectrum.
Stevens \etal ~study similar cases with the additional hypothesis of a  
$n=1$ power law   spectrum.
\end{itemize}

Both    papers offer convincing evidence that, given the adopted
hypotheses,    COBE data exclude MCM's with $\alpha > 0.75 -
0.8~L_{h} \approx 2~300-2~400~\hmpc$ (the given values  are with   $h =.5$).
Thus, with  these hypotheses, the only possible MCMs (for $\Omega =1$)
have very large identification lengths and  do not offer a great interest
from an observational point of view.  If one or two identification lengths are
infinite, interesting possiblities arise which, following Starobinskii, may lead
to peculiar symmetries recognizable in the CMB maps. 

The evidence for a $\Omega =1$ Universe is presently not very strong.  Although
similar   qualitative conclusions can be expected if $\Omega <1$, the precise
constraints probably differ because of the different geometry of the
universal covering, and of the different nature of the holonomies. The calculations
remain to be done and it is not certain   that  the CMB observations bring  tighter
contraints than those imposed by the geometry itself. 
 Moreover, if one is ready to envisage a multi--connected universe, the
question  of the  formation of large scale structures,  as well as  the details of
the cosmic history  should be addressed in a new fashion; 
the   interpretation of the origin and  characteristics of the CMB  might  
differ, and   the observational  constraints  derived above would not necessarily
hold. In our opinion, this maintains alive the hypothesis of a multi--connected
universe  with identification lengths smaller than horizon. In particular it
seems still justified to search for more direct constraints derived from the
apparent distribution of discrete objects at large scale.

\subsection{Cosmic magnetic fields} \label{magnetic}

Sokoloff \cite{Sok75} and collaborators  \cite{Ruz75} examined the relation between 
possible cosmic magnetic fields and multi--connectedness of space. The presence of
large scale intergalactic magnetic fields is suggested by the observed Faraday
rotation of distant extragalactic sources  (for a more recent review on
extragalactic fields, see \cite{Kro94}).

Effects due to  multi--connectedness are expected if there exists a
magnetic field of constant magnitude and direction over a very large cosmic scale
$\lambda$. The strength, which may be $\approx 10^{-9}$~G, plays no role in the
study, only the direction and the coherence scale are important. The main idea is 
that, since both multi--connectedness and a cosmic magnetic field   break the
isotropy of space,   some properties of   the   anisotropies introduced    should
coincide. 

Sokoloff  and collaborators  worked in the context of the barrel model, defined in
\secn{barrel},  which provides  a good local approximation of a MCM with negative
spatial curvature. The case $a =0$, corresponding  to a cylindrical universe
(thus with flat space),  is examined in  \cite{Ruz75}, the case  $a \neq 0$ in
\cite{Sok75}. The authors suggested to identify  the homogeneity
scale $\lambda$ of the magnetic field and the identification scale $h$ of the
barrel model. Also, they suggested that the principal directions
(the axis) of the barrel model may coincide with the direction of the magnetic
field; in their view,        multi--connectedness and cosmic
magnetic fields could have related origins  in    the early universe, what would
explain their coinciding  properties. In such a case,      
the observed  direction of  a cosmic magnetic field  would unveil  the principal
direction for multi--connectedness, and thus be a precious guide for a search for
ghosts.  The present observations of
extragalactic magnetic fields have however   not provided such   indice yet.

\section{Conclusion}

The hypothesis of a  multi--connected  space  widens considerably the variety
of universe models obeying the cosmological principle. Although most
characteristics of the usual \frl ~solutions  are preserved, new and original
effects appear. 
Interesting  consequences  fall into two categories. First, for an   identification
length   smaller than the horizon, direct observable effects are to be expected 
onto the appearance of the extragalactic universe. Second, the
theoretical interest remains fundamental  even if  the identification length is
comparable or greater than the horizon, in view of the fact that the topology has an
important influence on the states of quantum fields. 

 Directly observable effects  may be expected only if the identification length
is reachable by present observations. We have reviewed the present
observational constraints. The first one,  obtained from the distribution of
clusters or superclusters,  limits $\alpha$ -- the shortest
circumference of the universe -- to a few 10~$\hmpc$ and $\beta$ -- the
maximum dimension inscriptible in real space -- to a few 100~$\hmpc$. These 
limits leave room  for many observable effects in the
distribution of galaxies, clusters, superclusters, quasars, etc.  No  convincing
result has been obtained today  from the quasar distribution,
and no ``exotic effect'' is convincingly explained by this hypothesis. Other
limits, obtained from the \cmb ~observations, are   more stringent, since their
scales are  comparable to that of the horizon,  forbidding therefore  any
directly observable effects. However, given the hypothesis on which it relies,
there is some hope that it can be overpassed.  For the future, the only observational test
which can firmly establish a non--trivial topology is the statistical analysis of reciprocal 
distances between celestial objects in the universal covering space. 

In any case, none of the observational 
constraints limit the play of multi-connectedness in the early universe.  On one hand, the
perturbations which have led to the formation of galaxies and large cosmic
structures are thought to originate from primordial quantum fluctuations.  On the other
hand, the fundamental states of the fields themselves play an important role
in  cosmology. The most famous example is inflation, but other processes  only
marginally explored up to now, like the Casimir effect, may
also influence the cosmic dynamics. In all these aspects,
a non--trivial topology would have  major consequences. The conclusions from the
(still tentative) quantum cosmology are even stronger : they   favour  a multi--connected
rather than a simply--connected universe. Thus, at least from a theoretical point of view,
the field of cosmic topology  appears not closed but, on the contrary, in a
promising state of development.

\vspace{5cm}

{\bf Acknowledgements}  

We benefited from discussions with R. Hakim. We deeply thank D. Sokoloff for
his numerous suggestions and references, and the referee for his comments. Part of this
review is a continuation of the DEA project completed by Miss M.--A. Treyer at the
University of Paris~VII.


\end{document}